\documentclass[preprintnumbers, floatfix, letterpaper, twocolumn,aps,prd,epsfig,nofootinbib,natbib,longbibliography]{revtex4-1}

\usepackage{graphicx}
\usepackage{epstopdf}
\usepackage{latexsym}
\usepackage{amssymb}
\usepackage{amsmath}
\usepackage{color}
\usepackage{float}
\usepackage{mathrsfs}
\usepackage{multirow}
\usepackage{tabularx,booktabs}
\newcolumntype{Y}{>{\centering\arraybackslash}X}

\usepackage[center]{subfigure}
\usepackage{makecell}
\setcellgapes{4pt}
\begin{document}

  \renewcommand\arraystretch{2}
 \newcommand{\bq}{\begin{equation}}
 \newcommand{\eq}{\end{equation}}
 \newcommand{\bqn}{\begin{eqnarray}}
 \newcommand{\eqn}{\end{eqnarray}}
 \newcommand{\nb}{\nonumber}
 \newcommand{\lb}{\label}
 \newcommand{\cb}{\color{blue}}
    \newcommand{\cc}{\color{cyan}}
        \newcommand{\cm}{\color{magenta}}
\newcommand{\rc}{\rho^{\scriptscriptstyle{\mathrm{I}}}_c}
\newcommand{\rd}{\rho^{\scriptscriptstyle{\mathrm{II}}}_c} 
\newcommand{\PRL}{Phys. Rev. Lett.}
\newcommand{\PL}{Phys. Lett.}
\newcommand{\PR}{Phys. Rev.}
\newcommand{\CQG}{Class. Quantum Grav.}

\title{Qualitative dynamics and inflationary attractors in loop cosmology}
\author{Bao-Fei Li $^{1,2}$}
\email{Bao-Fei$\_$Li@baylor.edu}
\author{Parampreet Singh$^3$}
\email{psingh@phys.lsu.edu}
\author{Anzhong Wang$^{1, 2}$\footnote{The corresponding author}}
\email{Anzhong$\_$Wang@baylor.edu}
\affiliation{$^{1}$Institute for Advanced Physics $\&$ Mathematics,
Zhejiang University of Technology, Hangzhou, 310032, China\\
$^2$GCAP-CASPER, Department of Physics, Baylor University, Waco, TX, 76798-7316, USA\\
$^3$ Department of Physics and Astronomy, $\&$ Center for Computation and Technology, Louisiana State University, Baton Rouge, LA 70803, USA}


\begin{abstract}

Qualitative dynamics of three different loop quantizations of spatially flat isotropic and homogeneous models is studied using effective spacetime description of the underlying quantum geometry. These include the standard loop quantum cosmology (LQC), its recently revived modification (referred to as mLQC-I), and another related modification of LQC (mLQC-II) whose dynamics is studied in detail for the first time. Various features of LQC, including quantum bounce and pre-inflationary dynamics,  are found to be shared with the mLQC-I and mLQC-II models.  We study universal properties of dynamics for chaotic inflation,  fractional monodromy inflation, Starobinsky potential, non-minimal Higgs inflation, and an exponential potential. We find various critical points and study their stability, which reveal various qualitative similarities in the post-bounce phase for all these models. The pre-bounce qualitative dynamics of LQC and mLQC-II turns out to be very similar, but  is strikingly different from that of mLQC-I. In the dynamical analysis, some of the fixed points turn out to be degenerate for which center manifold theory is used. For all these potentials, non-perturbative quantum gravitational effects always result in a non-singular inflationary scenario with a phase of super-inflation succeeded by the conventional inflation. We show the existence of inflationary attractors, and obtain scaling solutions  in the case of the exponential potential. Since all of the models agree with general relativity at late times, our results are also of use in classical theory where qualitative dynamics of some of the potentials has not been studied earlier.

\end{abstract}

%
%

\maketitle

\section{Introduction}
\label{Intro}
\renewcommand{\theequation}{1.\arabic{equation}}\setcounter{equation}{0}

In classical cosmology, inflationary paradigm plays an important role in the resolution of many long-standing problems of the standard big-bang model,
 by assuming that the universe undergoes an exponential expansion sourced by a single or multiple scalar fields in the very early epoch. However, classical 
 general relativity (GR) is an incomplete theory, breaks down when spacetime curvature reaches the Planck scale, and inflation itself is past-incomplete \cite{borde}, 
 due to the big bang singularity where all physical quantities become infinite. 
Hence, the initial conditions of inflation in the classical theory are usually imposed at the onset of inflation. What  happened in the 
pre-inflationary phase becomes an interesting question,
 which can only be reliably addressed in a theory where the big-bang singularity is removed. In the last one and half decades, loop quantum gravity (LQG)  has been  rigorously applied to 
 understand singularity resolution in various cosmological spacetimes (for a review see Ref. \cite{asrev}) and black holes (see, e.g.  Ref. \cite{aos1}). Results are very encouraging, signaling a generic resolution of all strong curvature singularities in various cosmological spacetimes \cite{ps09,ps14}. The basic picture in loop quantum cosmology (LQC), a quantization of homogeneous and isotropic spacetimes using LQG turns out to be the following. At large volumes, there is an excellent agreement between GR and LQC, but as curvature approaches the Planck scale significant differences arise. Unlike GR, the big bang singularity is absent in LQC and  replaced by a quantum bounce due to non-perturbative quantum gravitational effects which bound the energy density by a universal value  \cite{aps,aps3,slqc}. At the fundamental level, the quantum dynamics in LQC is governed by a finite difference equation which for a wide variety of states can be approximated extremely well by an effective dynamics \cite{ps12,numlsu-2,numlsu-3,numlsu-4}. In LQC, the effective Hamiltonian results in modified Friedmann-Raychaudhuri (FR) equations which have quadratic terms in energy density. Implications of these corrections to GR equations have been widely studied to understand the Planck scale phenomenology (see, e.g. \cite{as} for a review).

Due to ambiguities in the quantization procedure, different effective Hamiltonians can result for homogeneous spacetimes in loop cosmology.\footnote{Here ``loop cosmology,''  refers collectively to various possible loop quantizations of cosmological spacetimes in the framework of LQG and should not be confused with LQC, here defined by the model developed in Refs. \cite{aps,aps3,slqc}, which is an example of such a quantization.} About a decade ago, Yang, Ding and Ma found two variants of LQC and obtained two different effective Hamiltonians. Both of these Hamiltonians were derived by treating the Lorentzian term in the Hamiltonian constraint independently from the Euclidean by using Thiemann's regularization of Hamiltonian constraint in LQG \cite{thiemann}. In contrast, for spatially flat models in the standard LQC, the Lorentzian term is written in the same form as Euclidean term and combined with the latter before quantization. We would refer to the  two modified versions of LQC found in Ref. \cite{YDM09} as mLQC-I and mLQC-II, respectively. These are different from each other in the way Lorentzian term is quantized. In mLQC-I, the extrinsic curvature in the Lorentzian term is directly expressed in terms of holonomies using an identity on the classical phase space. Whereas in mLQC-II, proportionality between extrinsic curvature and the Ashtekar-Barbero connection is used before expressing it in terms of holonomies. Unlike LQC where the quantum Hamiltonian constraint is a second order finite difference equation, mLQC-I and mLQC-II result in a fourth order quantum difference equation. For mLQC-I, all of the four roots are important for a consistent physical evolution where as for mLQC-II two of the roots are unphysical \cite{ss18b}.  

It should be noted that to the leading order the effective Hamiltonian of mLQC-I is identical to the one recently obtained from  the 
expectation values of the Hamiltonian operator in  LQG  with the help of complexifier coherent states in the homogeneous and isotropic 
Friedmann-Lema\^{i}tre-Robertson-Walker (FLRW) spacetime \cite{DL17,adlp}. Various aspects of the cosmological dynamics resulting from this 
effective Hamiltonian and modified FR equations were first obtained by the current authors in Ref. \cite{lsw2018}. These equations confirmed 
that the evolution of the universe in this model is asymmetric about the bounce, in contrast to the standard LQC. The existence of 
asymmetric bounce is tied to the necessity of two different branches corresponding to two pairs of physical roots of the quantum difference 
equation for a consistent evolution. Further, unlike the standard LQC, the modified FR equations include higher order terms than those 
quadratic ones in energy density \cite{lsw2018}.

Though mLQC-I has recently gained renewed attention \cite{DL17,adlp,lsw2018,agullo18}, little has been known so far about the dynamics of  
mLQC-II except for a scant observation that the bounce is symmetric \cite{YDM09}. In particular, important details of modified Friedmann 
dynamics and Planck scale physics have remained unknown. This gap is filled in our current work. In particular, a detailed study on the 
cosmological dynamics in mLQC-II is given in the next section where the modified  FR equations and various features of resulting dynamics 
are studied. In contrast to mLQC-I, we find the evolution in mLQC-II to be quite similar to LQC even though there exist higher order terms 
than quadratic in energy density in the modified FR equations. Our detailed investigations confirm that the quantum bounce is 
symmetric in mLQC-II. Unlike in mLQC-I where the pre-bounce branch results in a constant Planckian spacetime curvature regime, similar to a 
quantization of the Schwarzschild interior in LQC \cite{djs}, the pre-bounce branch in mLQC-II results in a classical universe as in LQC. 
The post-bounce branch of the three models all results in the classical GR limit at late times. 

Apart from understanding in detail the dynamics of mLQC-II and putting it on the same footing as LQC and mLQC-I, our main objective in this 
paper is to investigate pre-inflationary dynamics for various potentials in these three models. The potentials we consider are: chaotic 
($\phi^2$) inflation, fractional monodromy inflation, Starobinsky inflation, non-minimal Higgs inflation and inflation with an exponential 
potential.  Qualitative dynamics of chaotic and exponential potentials have been studied extensively in GR. Other potentials have not been 
well studied even in the classical theory. In LQC, the situation is similar with most of the works only on chaotic inflation  
\cite{svv,rs2012,aan, ashtekar-sloan, corichi, bg2016, ZWCKS17, zwcks2017, Shahalam,SSW18}, and remain to be so far addressed in mLQC-I  and 
mLQC-II. Since analytical solutions are difficult in presence of potentials, universal features of dynamics can be understood via phase 
space analysis and studying critical points. The qualitative dynamics reveals various features of these three models showing 
similarities in the post-bounce phases and differences between LQC and mLQC-II from mLQC-I. We study stabilities of fixed points in all the 
cases and show the existence of inflationary attractors for all three models. We show that all of the models considered here in loop 
cosmology make the past of inflation  complete. In addition, since dynamics of LQC, mLQC-I and mLQC-II agrees with GR in the post-bounce 
epoch, our investigation fills a gap in the classical theory on study of qualitative dynamics of fractional monodromy, Starobinsky potential 
and non-minimal Higgs potential.  In particular, various subtleties so far unnoticed for qualitative dynamics of Starobinsky and 
non-minimal Higgs potential are rigorously understood. 

The paper is organized as follows. In Sec. II we begin with obtaining the effective dynamics from effective Hamiltonian for mLQC-II. In particular, the modified FR equations are obtained for the first time. Sec. II also summarizes effective dynamics of LQC and mLQC-I, and compares numerical solutions with mLQC-II. In Sec. III, we focus on the qualitative analysis of dynamical systems resulting from the modified FR equations for $\phi^2$, fractional monodromy, Starobinsky, non-minimal Higgs and exponential potentials. We perform detailed phase space analysis, find various critical points and study their stabilities. We show the existence of inflationary attractors in the post-bounce epochs, and find scaling solutions for exponential potential. We summarize our results in Sec. IV. In the Appendix, we discuss details of stability analysis for non-linear systems which is relevant for fixed points in Starobinsky potential, as well as non-minimal Higgs potential.

\section{Effective Dynamics: Hamilton's and Friedmann-Raychaudhuri Equations}
\label{Section2}
\renewcommand{\theequation}{2.\arabic{equation}}\setcounter{equation}{0}

Due to the underlying quantum geometry, the fundamental description in loop cosmology is governed by a quantum difference equation which is the quantum Hamiltonian constraint. For states resulting in macroscopic universes at classical scales, it is possible to derive an effective spacetime description where the governing equations are a modified version of classical FR equations. 
The effective description has been rigorously tested in LQC using numerical simulations \cite{ps12} and turns out to be an excellent approximation for isotropic \cite{numlsu-2,numlsu-3} and anisotropic spacetimes \cite{numlsu-4}. Assuming the validity of the effective spacetime description, in the following we discuss the effective dynamics of LQC, mLQC-I and mLQC-II. We start with the latter as this is not yet derived in detail in the existing literature. With the effective Hamiltonian we shall obtain the modified FR equations via  Hamilton's equations. In the succeeding subsections, for the sake of reader's convenience,  we also summarize such equations for LQC and mLQC-I. For a detailed discussion of the effective dynamics of LQC, see Refs. \cite{ps06,aps2006}, while for mLQC-I we refer the reader to Ref. \cite{lsw2018}.

\subsection{Effective Dynamics of mLQC-II}

 In LQG,  the elementary classical phase space variables for the gravitational sector are the SU(2) Ashtekar-Barbero connection $A^i_a$ and  the conjugate triad $E^a_i$. With Gauss and spatial diffeomorphism constraints fixed, in the homogeneous  and isotropic universe the only relevant constraint is the Hamiltonian constraint whose vanishing yields the physical solutions. The gravitational part of the Hamiltonian constraint is a sum of the Euclidean and Lorentzian terms, given by
\bq
\mathcal{C}_{\mathrm{grav}} = {\cal C}_{\mathrm{grav}}^{(E)} - (1 + \gamma^2) {\cal C}_{\mathrm{grav}}^{(L)},~
\eq
with $\gamma$ as the Barbero-Immirzi parameter. To fix the value of $\gamma$ we follow Ref. \cite{immirzi} which fixes  
$\gamma \approx 0.2375$ by  matching the Hawking's semiclassical black hole entropy formula with the counting of the number of spin network 
states corresponding to the area of black hole horizon. 
With the choice of the lapse function being unity, the Euclidean part is 
\bq
\mathcal{C}_{\mathrm{grav}}^{(E)} = \frac{1}{2} \int \mathrm{d}^3 x \, \epsilon_{ijk} F^i_{ab} \frac{E^{aj} E^{bk}}{\sqrt{\mathrm{det}(q)}}, 
\eq
where $F_{ab}^i$ is the field strength of  connection $A^i_a$ and $|\mathrm{det}(q)|$ is the determinant of the spatial metric compatible with the triads. The Lorentzian part in the gravitational Hamiltonian is given by
\bq
\mathcal{C}_{\mathrm{grav}}^{(L)} =   \int \mathrm{d}^3 x \,   K^j_{[a} K^k_{b]}   \frac{E^{aj} E^{bk}}{\sqrt{\mathrm{det}(q)}}   ~,
\eq
where $K^i_a$  is the extrinsic curvature. Upon quantization, ambiguities can arise resulting from different treatments of the terms in the Hamiltonian constraint. 
 In LQC, the Euclidean and Lorentzian terms are treated on the equal footing due to the symmetry reduction at the classical level which makes the Lorentzian term a multiple of the Euclidean one.  However, this is not the case if Lorentzian term is treated independent from the Euclidean one in the quantization process which results in a different Hamiltonian than the standard LQC. Two resulting Hamiltonians, yielding mLQC-I and mLQC-II,  have been pointed out in literature, first in Ref. \cite{YDM09}. The detailed effective dynamics of mLQC-I was presented in Ref. \cite{lsw2018}, and for mLQC-II it will be  presented for the first time in this paper. A key difference in the effective dynamics of mLQC-I and mLQC-II in comparison to LQC is the presence of higher order terms than quadratic in energy density in the modified FR equations. Further, for both, the quantum Hamiltonian constraint is a fourth order difference equation \cite{YDM09,ss18b}, unlike the second order one in LQC.  The Hamiltonian in mLQC-II arises 
 from the substitution $K^i_a= A^i_a/\gamma$ in the Lorentzian term. Such a substitution is allowed due to the symmetries of spatially flat isotropic spacetime which permit treating extrinsic curvature as a connection. 
 Then, the resulting effective Hamiltonian ${\cal H} = ({\cal C}_{\mathrm{grav}} + {\cal C}_{\mathrm{M}})/8 \pi G$ is \cite{YDM09}:
 \bqn
\lb{Ha}
\mathcal H=-\frac{3|p|^{3/2}}{2\pi G\lambda^2\gamma^2}\sin^2\left(\frac{\bar \mu c}{2}\right)\Big\{1+\gamma^2\sin^2\left(\frac{\bar \mu c}{2}\right)\Big\}+\mathcal{H}_M.\nb \\
\eqn
 Here $c$ and $p$ are the symmetry reduced connection and triad variables which satisfy $\{c,p\}={8\pi G\gamma}/{3}$, 
$\bar \mu\equiv \sqrt{\Delta/|p|}$ \cite{aps3}, with $\Delta = \lambda^2 = 4\sqrt{3}\pi\gamma \ell_{\mathrm{Planck}}^2$ being the smallest nonzero
eigenvalue of the area operator in LQG, and $\mathcal{H}_M$ represents the Hamiltonian of the matter sector. The modulus sign on the triads arises because of their two orientations which will be fixed in the following to be positive. 

It turns out that a canonical transformation to a new set of gravitational phase space variables $b=c/p^{1/2}$ and $v=p^{3/2}=v_o a^3$ is particularly useful \cite{slqc}. Here $v_o$ is the fiducial volume of the fiducial cell introduced on the spatial manifold $\mathbb{R}^3$
 to define symplectic structure and $a$ denotes the scale factor of the universe. In terms of these variables, the effective Hamiltonian ${\cal H}$ becomes
\bq
\lb{2.1}
\mathcal H=-\frac{3v}{2\pi G\lambda^2\gamma^2}\sin^2\left(\frac{\lambda b}{2}\right)\Big\{1+\gamma^2\sin^2\left(\frac{\lambda b}{2}\right)\Big\}+\mathcal{H}_M.
\eq 
Then, the resulting Hamilton's equations are:
\bqn
\lb{LQG2a}
\dot v&=&\Big\{v, \mathcal H\Big\}=\frac{3v\sin(\lambda b)}{\gamma \lambda}\Big\{1+\gamma^2-\gamma^2\cos\left(\lambda b\right)\Big\},\\
\lb{LQG2b}
\dot b&=&\Big\{b, \mathcal H\Big\}=-\frac{6\sin^2\left(\frac{\lambda b}{2}\right)}{\gamma \lambda^2}\Big\{1+\gamma^2\sin^2\left(\frac{\lambda b}{2}\right)\Big\}-4\pi G\gamma P\nb\\
&& ~~~~~~~~~ = -4\pi G\gamma (\rho+P),
\eqn
where the pressure $P$ and energy density $\rho$ are defined respectively by $P\equiv -{\partial \mathcal H_M}/{\partial v}$ and $\rho \equiv \mathcal H_M/v$.   In order to obtain the modified FR equations, 
one can first express the Hubble parameter and acceleration of the scale factor in terms of the phase space variables $v$ and $b$. The Hubble parameter $H\equiv \dot a/a= \dot v/3v$, once plugged into Eq.(\ref{LQG2a}), yields
\bq
\lb{2.3}
H^2=\frac{\sin^2(\lambda b)}{\gamma^2\lambda^2}\left(1+\gamma^2-\gamma^2 \cos\left(\lambda b\right)\right)^2.
\eq
From the time derivative of Hubble parameter one finds, 
\bq
\lb{2.4}
\frac{\ddot a}{a}=H^2+\frac{\dot b}{\gamma}\Big\{\cos(\lambda b)+\gamma^2 \cos(\lambda b)-\gamma^2 \cos(2\lambda b)\Big\}.\nb\\
\eq
The vanishing of the Hamiltonian constraint yields,
\bq
\rho=4 \rho_c \sin^2\left(\frac{\lambda b}{2}\right)\left[1+\gamma^2\sin^2\left(\frac{\lambda b}{2}\right)\right],
\eq
which can be inverted to obtain,
\bq
\lb{2.5}
\sin^2(\lambda b_\pm/2)=\frac{-1\pm\sqrt{1+\gamma^2\rho/\rho_c}}{2\gamma^2},
\eq
where $\rho_c=3/[8\pi G \lambda^2\gamma^2]$ is the critical energy density in LQC \cite{aps3}. Only one of the roots $b_+$ is physically viable since $b$ must be real (Note that in the classical limit of Eq.(\ref{2.3}), $b$ is proportional to the Hubble rate). Therefore, the root $b_-$ will be discarded in the following analysis. The situation here is quite different from mLQC-I \cite{lsw2018}, where both of the branches are physically relevant to describe a  complete  and continuous evolution of the universe across the bounce.  Now,  substituting the $b_+$ branch of Eq.(\ref{2.5}) into Eqs.(\ref{2.3})-(\ref{2.4}), one immediately arrives at the modified FR equations:
\begin{widetext}
\bqn
\lb{2.6a}
H^2
&=&\frac{8\pi G \rho}{3}\left(1+\gamma^2 \frac{\rho}{\rho_c}\right)\left(1-\frac{(\gamma^2+1)\rho/\rho_c}{ (1+\sqrt{\gamma^2\rho/\rho_c+1})^2}\right),\\
\lb{2.6b}
\frac{\ddot a}{a}
&=&-\frac{4\pi G}{3}\left(\rho+3P\right)-4\pi G P\left[\frac{3\gamma^2+1-2\sqrt{1+\gamma^2\rho/\rho_c}}{1+\sqrt{1+\gamma^2 \rho/\rho_c}}\right]\frac{\rho}{\rho_c}\nb\\
&&-\frac{4\pi G \rho}{3}\left[\frac{7\gamma^2-4\gamma^2\rho/\rho_c-1+(5\gamma^2-3)\sqrt{1+\gamma^2\rho/\rho_c}}{(1+\sqrt{1+\gamma^2\rho/\rho_c})^2}\right]\frac{\rho}{\rho_c}.
\eqn
\end{widetext}
Using the above equations, one can show that 
 \bq
\lb{2.7}
\dot \rho+3H (\rho+P)=0, 
\eq
i.e.,  the matter-energy conservation law still  holds in mLQC-II without any change in properties of equation of state (as in LQC and mLQC-I).\footnote{The reason for this is tied to the fact that there is no quantum geometric contribution to the matter part of the Hamiltonian. Such a contribution can be included in addition, using inverse volume effects, which changes the equation of state but the matter-energy conservation law still holds albeit with quantum gravitational changes to $\rho$ and $P$ \cite{effectivew}.}  

From the modified FR equations we find that  the quantum bounce occurs at the critical density when $H=0$ and $\ddot a >0$, which fixes the value of  the critical density $\rho_c^{{\scriptscriptstyle{\mathrm{II}}}}$ to be 
\bq
\lb{2.8}
\rho_c^{{\scriptscriptstyle{\mathrm{II}}}}=4(\gamma^2+1)\rho_c.
\eq
Note that the critical density $\rho_c^{{\scriptscriptstyle{\mathrm{II}}}}$ should not be confused with the critical  density $\rho_c$ in LQC  and the one $\rho_c^{{\scriptscriptstyle{\mathrm{I}}}}$  in mLQC-I.
Due to the quantum bounce,  the big bang singularity is also resolved in this model, similar to LQC and mLQC-I.  Meanwhile, the bounce is accompanied by a phase of super-inflation, i.e. $\dot H > 0$. To determine the energy density for this phase, 
we first find  $\dot H$ using Eqs.(\ref{2.5}) and  (\ref{2.6a})-(\ref{2.6b}):
\begin{widetext}
\bq
\lb{2.8}
\dot H=\frac{4\pi G (\rho+P)}{\gamma^2}\left(2\gamma^2+2\gamma^2 \rho/\rho_c-3\gamma^2\sqrt{1+\gamma^2\rho/\rho_c}+3-3\sqrt{1+\gamma^2\rho/\rho_c}\right).
\eq 
\end{widetext}
From the vanishing of $\dot H$, we find that the super-inflation occurs for $\rho > \rho_s$ where  
\bq
\rho_s=\frac{\rho_c}{8 \gamma^2}\left(3(\gamma^2+1)\sqrt{1+2\gamma^2+9\gamma^4}+9\gamma^4+10\gamma^2-3\right).
\eq
For $\gamma=0.2375$, we find   $\rho_s=0.5132\rho_c^{{\scriptscriptstyle{\mathrm{II}}}}$.  

\begin{figure} 
{
\includegraphics[width=8cm]{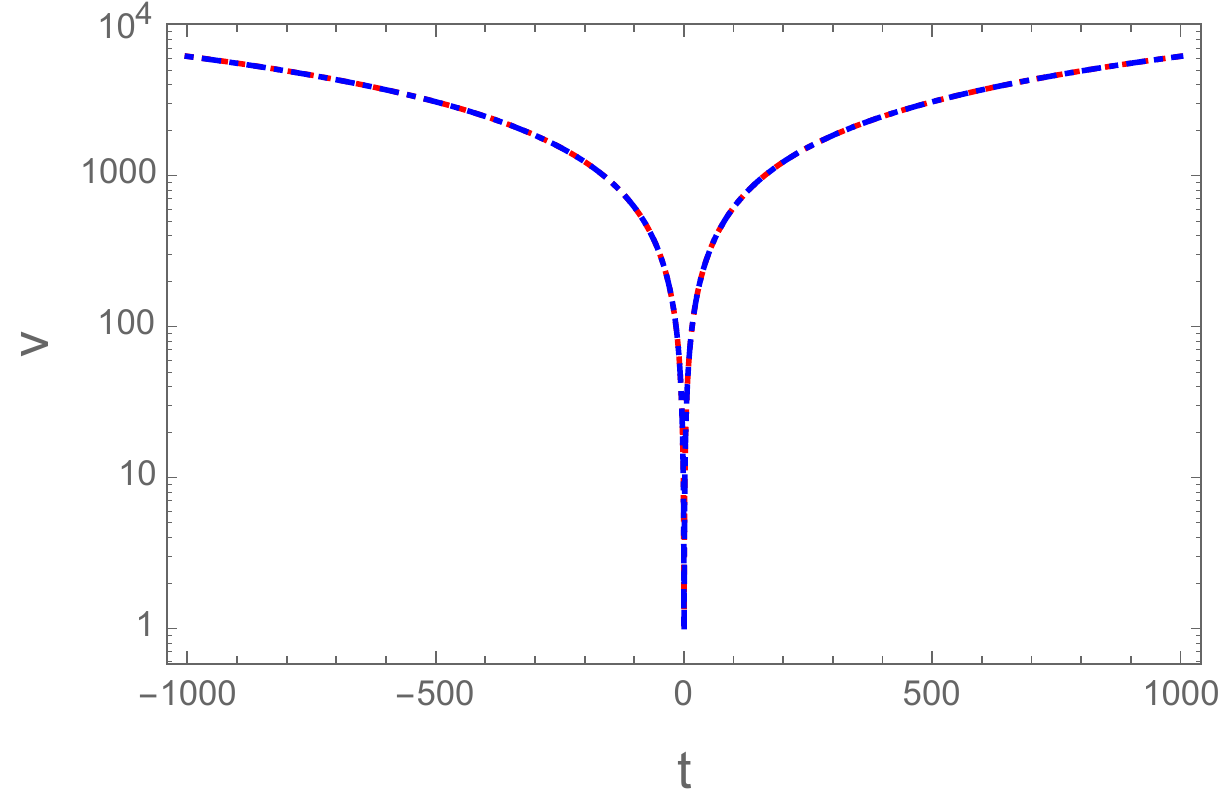}
\includegraphics[width=8cm]{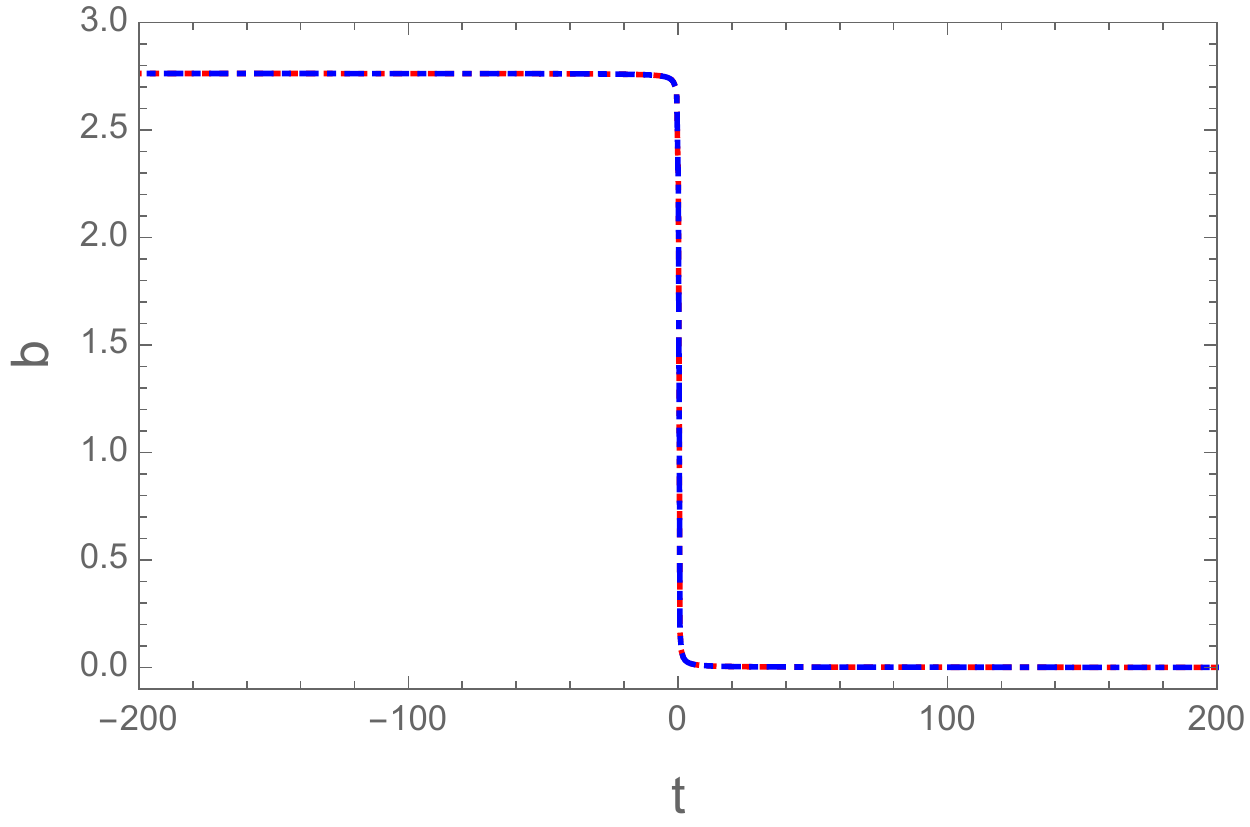}
\includegraphics[width=8cm]{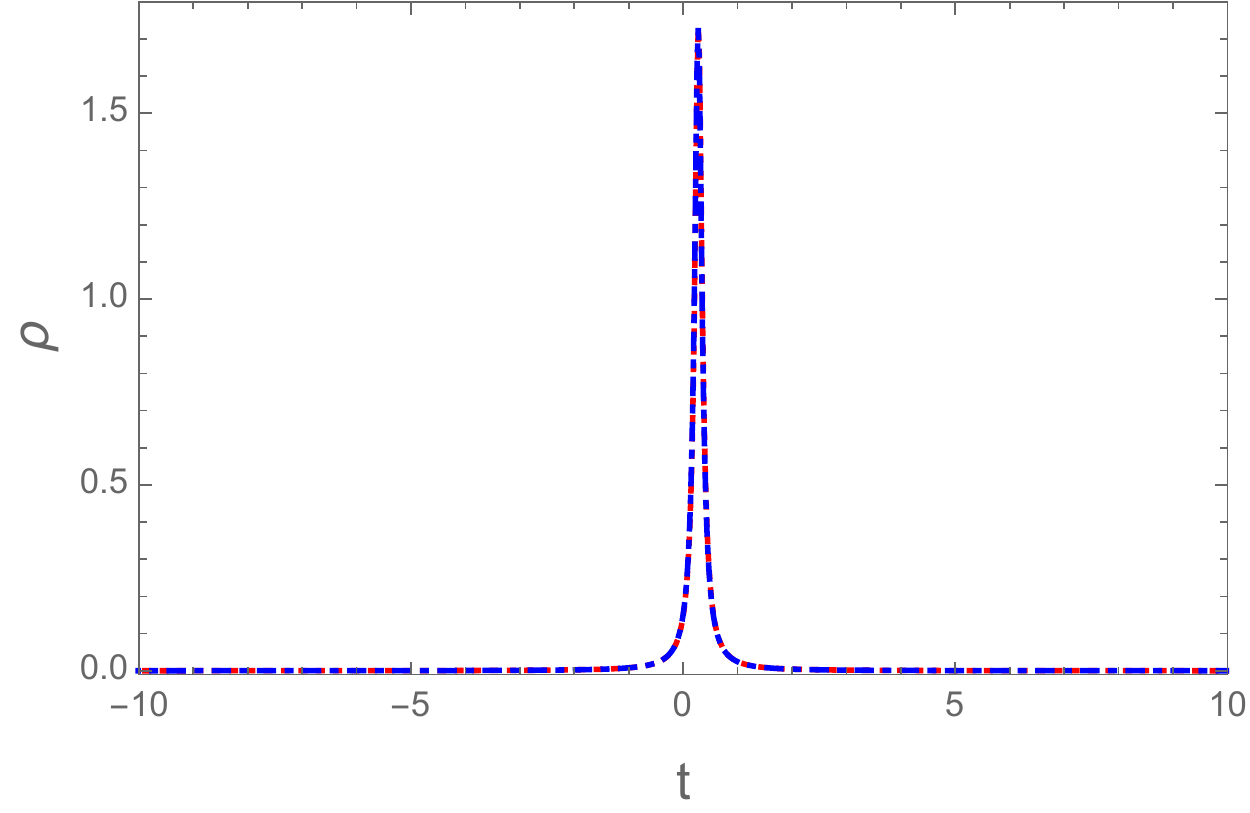}
}
\caption{Plots of variation in volume, $b$ and energy density for a massless scalar field in mLQC-II is shown. , Hamilton's equations (\ref{LQG2a})-(\ref{LQG2b}) are equivalent to the FR equations (\ref{2.6a})-(\ref{2.6b}). 
The initial conditions are chosen at $t_0=1000 t_{\text{Pl}}$ with the initial energy density $\rho_0=1.327\times 10^{-7} \rho_{\text{Pl}}$ and $\pi^0_\phi=1$. The bounce occurs at $t_B=0.28 t_{\text{Pl}}$. Numerics is performed using Hamilton's equations (\ref{LQG2a})-(\ref{LQG2b}), denoted by red dotted curves, as well as the modified FR equations (\ref{2.6a})-(\ref{2.6b}),  denoted by the blue dot-dashed lines. As expected they match exactly. }
\label{fig1}
\end{figure}

When $\rho \ll \rho^{{\scriptscriptstyle{\mathrm{II}}}}_c$, the modified FR equations (\ref{2.6a})-(\ref{2.6b}) yield their classical limits,
\bqn
H^2 &\approx& \frac{8\pi G}{3}\rho,\\
\frac{\ddot a}{a}&\approx&-\frac{4\pi G }{3}\left(\rho+3P\right).
\eqn
In order to uniquely determine the evolution of the universe, it is necessary to specify the matter content $ \mathcal H_M$. In this paper, we will mainly focus on the single scalar field inflation.
 Therefore, irrespective of the approaches applied when quantizing the background spacetime, the matter Hamiltonian in our analysis will always take   the following form 
\bq
\lb{2.9}
 \mathcal H_M=\frac{\pi^2_\phi}{2v}+v V(\phi),
\eq
where $\pi_\phi$ is the conjugate momentum of the scalar field and the potential energy term $V(\phi)$ will be further specified  in Sec. III. As a result, Hamilton's equations of the matter sector  read
\bqn
\lb{2.9a}
\dot \phi&=&\frac{\pi_\phi}{v},\\
\lb{2.9b}
\dot \pi_\phi&=&-vV_{,\phi},
\eqn
where $V_{,\phi}$ denotes the derivative of the potential  with respect to the scalar field.  Eqs.(\ref{LQG2a})- (\ref{LQG2b}) and  (\ref{2.9a})-(\ref{2.9b}) form a closed set of the first-order differential equations.
 In Fig. \ref{fig1}, we show the evolution in mLQC-II for the case of a massless scalar field, using both Hamilton's equations (\ref{LQG2a})-(\ref{LQG2b}), and the modified FR equations (\ref{2.6a})-(\ref{2.6b}). 
 As expected,  there is a perfect agreement between the two approaches. As in LQC, the quantum bounce in mLQC-II is also symmetric as can been seen from the volume variable $v$ near the bounce. 
 Besides, the energy density abruptly attains its peak right at the bounce which makes the momentum $b$ act like a step function,  since according to Eq.(\ref{LQG2b}), $b$ is an ever deceasing function of time 
 as long as the weak energy condition is satisfied. In the distant past, $b$ starts with a maximum value ($\approx 2.76$ in Planck units) and deceases slowly until the bounce where $b$ abruptly drops to a much smaller number 
 close to zero and keeps declining afterwards.   Both LQC and mLQC-II have the classical  limits in the pre- and post-bounce phases, while in mLQC-I, the classical limit only exists in the post bounce, 
 and the pre-bounce phase is asymptotically  de Sitter  with a large effective cosmological constant  \cite{lsw2018}.   

For completeness in the rest of this section, we shall   give a very brief summary over Hamilton's and FR equations for LQC and mLQC-I, and for details, we refer readers to \cite{ps06,aps2006,lsw2018}.

\subsection{Effective dynamics in LQC}

In the framework of LQC, the effective Hamiltonian in the spatially flat homogeneous and isotropic FLRW spacetime is given by \cite{ps06,aps2006}
\bq
\lb{2.10a}
\mathcal{H}=-\frac{3v\sin^2\left(\lambda b\right)}{8\pi G \gamma^2\lambda^2}+\mathcal H_M,
\eq
from which we  obtain the following Hamilton's equations for the variables $v$ and $b$
\bqn
\lb{LQCa}
\dot v&=&\frac{3v}{2\lambda \gamma}\sin(2\lambda b),  \\
\lb{LQCb}
\dot b&=&-\frac{3\sin^2\left(\lambda b\right)}{2 \gamma \lambda^2}-4\pi G\gamma P.
\eqn
Further, from the vanishing of the Hamiltonian constraint the energy density in LQC is given by 
\bq
\lb{2.11}
\rho=\rho_c \sin^2\left(\lambda b \right),
\eq
where  the critical energy density at the  bounce in LQC is denoted by $\rho_c$ ($=3 {\rho_{\text{Pl}}}/8 \pi G \lambda^2 \gamma^2\approx 0.41  \rho_{\text{Pl}}$).
The matter Hamiltonian and the corresponding equation of motion have the same forms as in mLQC-II, which are given by Eq.(\ref{2.9}) and (\ref{2.9a})-(\ref{2.9b}), respectively. Again, from Hamilton's equations (\ref{LQCa})-(\ref{LQCb}),
we find the modified FR equations,  
 \bqn
\lb{2.12a}
H^2 &=& \frac{8\pi G}{3}\rho\left(1 - \frac{\rho}{\rho_c}\right),  \\
\lb{2.12b}
\frac{\ddot a }{a}&=& -\frac{4\pi G}{3}\rho\left(1-\frac{4\rho}{\rho_c}\right)-4\pi G P\left(1-\frac{2\rho}{\rho_c}\right). ~~~~~~~~
\eqn
From these equations, it can be shown that the Klein-Gordon equation (\ref{2.7}) also holds here. The resulting dynamics is non-singular and results in a quantum bounce at $\rho \approx 0.41 \rho_{\mathrm{Pl}}$. In addition, we also have,
\bqn
\lb{2.12b}
\dot b=-4\pi G\gamma (\rho+P),
\eqn
which is also the case for mLQC-II as well as mLQC-I.

 \subsection{Effective dynamics in mLQC-I}
 
In mLQC-I, the effective Hamiltonian in  the  spatially flat homogeneous and isotropic FLRW spacetime is given by \cite{lsw2018}
\bqn
\lb{LQG1}
\mathcal {H}&=&\frac{3v}{8\pi G\lambda^2}\Big\{\sin^2(\lambda b)-\frac{(\gamma^2+1)\sin^2(2\lambda b)}{4\gamma^2}\Big\}\nb\\
&& ~~~~~~~~~ +\mathcal{H}_M,
\eqn
from which we  obtain  Hamilton's equations, 
\bqn
\lb{LQG1a}
\dot v&=&\Big\{v, \mathcal H\Big\}=\frac{3v\sin(2\lambda b)}{2\gamma \lambda}\Big\{(\gamma^2+1)\cos(2\lambda b)-\gamma^2\Big\}, \nb\\
\\
\lb{LQG1b}
\dot b&=&\Big\{b, \mathcal H\Big\}=\frac{3\sin^2(\lambda b)}{2\gamma \lambda^2}\Big\{\gamma^2\sin^2(\lambda b)-\cos^2(\lambda b)\Big\}\nb\\
&& ~~~~~~~~~~~~~~ -4\pi G\gamma P.
\eqn
The matter Hamiltonian and the equation of motion in the matter sector are also given  by Eq.(\ref{2.9}) and (\ref{2.9a})-(\ref{2.9b}). The novelty with respect to the effective Hamiltonian 
Eq.(\ref{LQG1}) lies in the emergence of two asymmetric branches when one tries to solve for the energy density from the Hamiltonian in terms of $b$, given by, 
\bqn
\lb{2.13}
\sin^2(\lambda b_{\pm})= \frac{1\pm\sqrt{1-\rho/\rho^{\scriptscriptstyle{\mathrm{I}}}_c}}{2(\gamma^2+1)},  
\eqn
where $\rho^{\scriptscriptstyle{\mathrm{I}}}_c$ represents the critical energy density in mLQC-I, which is given by  $\rho^{\scriptscriptstyle{\mathrm{I}}}_c \equiv \rho_c/[4(\gamma^2+1)]$. 
 Similarly, using the technique of Sec.II.A,  from Hamilton's equations (\ref{LQG1a})-(\ref{LQG1b}), one can find the corresponding modified FR equations which are made up with two separate branches  $b_{\pm}$.
  In the $b_{-}$ branch  the modified FR equations are given by 
\begin{widetext}
\bqn
\lb{LQG1c}
H^2 &=&\frac{8\pi G \rho}{3}\left(1-\frac{\rho}{\rho^{\scriptscriptstyle{\mathrm{I}}}_c}\right)\Bigg[1  +\frac{\gamma^2}{\gamma^2+1}\left(\frac{\sqrt{\rho/\rho^{\scriptscriptstyle{\mathrm{I}}}_c}}{1 +\sqrt{1-\rho/\rho^{\scriptscriptstyle{\mathrm{I}}}_c}}\right)^2\Bigg],\\
\lb{LQG1d}
\frac{\ddot a}{a} &=&-\frac{4\pi G}{3}\left(\rho + 3P\right)
  + \frac{4\pi G \rho}{3}\left[\frac{\left(7\gamma^2+ 8\right) -4\rho/\rho^{\scriptscriptstyle{\mathrm{I}}}_c+\left(5\gamma^2 +8\right)\sqrt{1-\rho/\rho^{\scriptscriptstyle{\mathrm{I}}}_c}}{(\gamma^2 +1)\left(1+\sqrt{1-\rho/\rho^I_c}\right)^2}\right]\frac{\rho}{\rho^{\scriptscriptstyle{\mathrm{I}}}_c}\nb\\
  &&  + 4\pi G P \left[\frac{3\gamma^2+2+2\sqrt{1-\rho/\rho^{\scriptscriptstyle{\mathrm{I}}}_c}}{(\gamma^2+1)\left(1+\sqrt{1-\rho/\rho^{\scriptscriptstyle{\mathrm{I}}}_c}\right)}\right]\frac{\rho}{\rho^{\scriptscriptstyle{\mathrm{I}}}_c},
\eqn
\end{widetext}
which are valid only in the post-bounce phase, in order to have a viable cosmological model \cite{lsw2018}. On the other hand, the modified FR equations in the pre-bounce phase are given by the  $b_+$ branch solution, and read 
\begin{widetext}
\bqn
\lb{LQG1e}
H^2 &=&\frac{8\pi G\alpha  \rho_\Lambda}{3}\left(1-\frac{\rho}{\rho^{\scriptscriptstyle{\mathrm{I}}}_c}\right)\left[1+\left(\frac{1-2\gamma^2+\sqrt{1-\rho/\rho^{\scriptscriptstyle{\mathrm{I}}}_c}}{4\gamma^2\left(1+\sqrt{1-\rho/\rho^{\scriptscriptstyle{\mathrm{I}}}_c}\right)}\right)\frac{\rho}{\rho^{\scriptscriptstyle{\mathrm{I}}}_c}\right], \\
\lb{LQG1f}
\frac{\ddot a}{a} &=&- \frac{4\pi \alpha G}{3}\left(\rho + 3P - 2\rho_\Lambda \right)  +4\pi G\alpha P\left(\frac{2-3\gamma^2 +2\sqrt{1-\rho/\rho^{\scriptscriptstyle{\mathrm{I}}}_c}}{(1-5\gamma^2)\left(1+\sqrt{1-\rho/\rho^{\scriptscriptstyle{\mathrm{I}}}_c}\right)}\right)\frac{\rho}{\rho^{\scriptscriptstyle{\mathrm{I}}}_c}\nb\\
&& - \frac{4\pi G\alpha \rho}{3}\left[\frac{2\gamma^2+5\gamma^2\left(1+\sqrt{1-\rho/\rho^{\scriptscriptstyle{\mathrm{I}}}_c}\right)-4\left(1+\sqrt{1- \rho/\rho^{\scriptscriptstyle{\mathrm{I}}}_c}\right)^2}{(1-5\gamma^2)\left(1+\sqrt{1-\rho/\rho^{\scriptscriptstyle{\mathrm{I}}}_c}\right)^2}\right]\frac{\rho}{\rho^{\scriptscriptstyle{\mathrm{I}}}_c},
\eqn
\end{widetext}
where $\alpha\equiv ({1-5\gamma^2})/({\gamma^2+1})$ and $\rho_\Lambda \equiv{3}/[{8\pi G\alpha \lambda^2(1+\gamma^2)^2}]$.
 We note that in this case the momentum $b$ also satisfies Eq.(\ref{2.12b}), as in the LQC and mLQC-II models. In fact, using this property one can see clearly   the matching of  the two different branches at the quantum bounce 
  in mLQC-I model, and all other matchings are mathematically inconsistent \cite{lsw2018}. 
 
\begin{figure} [h!]
{
\includegraphics[width=8cm]{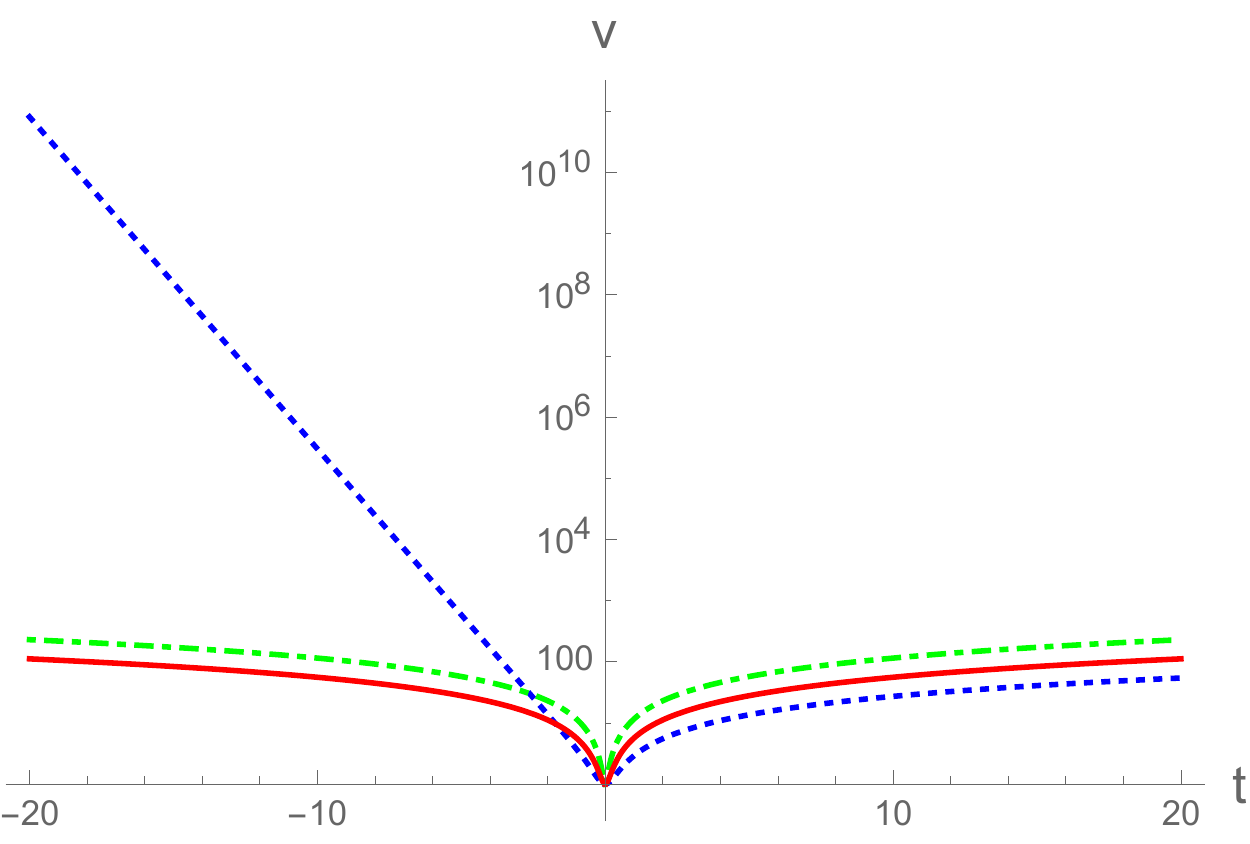}
\includegraphics[width=8cm]{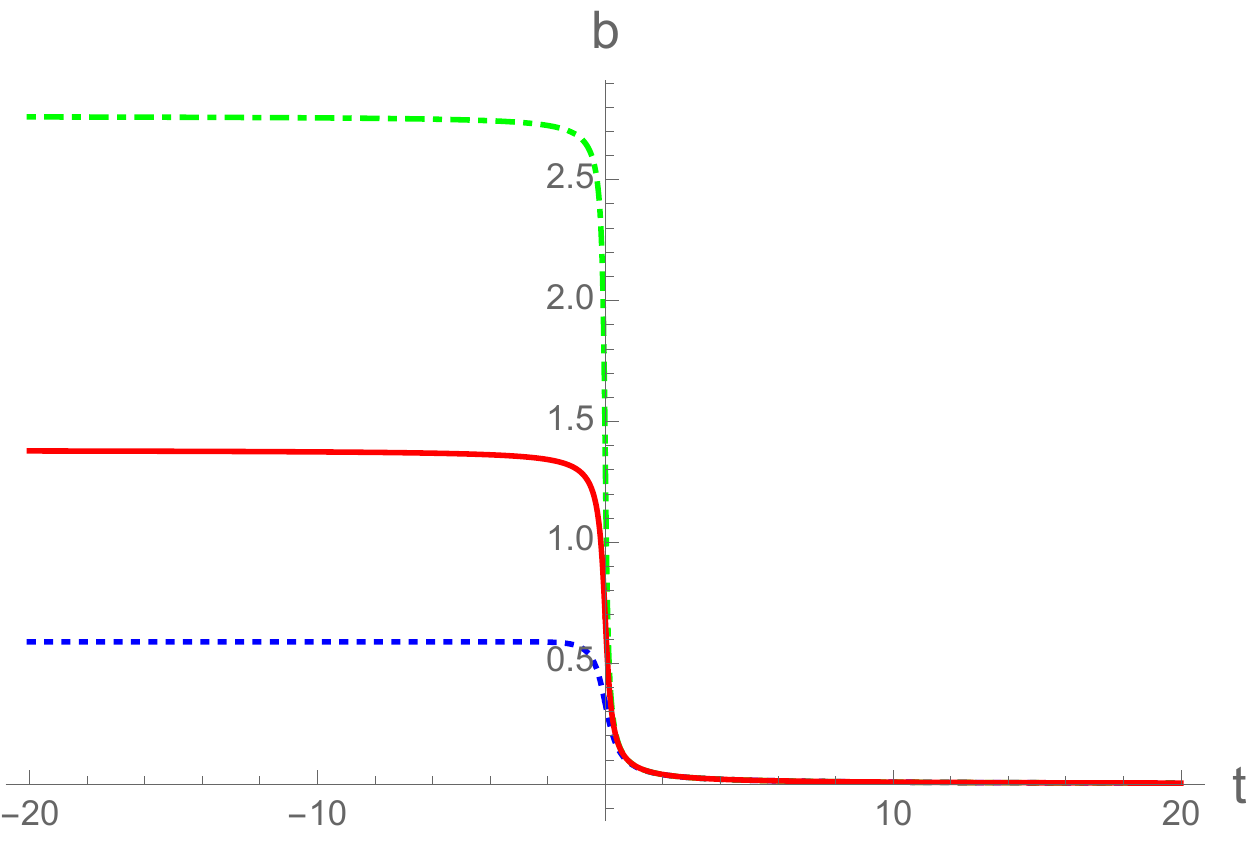}
\includegraphics[width=8cm]{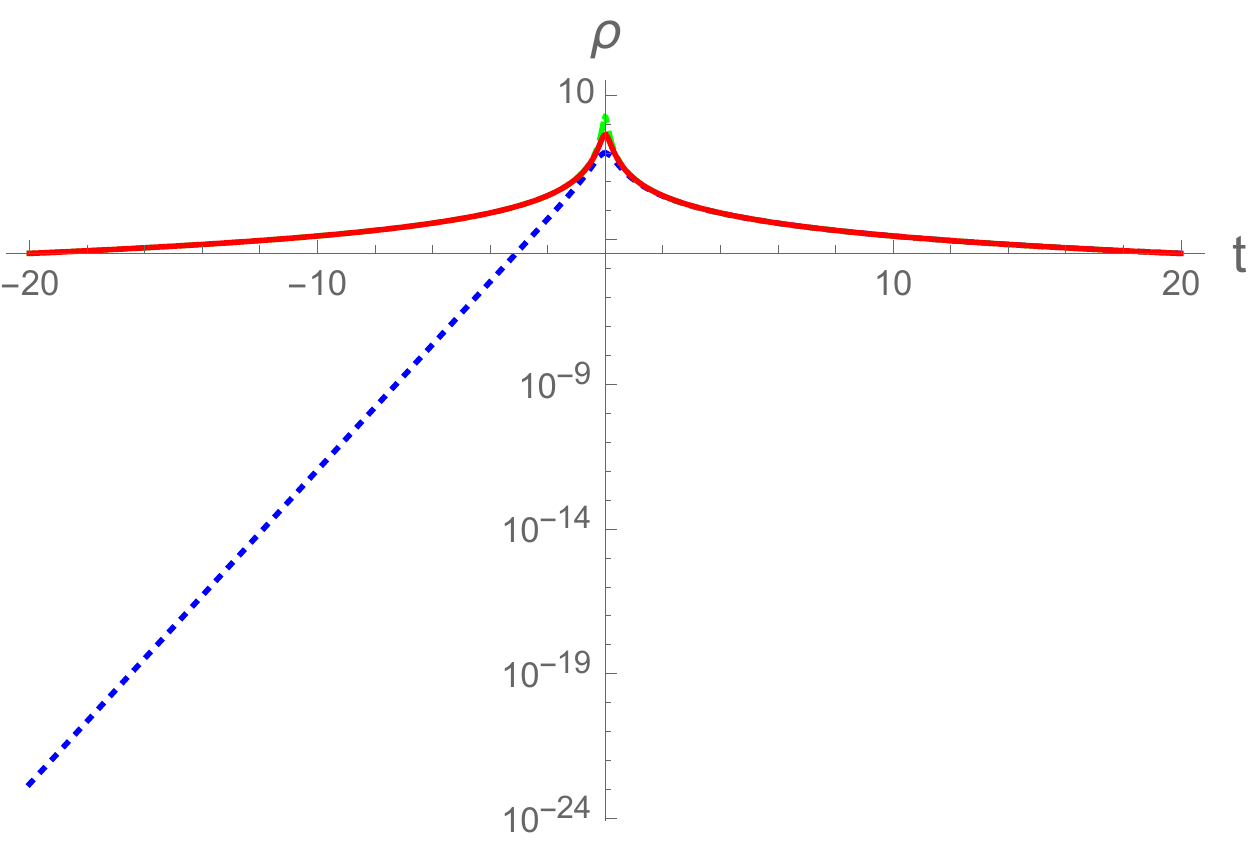}
}
\caption{Comparison of three models, LQC (red solid line), mLQC-I (blue dotted line) and mLQC-II (green dot-dashed line) for  a massless scalar field. 
The initial conditions are chosen at the bounce with $\phi_B$ set to zero and a positive $\dot \phi_B$.}
\label{compare}
\end{figure}

In order to qualitatively compare the three models discussed in this section, we carry out the numerical simulations for a massless scalar field. The results are shown in Fig. \ref{compare},
 from which we can see clearly  the following: (i) The evolution of the universe is symmetric with respect to the bounce in LQC and mLQC-II,  while it is asymmetric in mLQC-I,
 as can be seen  clearly from the volume and energy density. (ii) In the post-bounce  phase, the volume of the universe in mLQC-II is always larger than that in LQC which, in turn, is greater than
 that in  mLQC-I. This behavior is related to the relative magnitude of the critical density in each model. Basically, $\rho^{\scriptscriptstyle{\mathrm{II}}}_c>\rho_c>\rho^{\scriptscriptstyle{\mathrm{I}}}_c$. 
 (iii) A de-Sitter spacetime in the contracting phase of mLQC-I can be observed due to the exponential growth of the volume. (iv) In the second subfigure, $b$  assumes different asymptotic values at early stages of evolution in
 the  three models. This behavior can be accounted for by different relationships between energy density and $b$ in these models. In LQC, mLQC-I and mLQC-II, $b$ and $\rho$ satisfy Eq. (\ref{2.11}), (\ref{2.13}) and (\ref{2.5}),
  respectively. More similarities and distinctions among these models will be unraveled in the next section where we focus on qualitative dynamics of these models with  a variety of potentials.

\section{Dynamical Systems and Phase Space Portraits}
\label{Section3}
\renewcommand{\theequation}{3.\arabic{equation}}\setcounter{equation}{0}

In this section, we discuss the qualitative behavior of the  spatially flat FLRW universe in the three different models of loop cosmology: LQC, mLQC-I and mLQC-II, by using dynamical system analysis. We consider five different potentials: chaotic, fractional monodromy, Starobinsky, non-minimal Higgs, and exponential. Qualitative dynamics plays an important role whenever analytical solutions are hard 
to obtain, as in the present cases, and reveals details of the existence of attractors and asymptotic behavior.
Studying phase space portraits of dynamical variables, one can easily identify the slow-roll inflationary separatrices before the homogeneous scalar field enters  a phase of fast oscillations, 
also known as the reheating phase. In LQC, qualitative dynamics has been studied only for $\phi^2$ potential in Ref.  \cite{svv} and for the exponential potential in \cite{rs2012}. Further, stability of critical points even in these cases has not been investigated. 
In the following we shall consider several popular potentials, study  their detailed qualitative dynamics including stability of critical points not only in LQC but also in mLQC-I and mLQC-II.

In all the considered models, quantum geometric effects only influence the gravitational sector of the Hamiltonian. Thus, the Klein-Gordon equation and the energy conservation law remain unchanged.  
Starting from the Klein-Gordon equation of the scalar field in the spatially flat homogeneous and isotropic FLRW spacetime,
\bq
\lb{3.1}
\ddot \phi+3H^i \dot \phi+V_{,\phi}=0,
\eq
where $H^i$ denotes the Hubble rate for LQC, mLQC-I or mLQC-II,  
one can define two dimensionless variables for a positive potential $V$:
\bq
\lb{3.2}
X=\epsilon_\phi \sqrt{\frac{V}{\rho^i_c}}, \quad \mathrm{and} \quad Y=\frac{\dot \phi}{\sqrt{2\rho^i_c}},
\eq
which satisfy, 
\bq
\lb{3.3}
X^2+Y^2=\frac{\rho}{\rho^i_c} .
\eq
Here $\epsilon_\phi=\pm 1$,   and $\rho^i_c$  denotes the energy density at the bounce for different models, with $\rho_c^i = \rho_c$ for LQC, 
$\rho_c^i = \rc$ for mLQC-I and $ \rho_c^i = \rd$ for mLQC-II. As a result, all the trajectories in the phase space are confined within the unit circle,
\bq
\lb{3.4}
X^2+Y^2\le 1.
\eq
With the help of the modified Friedmann equation in each model, we can now turn the evolution equation Eq.(\ref{3.1}) into an autonomous system whose equations of motion are given by  two first-order ordinary differential equations
\bqn
\lb{3.5}
\dot X&=&\frac{\epsilon_\phi V_{,\phi} Y }{\sqrt{2 V}},\\
\lb{3.6}
\dot Y &=&-3H^i Y -\frac{V_{,\phi}}{\sqrt{2\rho^i_c} } ~.
\eqn
Here both $V_{,\phi}$ and $V$ are regarded as functions of $X$ and $H^i$. 
For all the models considered in this manuscript, Eqs.(\ref{3.5})-(\ref{3.6}) form a closed set because $H^i$ is a function of the energy density, and hence a function of $X$ and $Y$.  It is useful to express the equation of state $\omega_\phi:=P/\rho$ in terms of $X$ and $Y$:
\bq
\lb{3.7}
w_{\phi}  \equiv\frac{\dot\phi^2/2 - V(\phi)}{\phi^2/2  + V(\phi)} = 1-\frac{2}{1+(Y/X)^2}.
\eq

Fixed points are determined by setting  $\dot X=\dot Y=0$, and can be characterized by the behavior of the equation of state since it depends on the slope of the separatrix when the solutions converge to the fixed points. In particular, if the fixed point is a focus, the equation of state is generally undefined due to the oscillating nature of the scalar field as it approaches the focus. If the  dynamical system contains a slow-roll inflationary phase, one would expect that there will exist one or two separatrices located on the $X$ axis,  so that the resulting effective equation of state is  approximately $-1$,
 which ensures the exponential expansion of the universe. The stability of a given fixed point can be analyzed by inducing a small perturbation around the fixed point and analyzing the characteristic eigenvalues of the resulting equation of motion for the perturbation \cite{b1985}. Note that in various papers in classical cosmology, such as Ref. \cite{clw1998}, dynamical systems are studied with respect to $N = \ln(a)$. In the pre-bounce era, $N$ decreases as $t$ increases. Due to this a fixed point which is a repeller for trajectories evolving forward in time $t$ in the pre-bounce epoch, will be an attractor for trajectories evolving with increasing $N$. In the following analysis, the attractor and repeller behavior will always be for forward time evolution, i.e. towards the bounce surface in the pre-bounce epoch and away from the bounce surface in the post-bounce epoch.

\begin{widetext}

\begin{figure} 
{
\includegraphics[width=6cm]{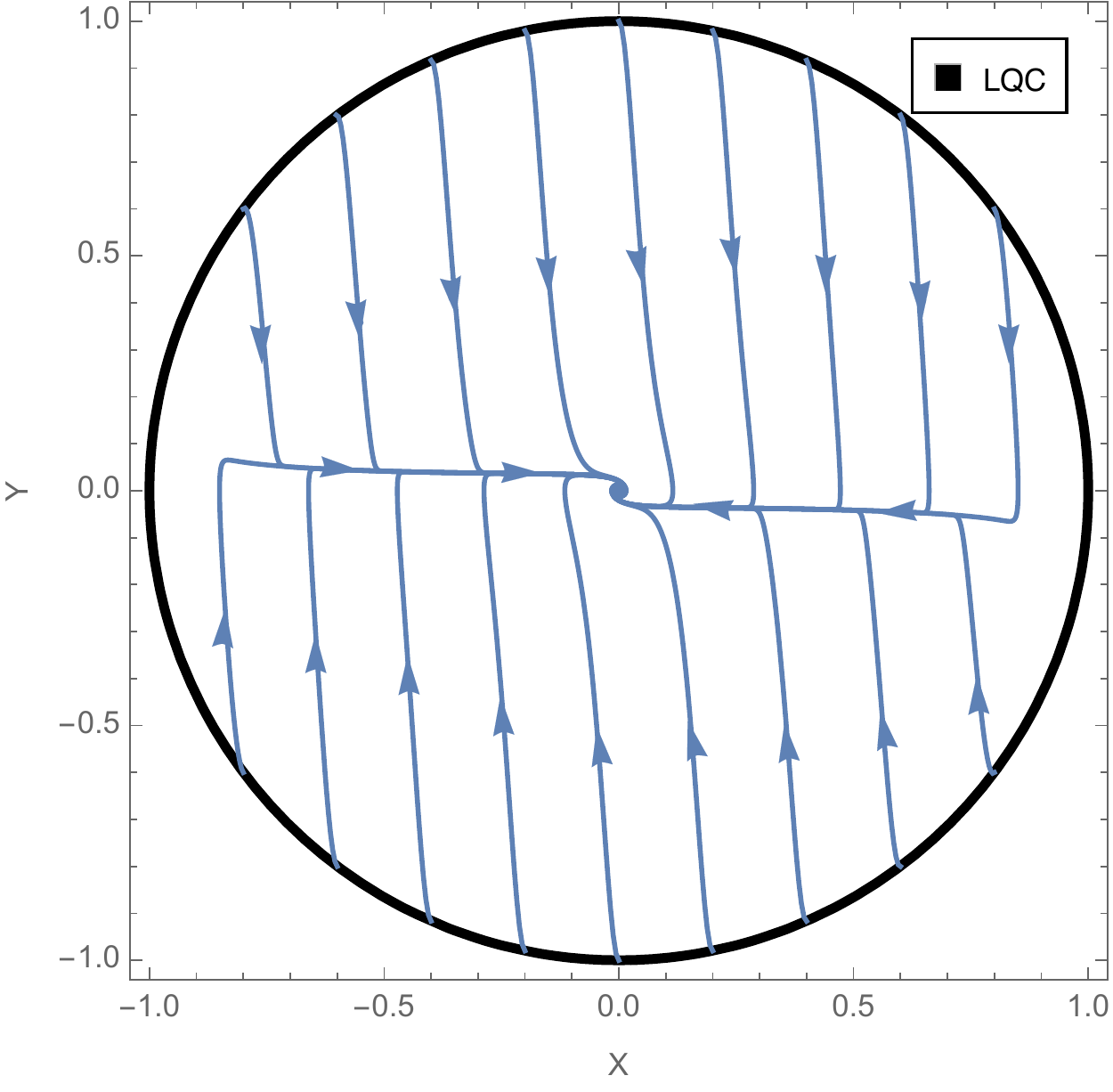}
\includegraphics[width=6cm]{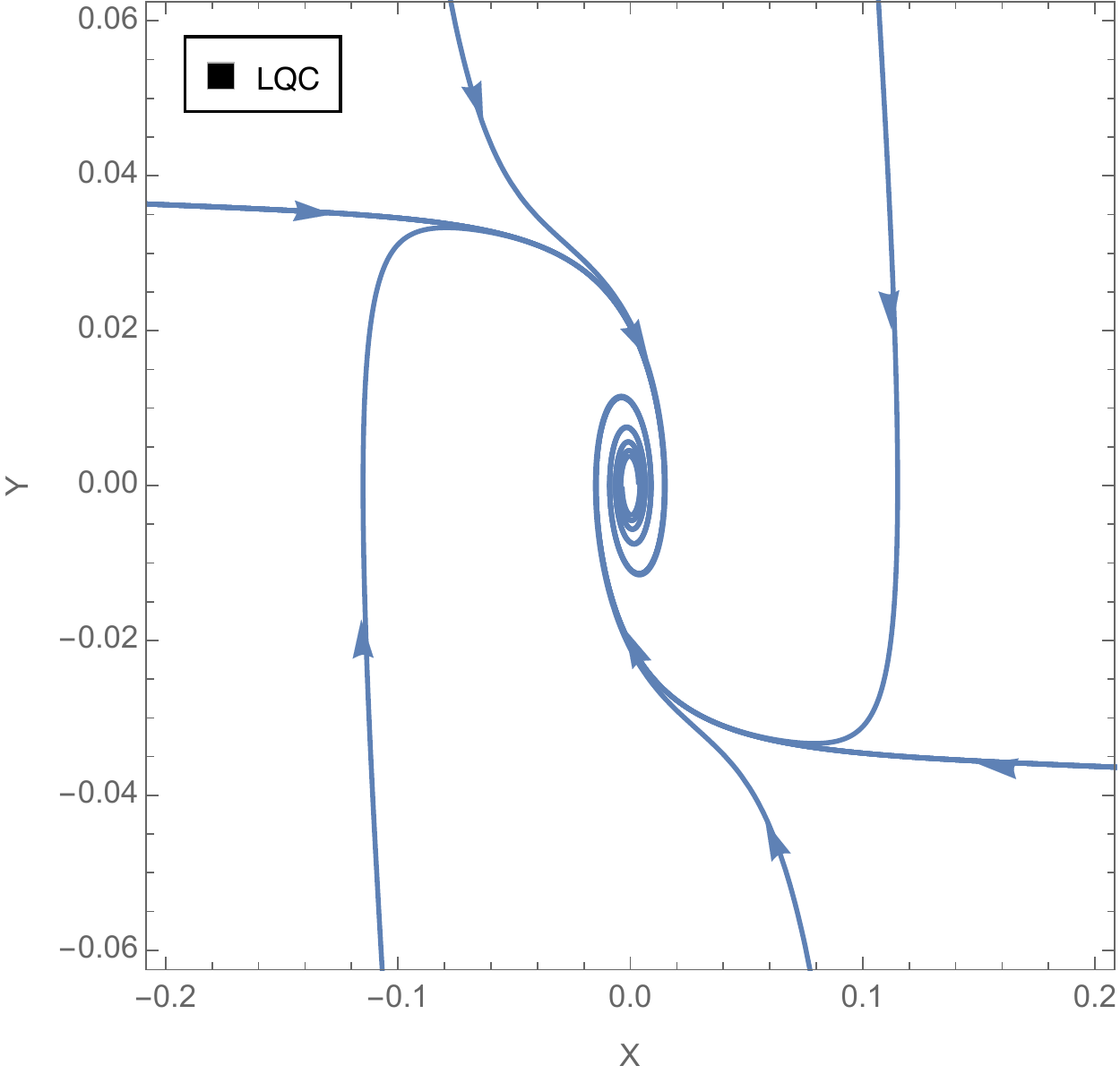} \\
\includegraphics[width=6cm]{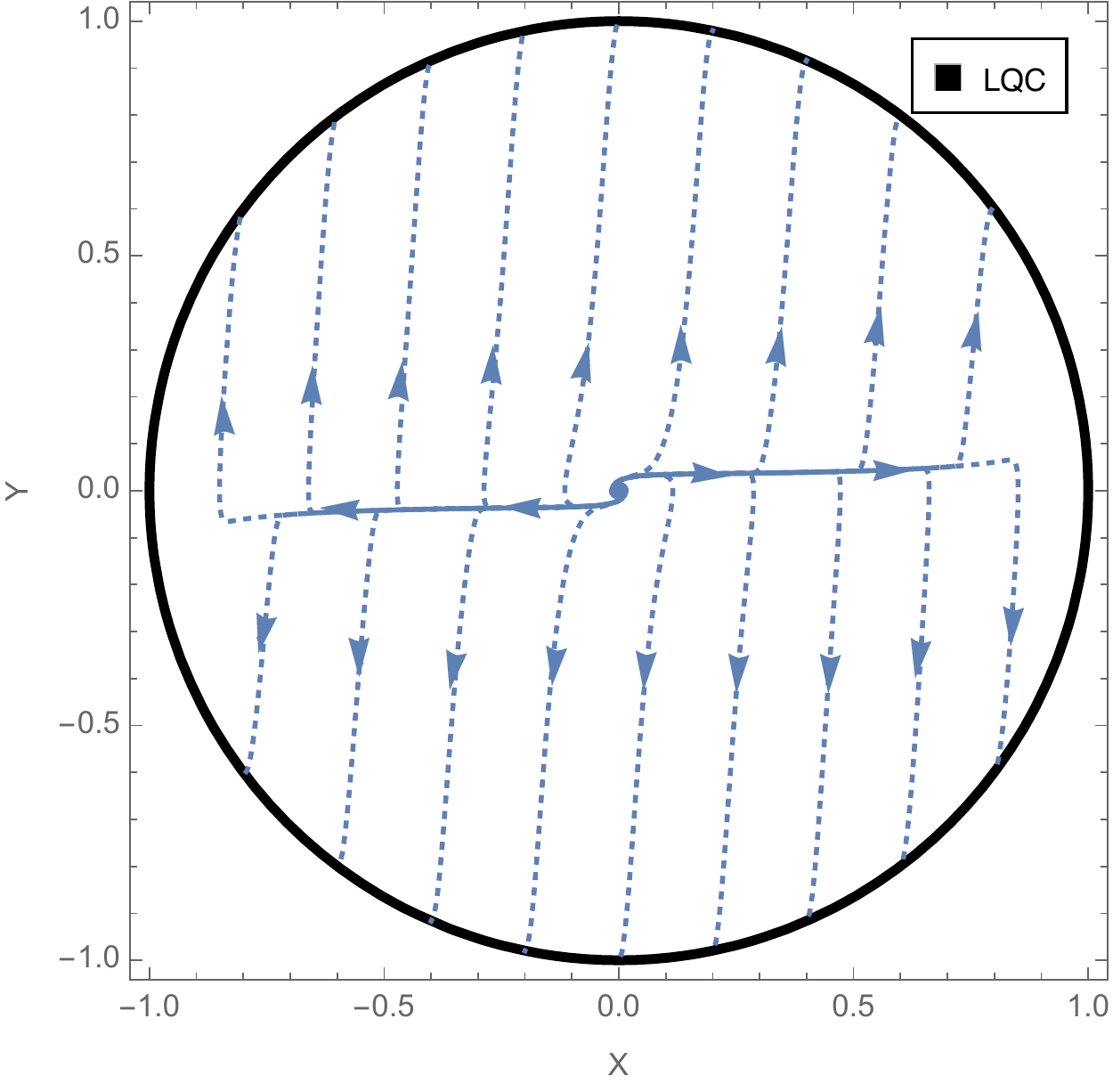}
\includegraphics[width=6cm]{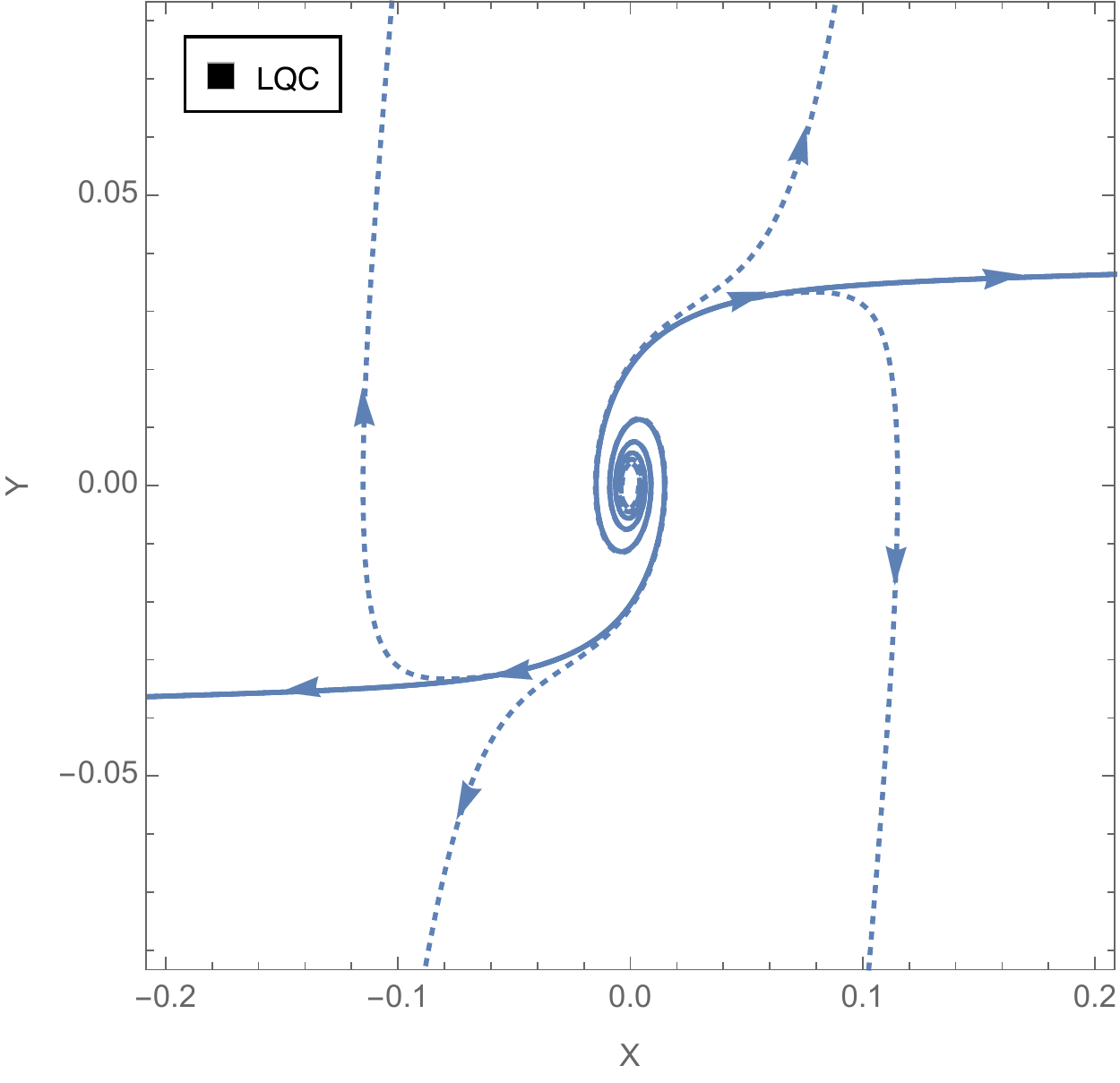}
}

\caption{Phase space portraits for the $\phi^2$ potential in LQC. For visualizations we have chosen  $m = 0.2$ in the Planck units. Solid (dashed) curve indicates the post-bounce (pre-bounce) evolution. Unit circle denotes the bounce surface. Arrows on curves indicate forward time evolution.  There are two inflationary separatrices readily seen in the top panel, which imply the inflationary solution is an attractor for the chaotic potential in the post-bounce regime. In the second panel, a zoom-in view near the origin is given to reveal the characteristic spiral structure of the phase portraits when the inflaton continuously loses its energy in the reheating phase. }
\label{fig3.1}
\end{figure}

\begin{figure} 
{
\includegraphics[width=6cm]{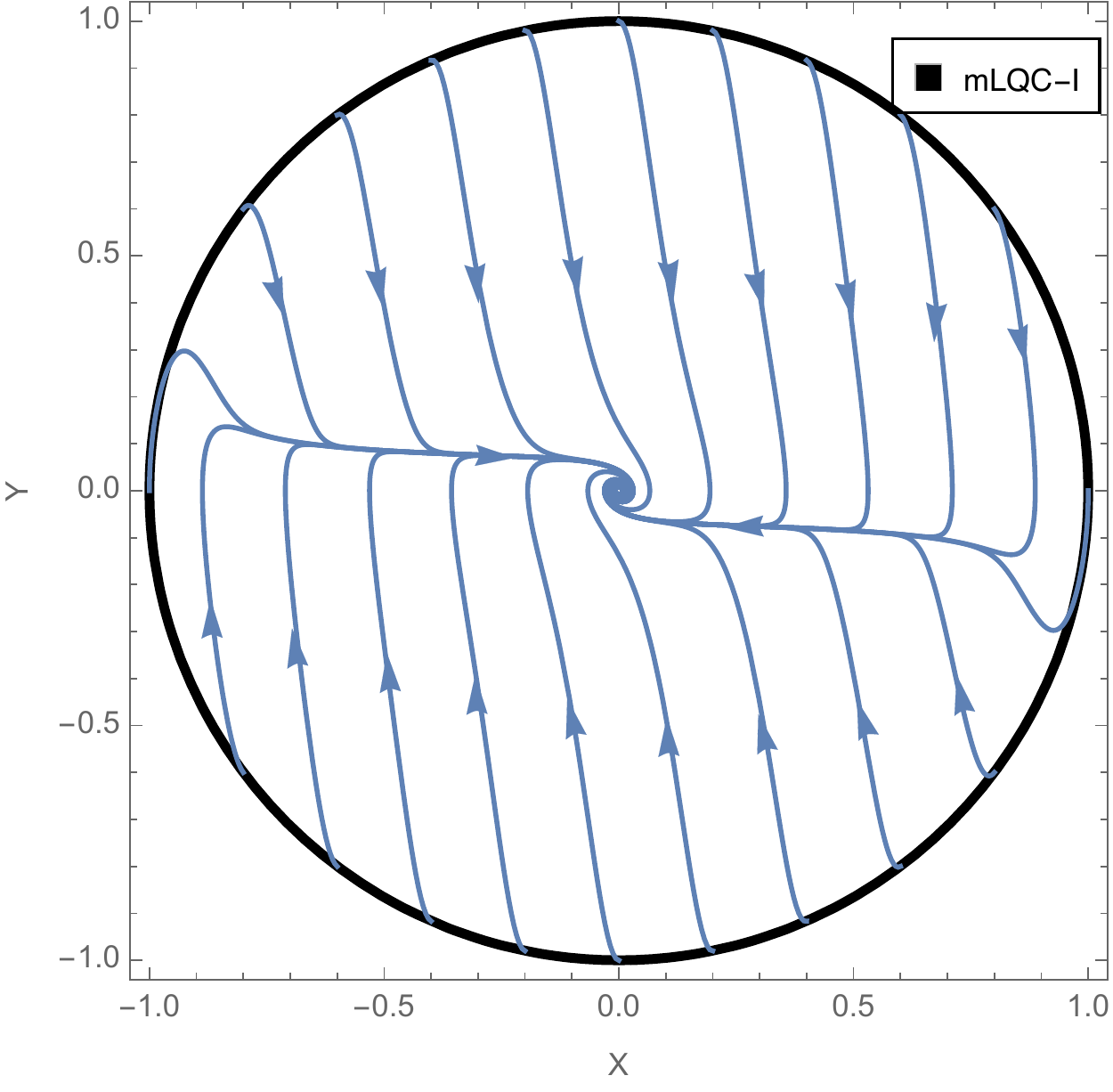}
\includegraphics[width=6cm]{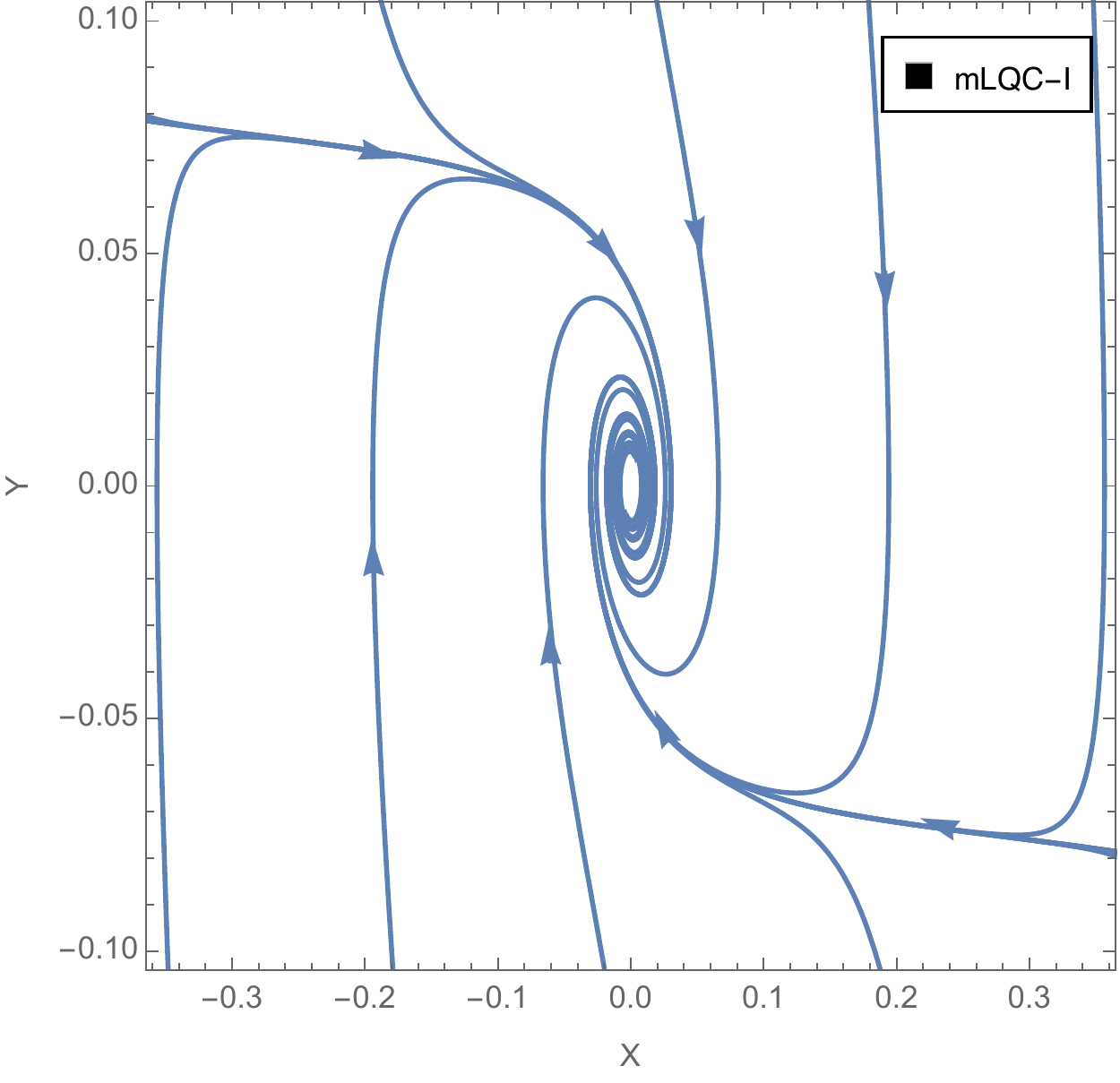}\\
\includegraphics[width=6cm]{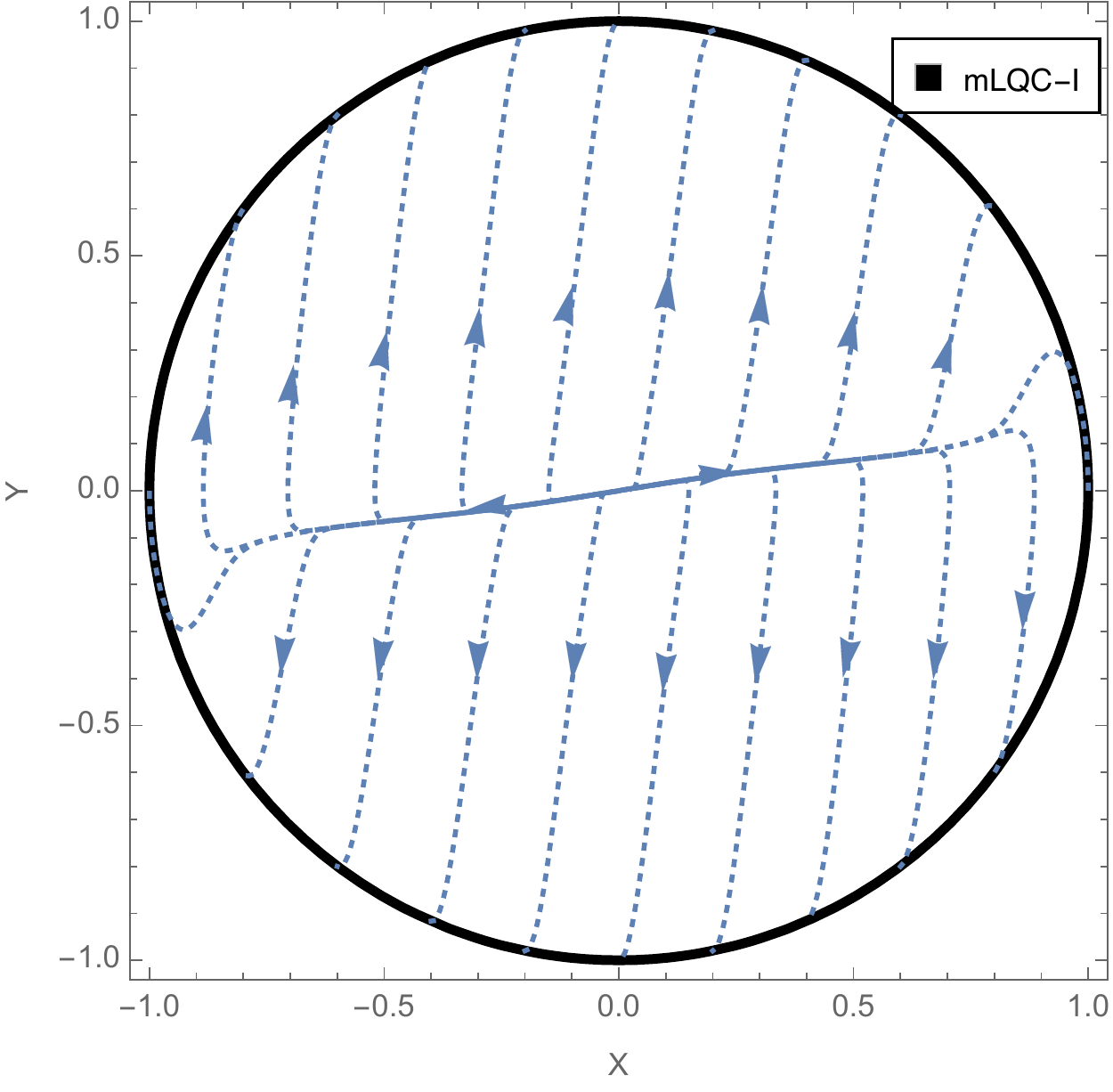}
\includegraphics[width=6cm]{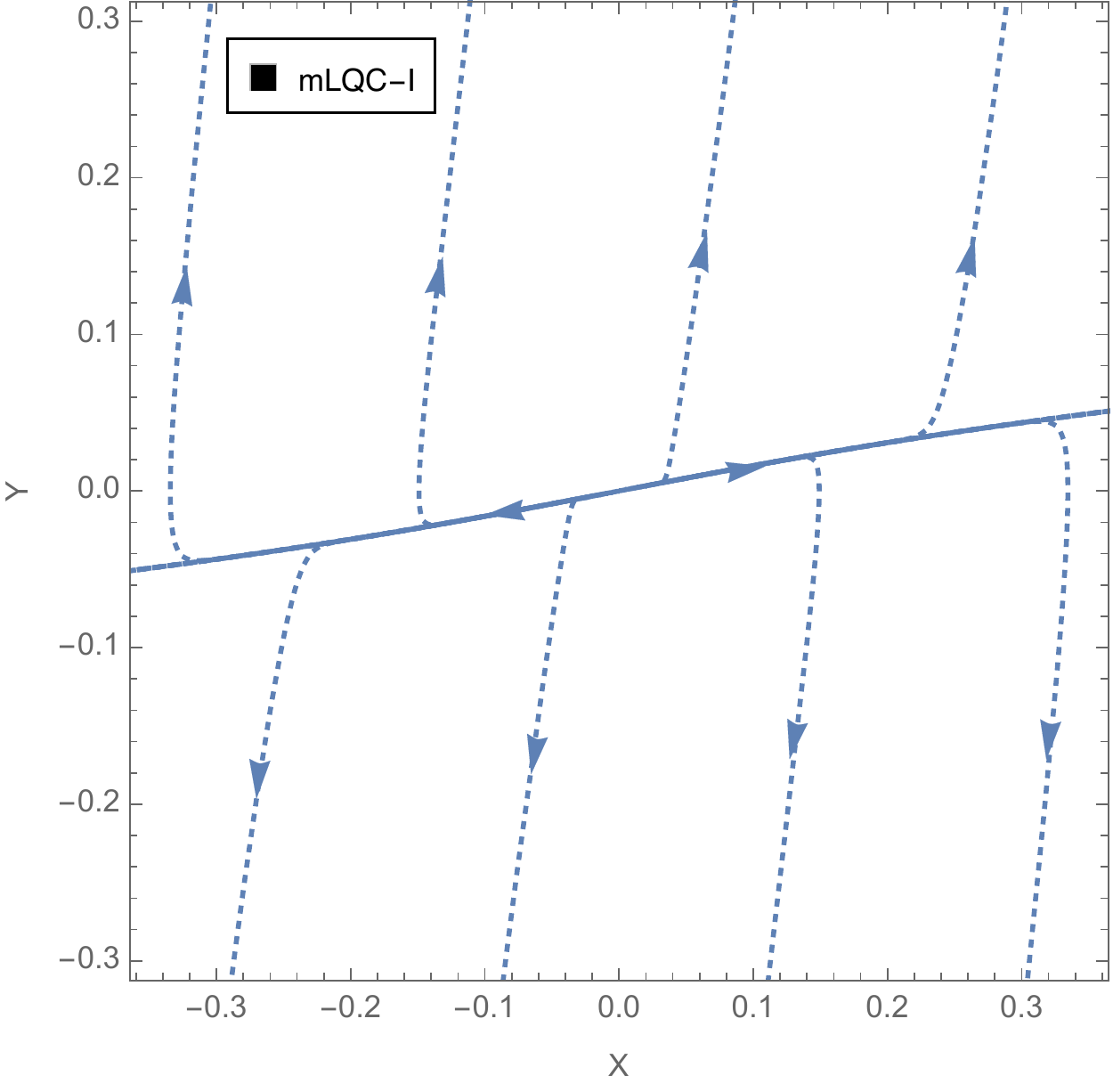}
}
\caption{Phase space portraits for the $\phi^2$ potential in mLQC-I.  With the same mass ($m = 0.2$)  pre- and post-bounce phases are  denoted respectively by solid and dashed lines. Analogous to Fig. \ref{fig3.1}, inflationary solutions represented by two separatrices are still the attractors for the chaotic potential in mLQC-I. One important difference that can be seen directly from the graph lies in the  two bottom panels in which the absence of the spiral signifies a large Hubble parameter (zeroth order compared to the perturbations around the origin) shows up in early times.} 
\label{fig3.2}
\end{figure}

\begin{figure} 
{
\includegraphics[width=6cm]{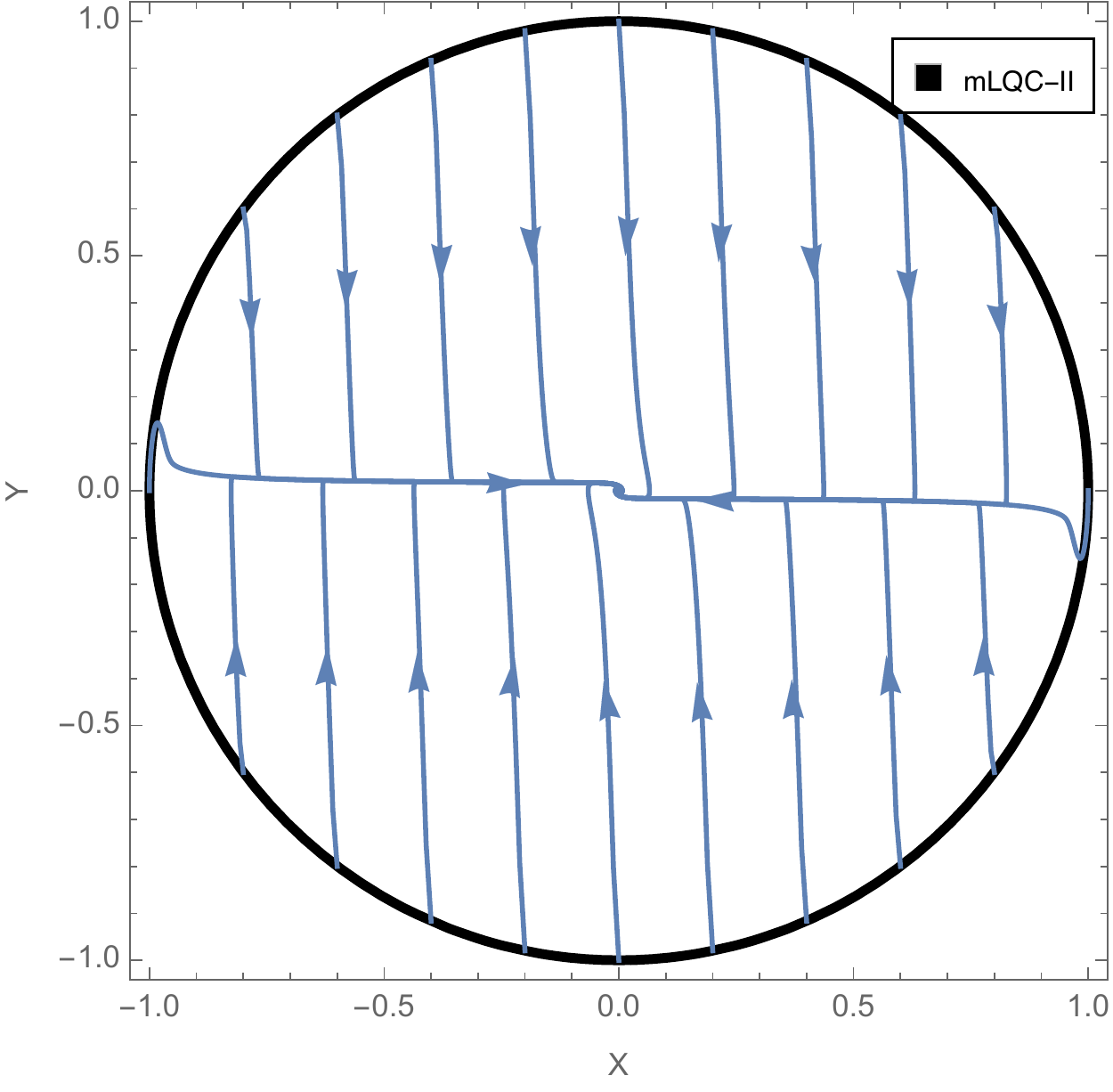}
\includegraphics[width=6cm]{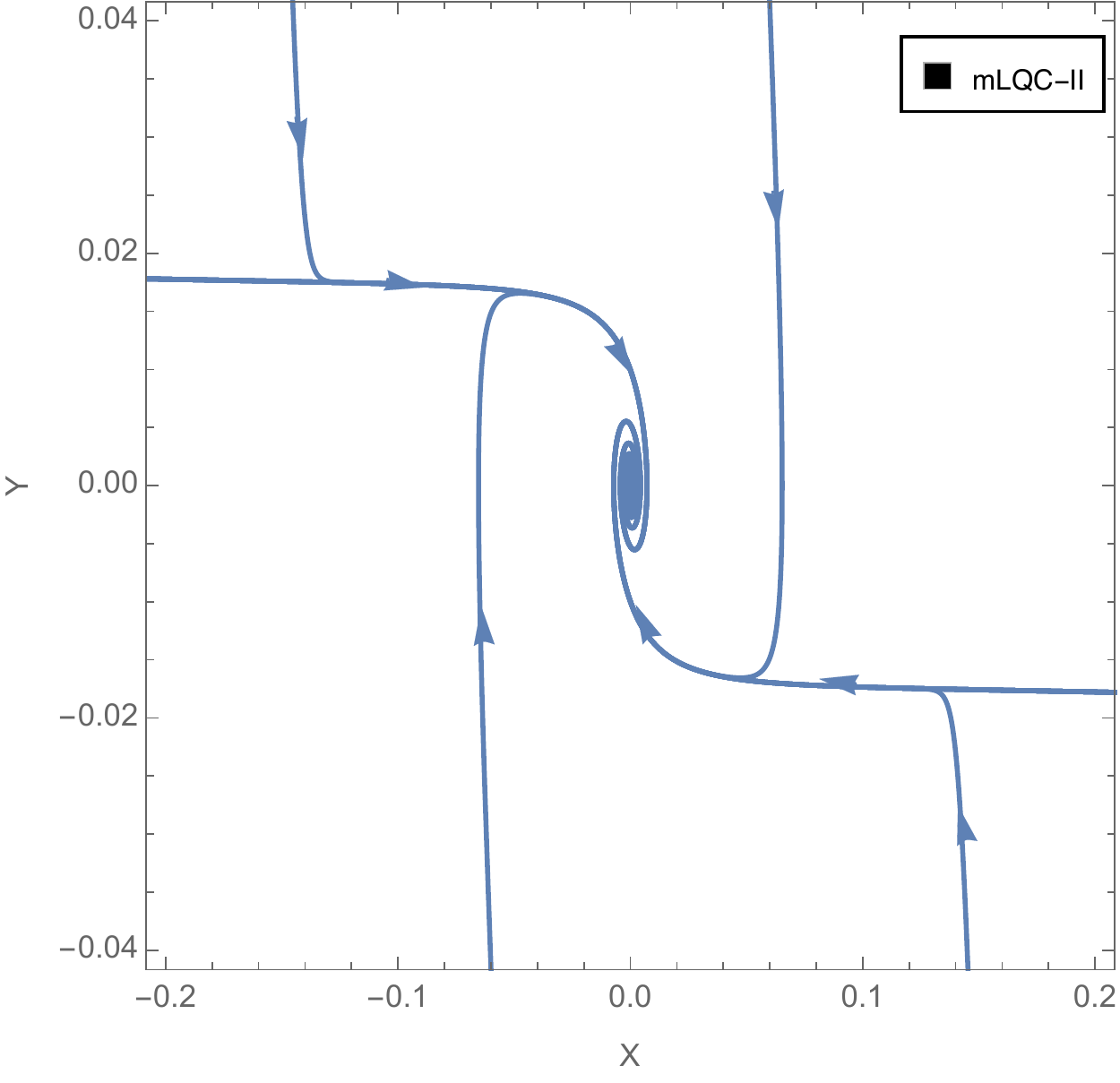}\\
\includegraphics[width=6cm]{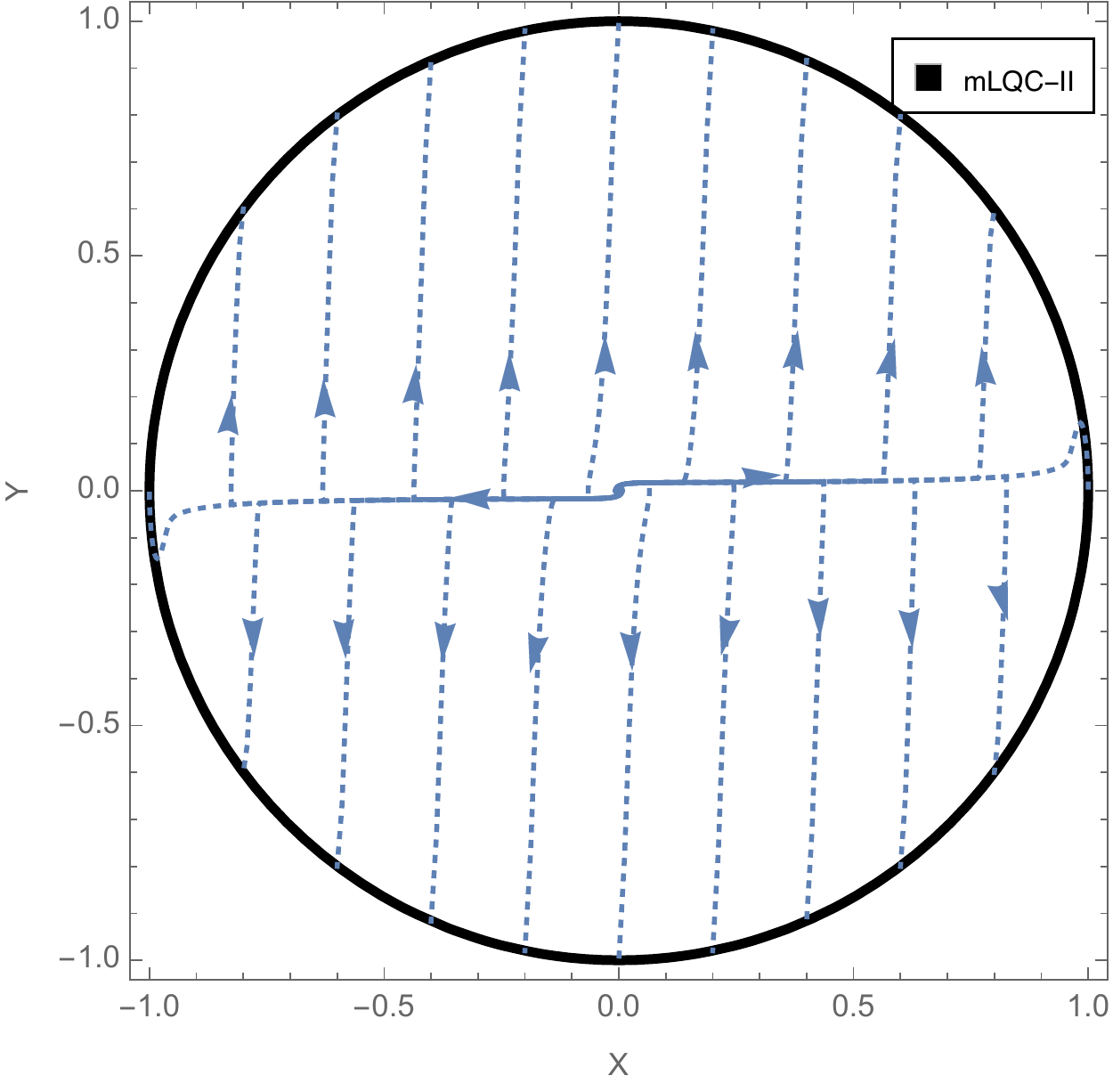}
\includegraphics[width=6cm]{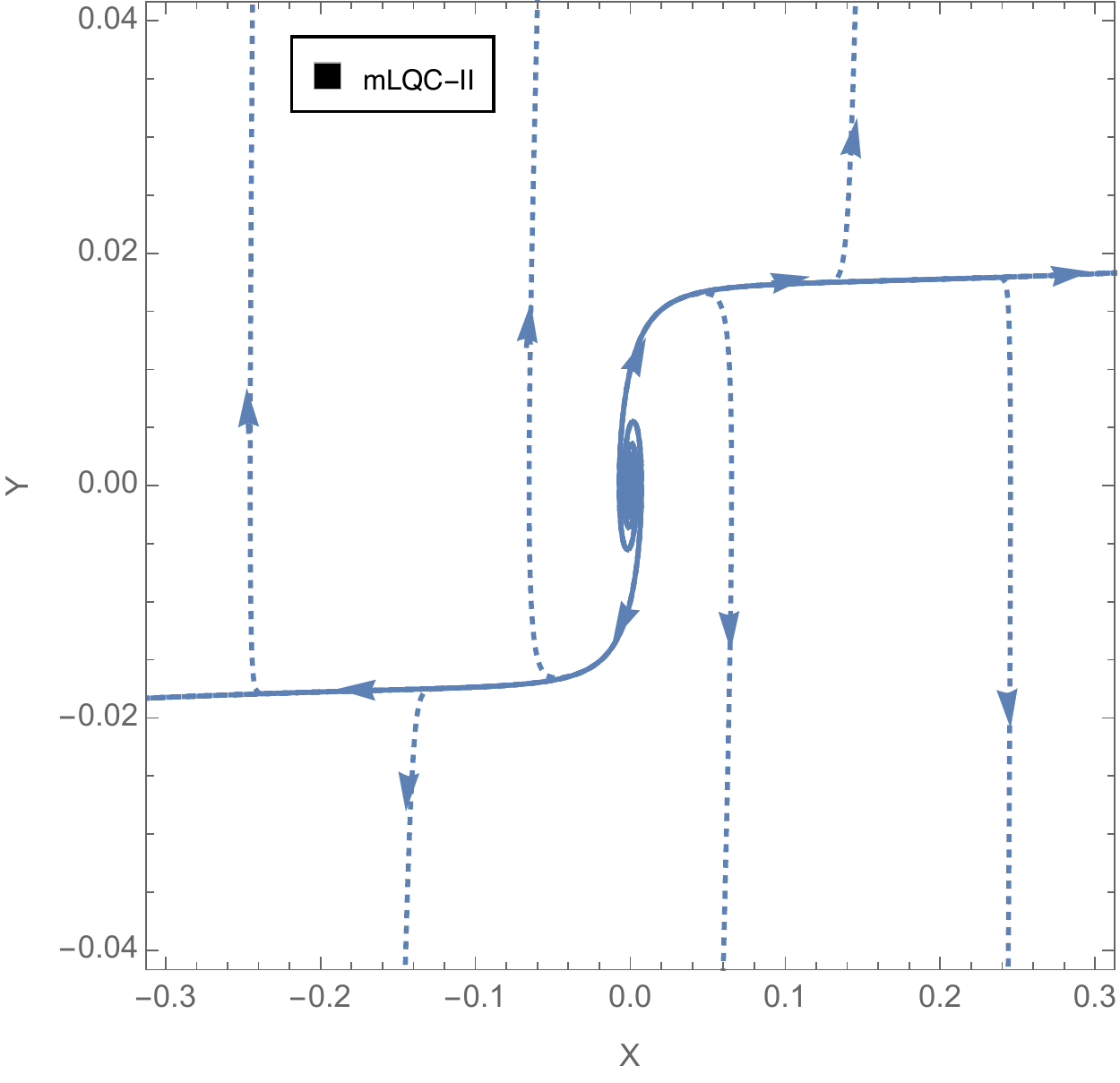}
}
\caption{Phase space portraits for the $\phi^2$ potential in mLQC-II.   The same initial conditions and mass as in Fig. \ref{fig3.1} are employed. The patterns in these figures resemble those in LQC for the same chaotic potential. In particular, 
 the evolution of the universe in mLQC-II is also symmetric with respect to the quantum bounce.  The oscillations of the trajectories can be observed in both pre- and post-bounce phases and this behavior is independent of the mass parameter.}
\label{fig3.3}
\end{figure}

\end{widetext}

\subsection{$\phi^2$ Potential}

One of the most popular  inflationary models  is the chaotic inflation \cite{l1983},  in which the self-interacting potential has the form
\bq
\lb{3.8a}
V(\phi)=\frac{1}{2}m^2 \phi^2.
\eq
Consequently,  with the phase space variables taken as 
\bq
X=\frac{m \phi}{\sqrt{2\rho^i_c}} \quad \quad Y=\frac{\dot \phi}{\sqrt{2\rho^i_c}},
\eq
the equations of motion (\ref{3.5})-(\ref{3.6}) can be simplified to 
\bqn
\lb{3.8}
\dot X &=&m Y ,\\
\lb{3.9}
\dot Y&=&- m X-3 H^i Y . 
\eqn
 These indicate that the only fixed point  on the $X\text{-}Y$ plane is the origin, $(X, Y) = (0, 0)$. In Figs. \ref{fig3.1} - \ref{fig3.3}, we provide the phase space portraits of LQC, mLQC-I and mLQC-II for the chaotic inflationary potential. In these figures, all the phase space trajectories are confined within the unit circle on which  the bounce occurs. The arrows in the figures indicate the direction of the forward-time evolution.  In all the models, we find the existence of inflationary attractors in the post-bounce phase.  In Fig. \ref{fig3.1},  corresponding to LQC dynamics, the top subfigure illustrates the behavior of the solutions in the post-bounce phase, in which  all the trajectories commencing from the unit circle  converge to one of the separatrices (both of which appear parallel to $X$ axis with a very small value of $Y$ as the scalar field is almost constant during the slow-roll inflation) before oscillating about the  fixed point (origin). In the second subfigure, a zoom-in view near the origin shows that the two separatrices can be approximated by  $Y \approx \pm0.035$, which together with Eq.(\ref{3.7}) indicates that in order to ensure an inflationary phase, the absolute value of  $|X|$ at the bounce must be greater than a certain value, which  in the present case can be approximated  as $0.05$ (only for these compatible values we have  $w_{\phi} < -1/3$, which corresponds to an inflationary phase) \footnote{It should be noted that this is only a necessary condition to have a viable inflationary phase. To be consistent with observations, additional conditions are needed, such as at least $60$ e-folds. These conditions will be considered in detail in Ref. \cite{lsw2018c}.}.  The last two subfigures of Fig.  \ref{fig3.1} show the evolution in the pre-bounce stage where the situation is similar to what  is seen in the post-bounce stage owing to the time reversal symmetry in LQC. 

Figs. \ref{fig3.2} - \ref{fig3.3} show the phase space trajectories for mLQC-I and mLQC-II, respectively.  We see that the evolution in mLQC-II is very similar to that in LQC in both of the pre-bounce and post-bounce phases. In contrast, both of these models have  dynamics quite different from that of mLQC-I in the pre-bounce phase, in which  a repelling node at the origin can be clearly discerned due to the absence of the spiral near the origin in the  last two subfigures of Fig. \ref{fig3.2}. As in the case of LQC, we can estimate the minimal value of $X$ at bounce to ensure an inflationary phase in the post-bounce regime.  In particular, the inflationary separatrix in the post-bounce stage of mLQC-I can be roughly approximated by $Y \approx \pm 0.08$ as shown in the second subfigure of Fig. \ref{fig3.2}. Consequently, the rough estimate on the minimal value of $|X|$ at the bounce to ensure the occurrence of inflationary phase is approximately $0.1$. 
Similarly, for the inflationary separatrix in mLQC-II, the second subfigure of Fig. \ref{fig3.3} provides an estimate on $Y$ as $Y \approx \pm 0.017$, which sets a lower bound on $|X|$ at the bounce ($|X|\geq 0.024$) in order for the scalar field to allow an inflationary phase in the post-bounce stage.

In the post-bounce expanding phase, all the three models, LQC, mLQC-I and mLQC-II, are approximated by classical GR at large volumes for matter which obeys the weak energy condition.\footnote{For matter violating the weak energy condition, such as phantom \cite{ssd}, LQC shows non-trivial departures from GR resulting in a resolution of the big rip singularity  \cite{ps09,sst}. The same conclusion holds for mLQC-I and mLQC-II \cite{ss18a}.} For such matter, as the scale factor increases, the Hubble rate decreases. Indeed, after the slow roll inflation in the post-bounce phase in all of these models, the value of the Hubble rate, which is of the  same order as $X$ and $Y$, becomes very small.  Therefore, to the first-order in perturbations, only the first term on the right-hand side of  Eq.(\ref{3.9}) contributes to the equations of motion of the perturbations. If $\mu$ and $\nu$ denote the corresponding perturbations of $X$ and $Y$, using Eqs.(\ref{3.8}) and (\ref{3.9}), the perturbation equations are given by
\bqn
\lb{3.10a}
\dot  \mu&=&m \nu,\\
\lb{3.10b}
\dot  \nu&=&- m \mu .
\eqn
The characteristic eigenvalues of Eqs.(\ref{3.10a})-(\ref{3.10b}) turn out to be $\pm i m$. Since both of the eigenvalues are purely imaginary,   to the first-order in perturbations, the fixed point of dynamical equations in LQC, mLQC-I and mLQC-II in the post-bounce phase can be viewed as  a center. This translates to an oscillatory behavior of solutions at the end of inflationary phase, that is  a common feature shared by all three models as reflected in the top  panels of Figs. \ref{fig3.1}-\ref{fig3.3}.  However, in the post-bounce stage of the phase portraits of Fig. \ref{fig3.1}-\ref{fig3.3},  instead of a center, there exists a stable spiral at the origin. The origin of this spiral is actually due to the friction term  $3 H^i \dot \phi $ in the Klein-Gordon equation (which is $3 H^i Y$ in (\ref{3.9})). Note that when only the first-order terms are taken into account, we obtain Eqs. (\ref{3.10a})-(\ref{3.10b}) corresponding to two Harmonic oscillators.  However, once the friction term is considered at the level of the second-order perturbations, it results in a diminishing magnitude of the oscillations around the origin. As a result, the origin in the post-bounce phase of all three models is a stable spiral and a late time attractor. This is easy to understand since all three models have the same classical  limit at late times. One can think of the oscillations as the main effect from the first-order perturbation while a decease in the magnitude of the oscillation is an additional due to the second-order perturbations.  In Appendix A, we show a rigorous treatment of the origin based on the Lyapunov theorem. The result is consistent with our analysis here.

  In the pre-bounce phase, as is evident from the phase space plots, the situation is similar for LQC and mLQC-II since classical GR is recovered on both sides of the bounce. The fixed point again turns out to be a center if only the first-order in perturbations is considered.  The Hubble rate is negative in the contracting phase, making  $3 H \dot \phi$ term anti-frictional which serves as a driving force to push any perturbations away from the origin. This makes the origin in the pre-bounce phase as an early time repeller for the forward time evolution.  Analogous to the post-bounce case, one can also treat oscillations as a first-order effect and an increase in the magnitude of oscillations as a second-order effect. 
   The situation is different for mLQC-I where the bounce is asymmetric and the pre-bounce phase is dominated by the cosmological constant of the Planck scale which has its origins in quantum geometry. In the pre-bounce phase of mLQC-I,  the modified Friedmann equation is given by Eq.(\ref{LQG1e}),  from which it is easy to see that the Hubble rate  does not become small when energy density of inflaton becomes small at large volumes at very early times. This is essentially because of the constant value of the Hubble rate due to the cosmological constant dominated dynamics in this phase.  Therefore, the equations of motion of the perturbations  (\ref{3.10a})-(\ref{3.10b}) in the contracting phase in mLQC-I receive an extra contribution from the second term on the right-hand side of Eq.(\ref{3.9}),  which changes Eqs.(\ref{3.10a})-(\ref{3.10b}) to the following:
\bqn
\dot \mu&=&m \nu, \\ 
\dot \nu&=&-m \mu+\frac{3}{\lambda (1+\gamma^2)}\nu.
\eqn
The characteristic eigenvalues are,
\bq
\lb{node1}
m_\pm=\frac{1}{2\lambda(1+\gamma^2)}\left(3\pm\sqrt{9-4m^2\lambda^2(1+\gamma^2)^2}\right),
\eq
both of which have a positive real part irrespective of the mass. If $m \leq 3/(2 \lambda (1+\gamma^2)) \approx 0.62$ then both the eigenvalues are positive and real and therefore the  origin is an unstable (repelling) node in the pre-bounce phase at the level of the first-order perturbations.    It is also worthwhile to note that if the mass is chosen to be large enough, $m > 0.62$ in Planck units, 
the characteristic eigenvalues will develop an imaginary part which will cause the growing oscillations of phase space trajectories while they are moving away from the origin in the pre-bounce stage of mLQC-I. The fixed point will be an unstable spiral. However, note that the magnitude of the scalar curvature perturbation is of the order of $10^{-9}$, which phenomenologically sets the mass of the inflaton to the order of $10^{-6}$ ($\ll 0.62$). For such a value of mass, the two eigenvalues are $m_+ \approx \frac{3}{\lambda (1+\gamma^2)}$ and $m_- \approx 0^+$. Therefore, any perturbations around the origin grow at an exponential rate.

\begin{figure} [h!]
{
\includegraphics[width=6cm]{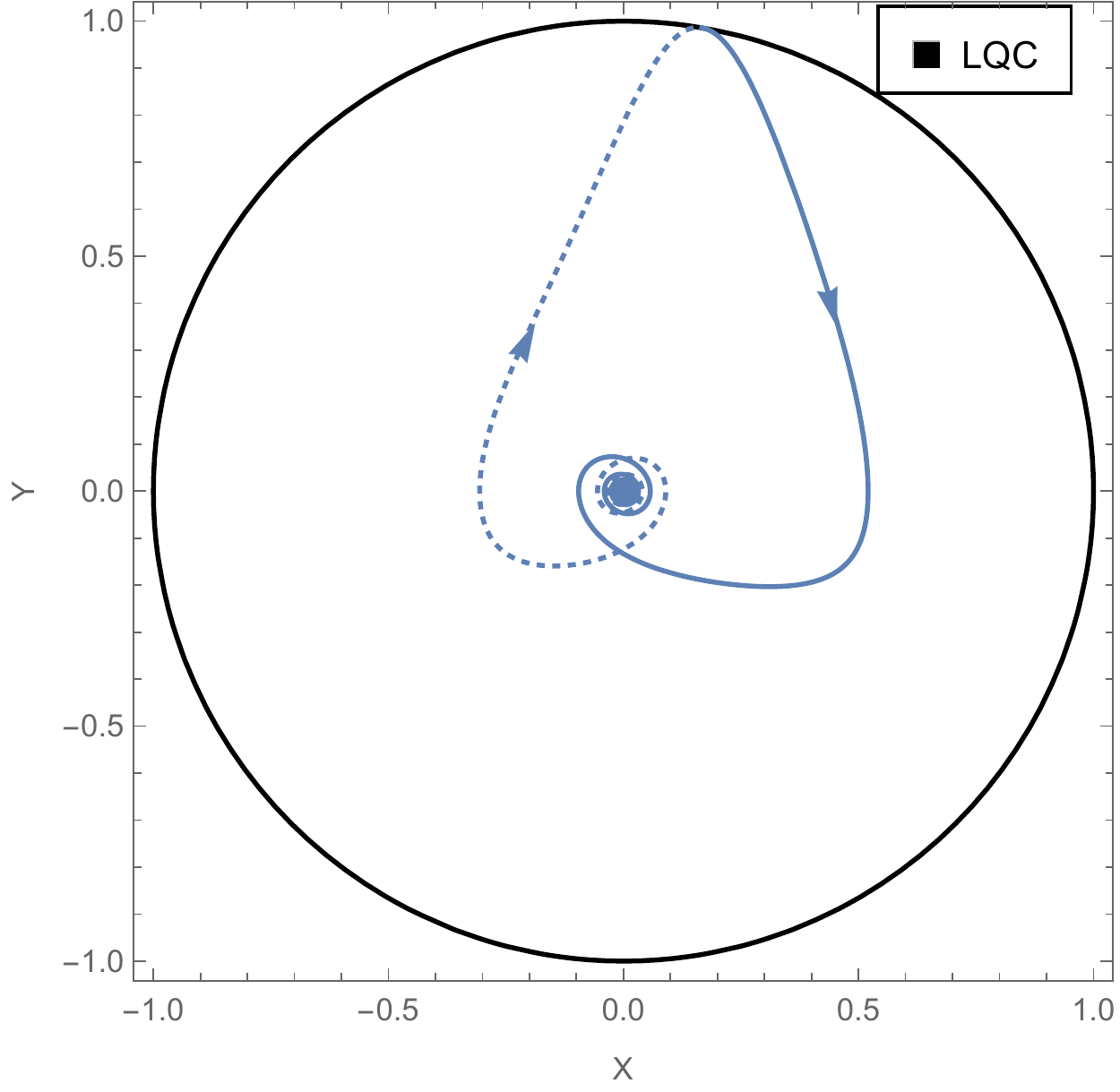}
\includegraphics[width=6cm]{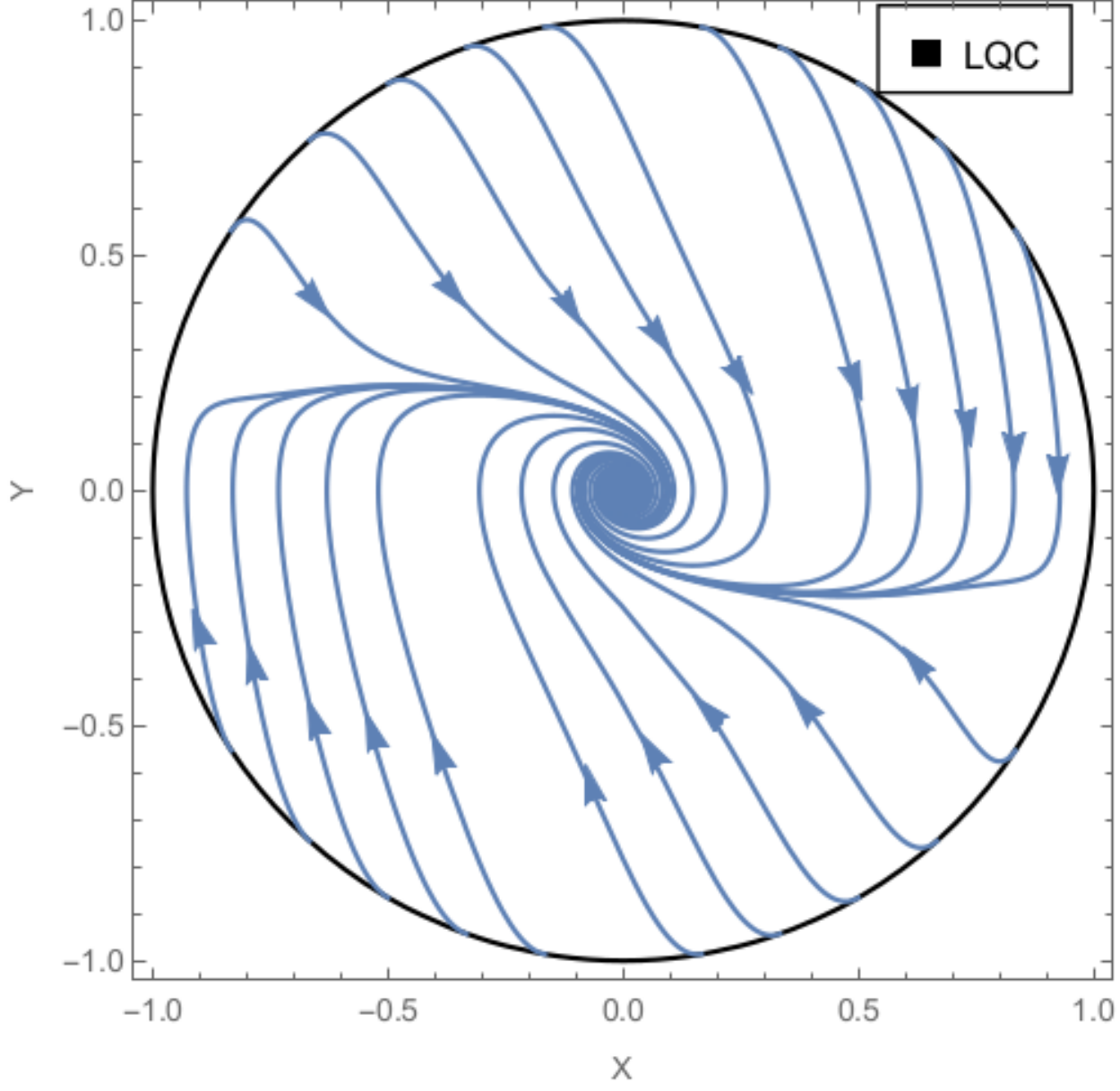}
\includegraphics[width=6cm]{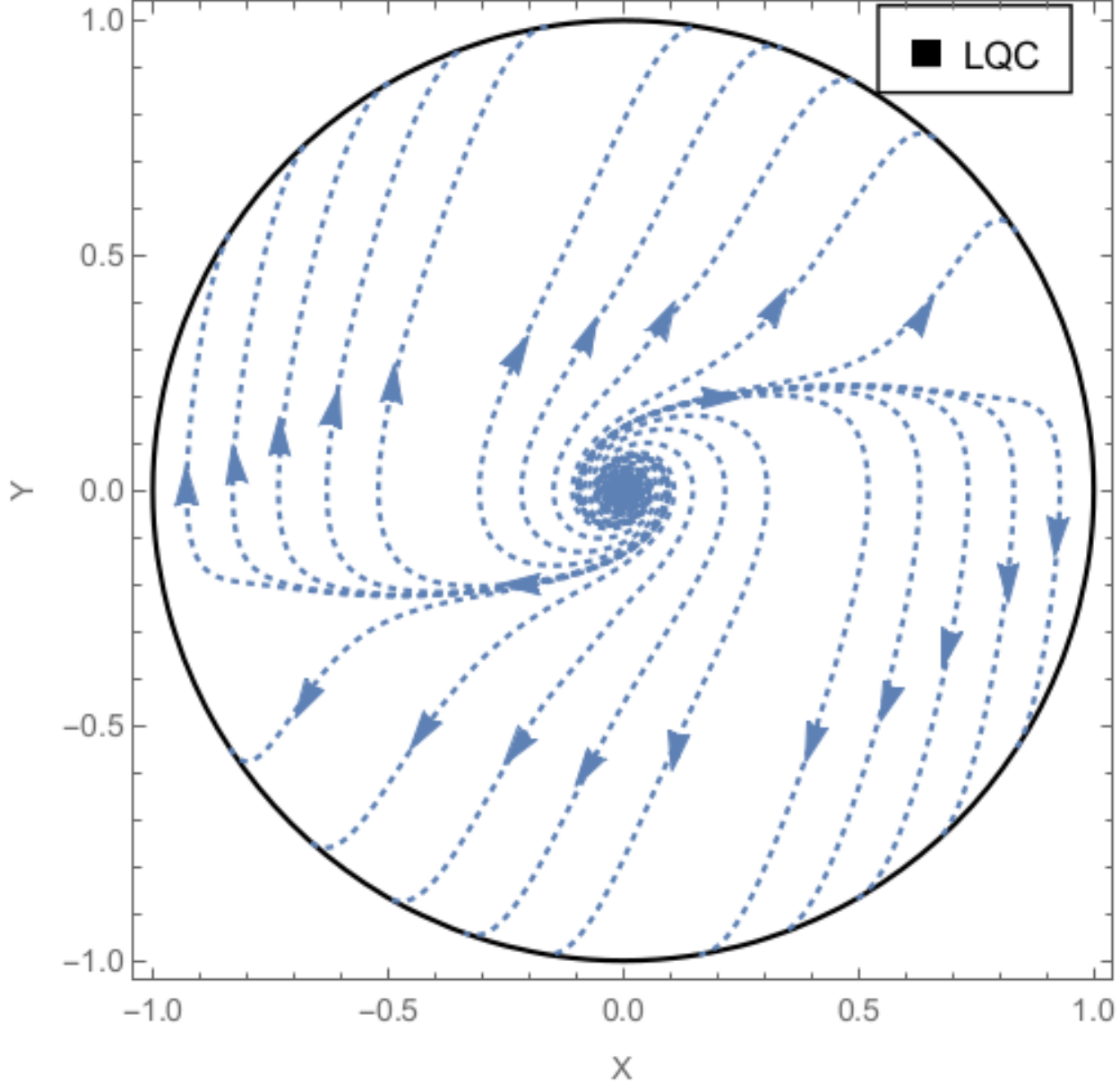}
}
\caption{Phase space portraits for the monodromy potential in LQC with $p=2/3$, $n=4$ and $\phi_0 = 0.6$ (in Planck units) in Eq. (\ref{3.7c}).  The arrows indicate  forward-time evolution of the phase space trajectories. Solid lines stand for 
trajectories in the post-bounce phase,  while the  dotted lines for those in the pre-bounce phase. The top panel shows a complete  trajectory of one particular solution from the past to the future. In the middle panel, 
more initial data are chosen at the bounce in order to show explicitly the existence of two inflationary trajectories in the post-bounce phase. Finally, the bottom panel shows the results in the pre-bounce phase.\label{fig29}}
\end{figure}

\begin{figure}  [h!]
{
\includegraphics[width=6cm]{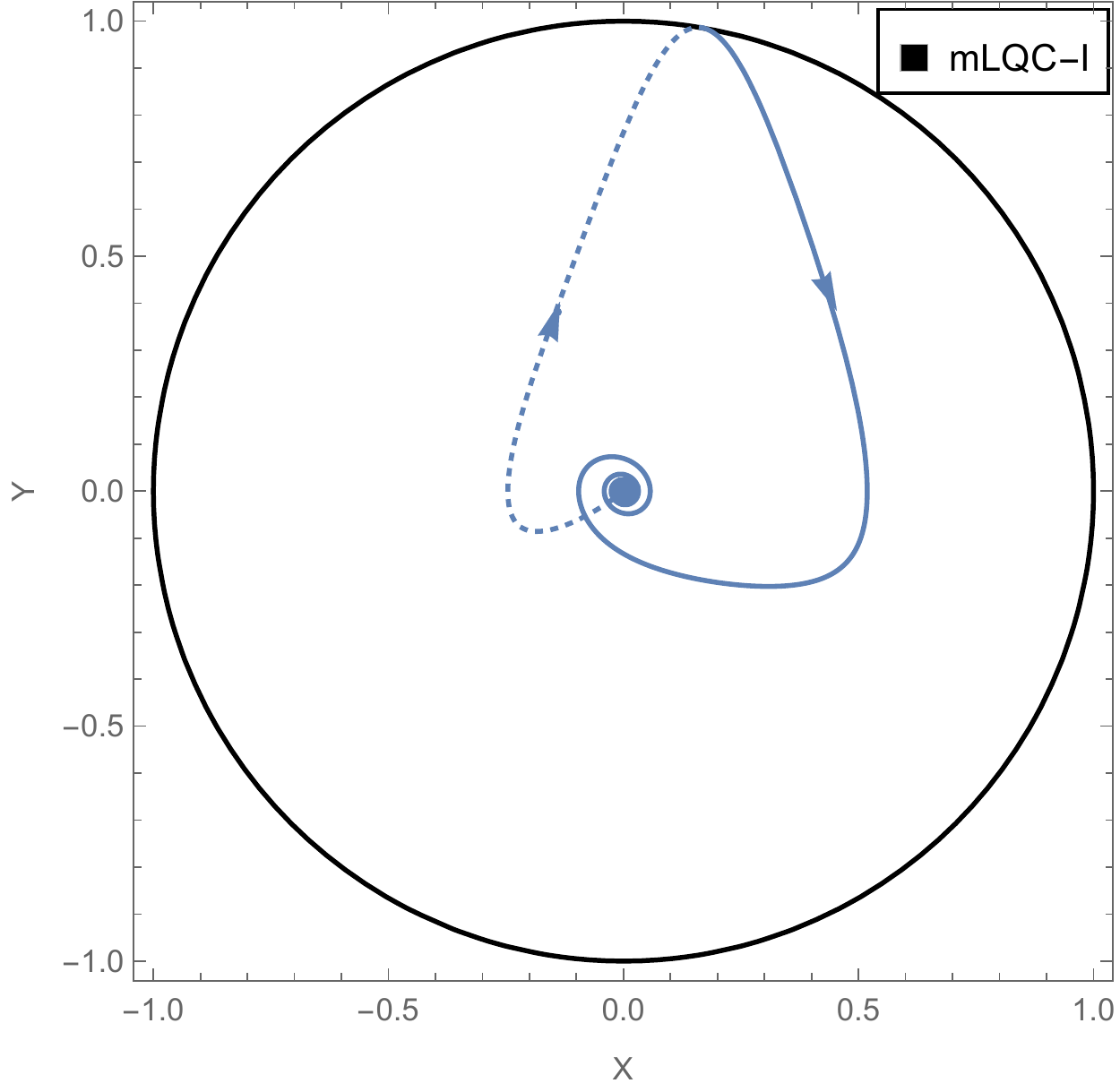}
\includegraphics[width=6cm]{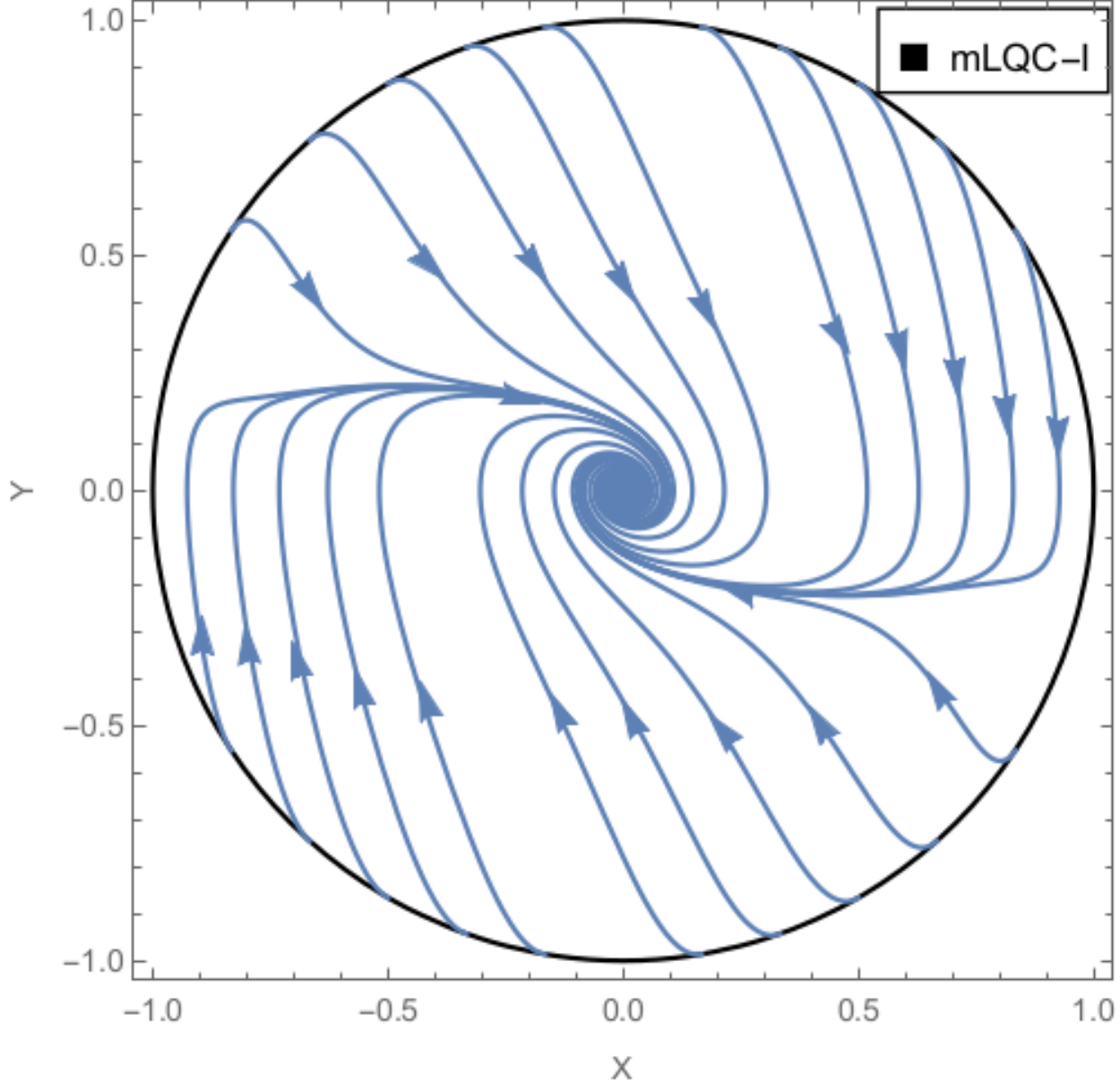}
\includegraphics[width=6cm]{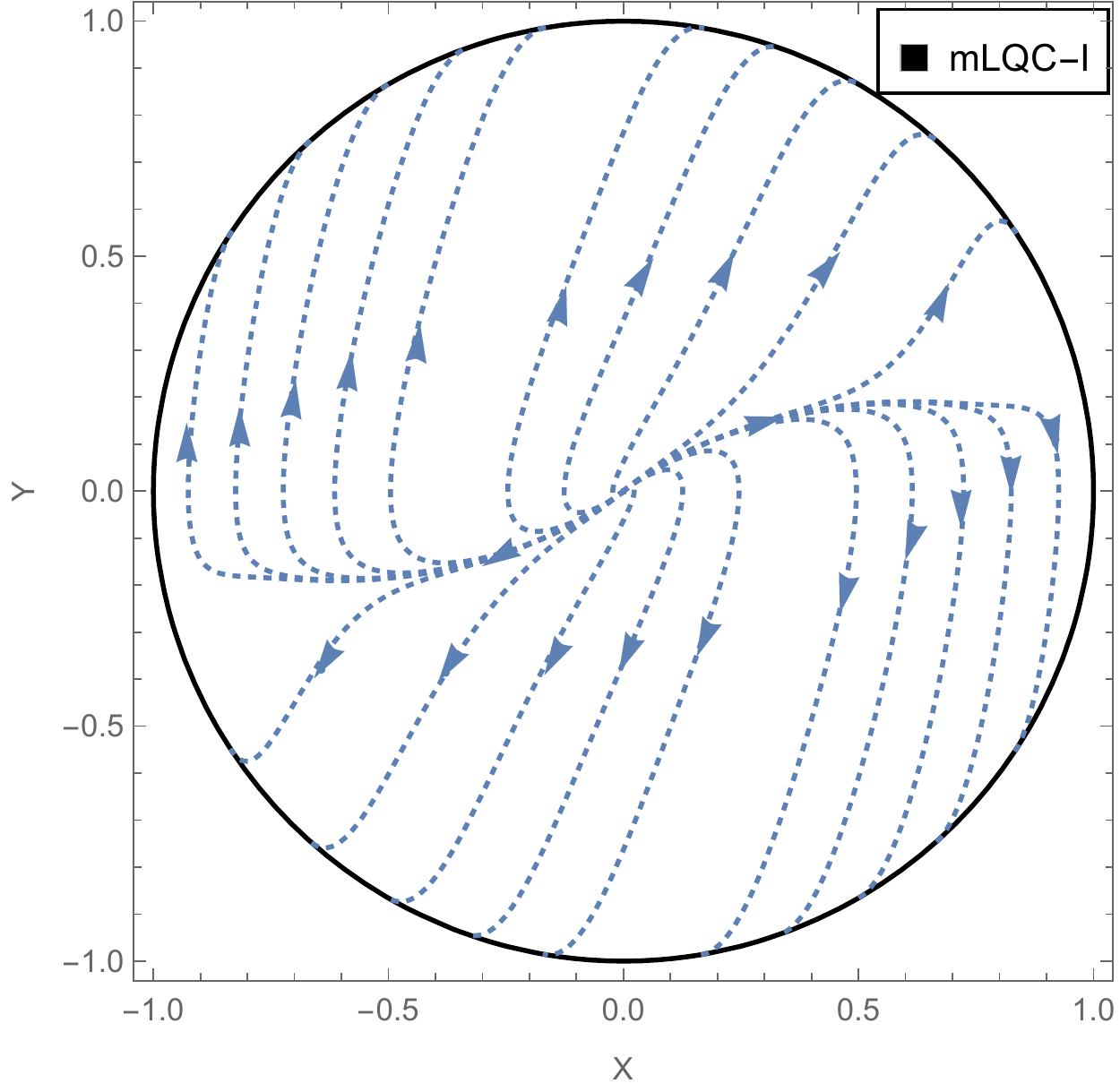}
}
\caption{ {Phase space portraits for the monodromy potential in mLQC-I with $p=2/3$, $n=4$ and $\phi_0 = 0.6$ (in Planck units) in Eq. (\ref{3.7c}).  All the trajectories in the pre-bounce phase quickly converge to the origin in the 
backward evolution of the universe. As compared with LQC and mLQC-II, there are no oscillations, when they are approaching the origin in the pre-bounce phase. }
}
\label{fig28}
\end{figure}

\begin{figure} [h!]
{
\includegraphics[width=6cm]{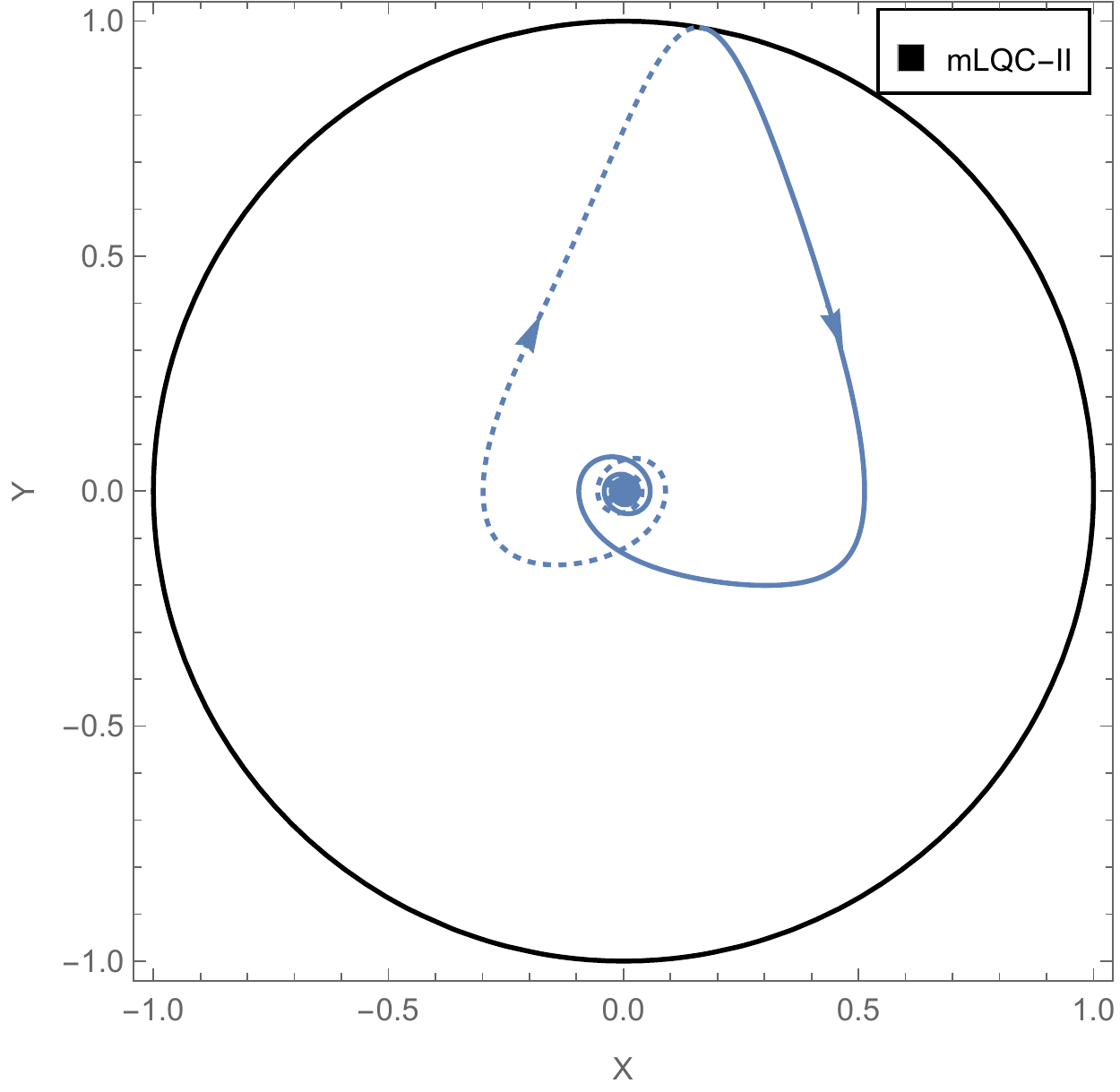}
\includegraphics[width=6cm]{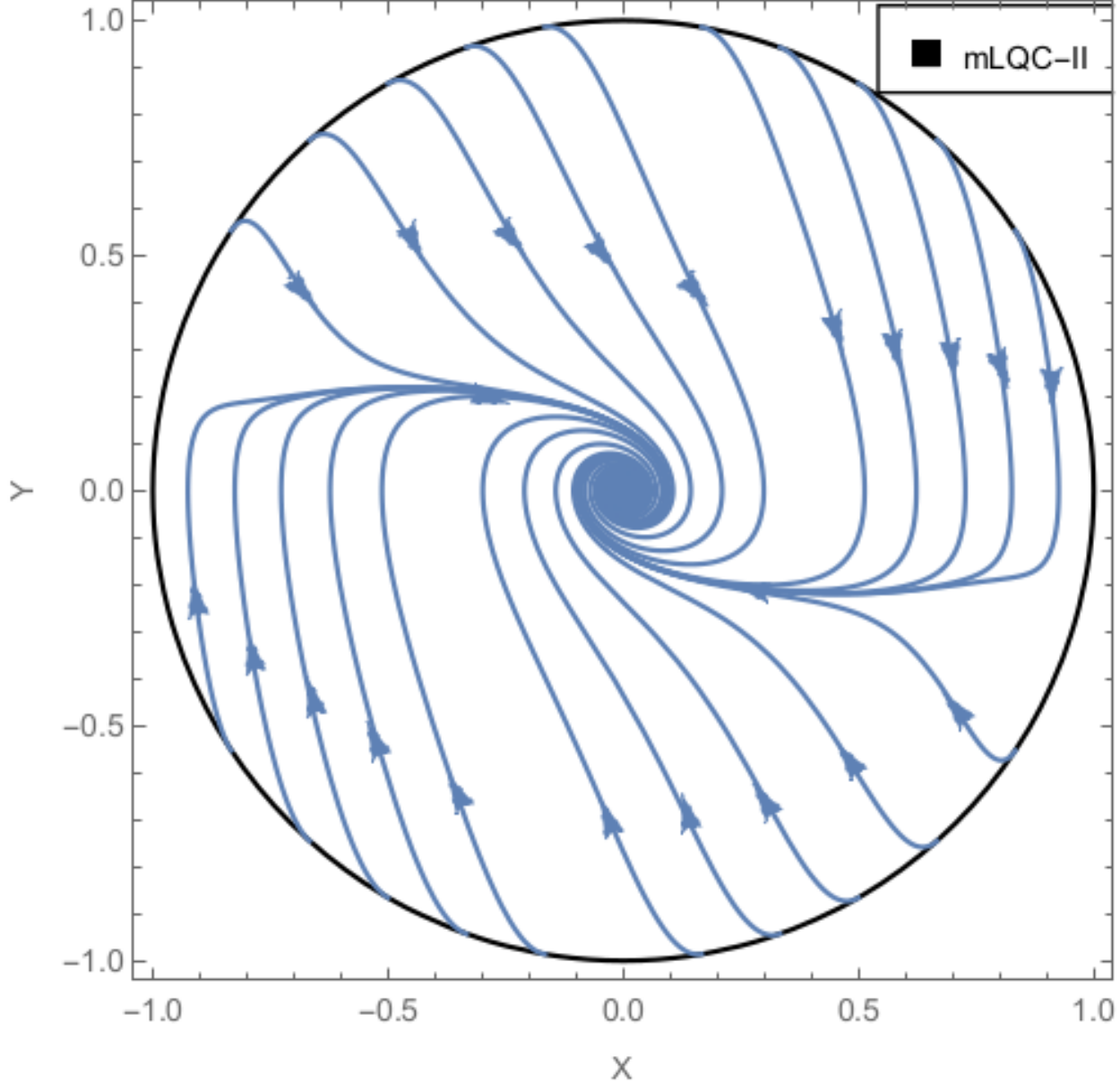}
\includegraphics[width=6cm]{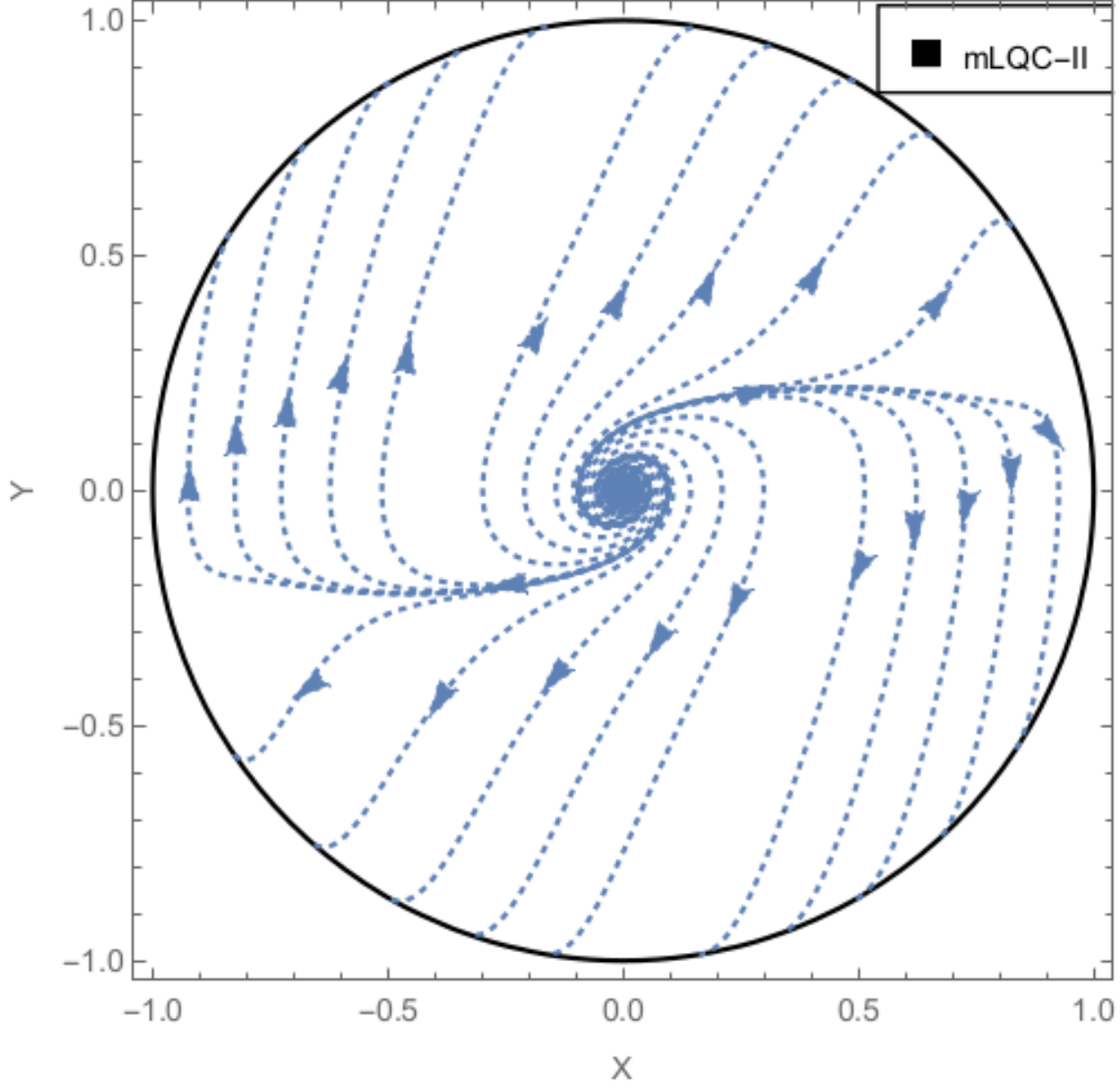}
}
\caption{ {Phase space portraits for the monodromy potential in mLQC-II with $p=2/3$, $n=4$ and $\phi_0 = 0.6$ (in Planck units) in Eq. (\ref{3.7c}). Since the evolution of the universe is symmetric about the bounce both in LQC and mLQC-II, the patterns in these plots resemble those in Fig. \ref{fig29}.
}}
\label{fig30}
\end{figure}

\subsection{Fractional Monodromy Potential }

Besides the chaotic potential, in the family of power law potentials, the monodromy potential inspired by string/M-Theory and supergravity also receives lots of attention,  as it fits the CMB data quite well \cite{Planck2015}. In this model,  the potential is  given by 
\bq
\lb{3.7b}
V=V_0 \left|\frac{\phi}{m_{\text{Pl}}}\right|^p,
\eq
where $0< p \le 1$.  The case with $p=1$ is referred to as the linear monodromy potential,  while the one with $0<p< 1$ as the  fractional monodromy potential. In this paper, we focus on the latter with a particular choice $p=2/3$. 
 A modification of the potential (\ref{3.7b}) has been proposed \cite{mst2018}:
\bq
\lb{3.7c}
V=V_1  \left|\frac{\phi}{\phi_0}\right|^p\frac{1}{[1+(\frac{\phi_0}{\phi})^n]^{\frac{2-p}{n}}},
\eq
which alleviates the problems of the monodromy potential in the reheating phase. In the modified form of the potential,  
$n$ is an integer larger than unity. (Following Ref. \cite{mst2018} we will take $n=4$). Two limits of Eq.(\ref{3.7c}) should be noticed. When $|\phi| \gg |\phi_0|$, $V\approx V_1  |\phi/\phi_0|^p$ which leads to $V_1=V_0 |\phi_0/m_{\text{Pl}}|^p$.  On the other hand, when $|\phi| \ll |\phi_0|$, $V\approx V_1  \left(\phi/\phi_0\right)^2$.  Before reheating, the scalar field assumes a large value (much larger than $\phi_0$), and in this stage the universe is governed by the fractional monodromy potential.  However,  towards the end of inflation,  the scalar field  becomes almost zero and the potential resembles the chaotic inflationary one,  which ensures a continuous transition from inflation to reheating.  One then chooses $\phi_0$  equal to the value of the scalar field at the end of inflation. For the phase before reheating,  Eq.(\ref{3.7b}) sets $\phi_0=p/\sqrt{16\pi}m_{\text{Pl}}$,  which is about $0.094m_{\text{Pl}}$ for $p=2/3$. One may apply the modified potential Eq.(\ref{3.7c}) with all the parameters fixed as stated above, except $V_0$, which is directly fixed by the amplitude of the scalar curvature perturbation from CMB data \cite{mst2018}. However, for the phase space portraits, we choose a different set of parameters for a faster convergence. In Figs. \ref{fig29}-\ref{fig30},  we simply choose  $V_1=\rho^i_c (\phi_0/m_\text{Pl})^{2/3}$ with  $\phi_0=0.6m_{\text{Pl}}$.  The phase space variables are defined by
\bq
  X=\frac{\phi}{m^{1/3}_\text{Pl}\left(\phi_0^4+\phi^4\right)^{1/6}}, \quad Y=\frac{\dot \phi}{\sqrt{2\rho^i_c}},
\eq
 such that all the physical trajectories are confined within the unit circle.  Following the above parameterization, the Klein-Gordon equation of the scalar field is equivalent to two first-order differential equations which read
\bqn
\lb{monoxy1}
\dot X&=&\frac{\left(3\phi_0^4+\phi^4\right)\sqrt{2\rho^i_c}Y}{3m^{1/3}_\text{Pl}\left(\phi^4_0+\phi^4\right)^{7/6}}, \\
\lb{monoxy2}
\dot Y &=& -3 H^i Y-\frac{V_{,\phi}}{\sqrt{2\rho^i_c}},
\eqn 
where $V_{,\phi}$ is explicitly given by 
\bq
V_{,\phi}=\frac{2 \rho^i_c \phi \left(3\phi^4_0+\phi^4\right)}{3 m^{2/3}_\text{Pl}\left(\phi^4_0+\phi^4\right)^{4/3}}.
\eq
The system governed by Eqs. (\ref{monoxy1})-(\ref{monoxy2}) is closed, since $\phi$ is a function of $X$, and the Hubble parameter $H^i$ only depends on the energy density which in turn is a function of $X$ and $Y$. As a result, the fixed point of the system can be directly read off from the right-hand side of Eqs. (\ref{monoxy1})-(\ref{monoxy2}) with $\dot X=\dot Y=0$, which has only one solution $(X, Y)=(0, 0)$. To find out properties of the fixed point, we can simply perturb the system around the origin with $X=\mu$ and $Y=\nu$. 

To understand the nature of this fixed point, let us perform a perturbation analysis. 
The linearized equations of motion for perturbations $\mu$ and $\nu$ in the post-bounce stage of LQC, mLQC-I  and mLQC-II assume the form:
\bqn
\dot \mu&=&\delta \nu,\\
\dot \nu &=&-\delta \mu,
\eqn
where $\delta= \sqrt{2\rho^i_c}m^{-\frac{1}{3}}_\text{Pl}\phi_0^{-\frac{2}{3}}$ with the characteristic eigenvalues given by $\pm i \delta$. Therefore, similar to the fixed point in the  chaotic potential,   if only the  first-order perturbations are taken into account, the origin in the fractional monodromy potential appears to be a center. However, the friction term $3 H^i Y$ in Eq. (\ref{monoxy2}) again plays the role of diminishing the magnitude of the oscillations gradually. Therefore, the origin turns out to be a late time stable spiral which is an attractor as evident in the second panels of Fig. \ref{fig29}-\ref{fig30}. 
  This is obvious because the potential we use in Eq. (\ref{3.7c}) reduces to the chaotic potential in the reheating phase. As a result, all the conclusions about this fixed point  in the chaotic potential can be carried over to the fractional monodromy potential. In particular, the origin in the pre-bounce phase of LQC and mLQC-I is an early time repeller at the level of the second-order perturbations, consistent with the behavior of the phase space trajectories in the last panels of  Figs. \ref{fig29} and \ref{fig30}. Meanwhile, due to the quantum cosmological constant in the pre-bounce stage of mLQC-I, the evolution equations for  the perturbations $\mu$ and $\nu$ take the form 
\bqn
\dot \mu&=&\delta_{\scriptscriptstyle{\mathrm{I}}} \nu,\\
\dot \nu &=&-\delta_{\scriptscriptstyle{\mathrm{I}}}  \mu+\frac{3\nu}{\lambda \left(1+\gamma^2\right)},
\eqn
from which one can easily find the following two characteristic eigenvalues 
\bq
\lb{node2}
m_\pm=\frac{1}{2\lambda(1+\gamma^2)}\left(3\pm\sqrt{9-4 \delta^2_{\scriptscriptstyle{\mathrm{I}}}\lambda^2(1+\gamma^2)^2}\right),
\eq
with $\delta_{\scriptscriptstyle{\mathrm{I}}}= \sqrt{2\rho^{\scriptscriptstyle{\mathrm{I}}}_c}m^{-\frac{1}{3}}_\text{Pl}\phi_0^{-\frac{2}{3}}$. 
The eigenvalues are real and positive if 
\bq
\phi_0 \ge \left\{ \frac{2 \lambda \sqrt{2 \rho^{\scriptscriptstyle{\mathrm{I}}}_c} (\gamma^2+1) }{3 m^{1/3}_\text{Pl} }\right\}^{3/2} \approx 0.59 m_\text{Pl} ~.
\eq
In this case the fixed point is an unstable node as we can see in the last subfigure of Fig. \ref{fig28}. However, if $\phi_0$ is chosen to be a smaller number than above,  the square root term in Eq. (\ref{node2}) would become  purely imaginary.  Thus,   the phase space trajectories would also exhibit oscillating behavior as they are moving away from the origin. The fixed point turns out to be an unstable spiral.

  Compared with the phase portraits in chaotic potential, Figs. \ref{fig29} - \ref{fig30} also exhibit  two separatrices in the post-bounce stage of all three models and the pre-bounce phase of LQC and mLQC-II. This indicates that the inflationary solutions are also attractors in the fractional monodromy potential in LQC, mLQC-I and mLQC-II in the post-bounce regimes. Besides, as we employ the modified potential in our simulations, at small $\phi$, the fractional monodromy potential coincides with the chaotic potential. As a result, the properties of the origin in all three models are the same as in the chaotic potential. Generic solutions start from the origin in the pre-bounce phase, go through the quantum bounce on the unit circle and then approach the inflationary separatrices, and finally converge to the origin in the post-bounce phase. 

\begin{figure}[h!] 
{
\includegraphics[width=6cm]{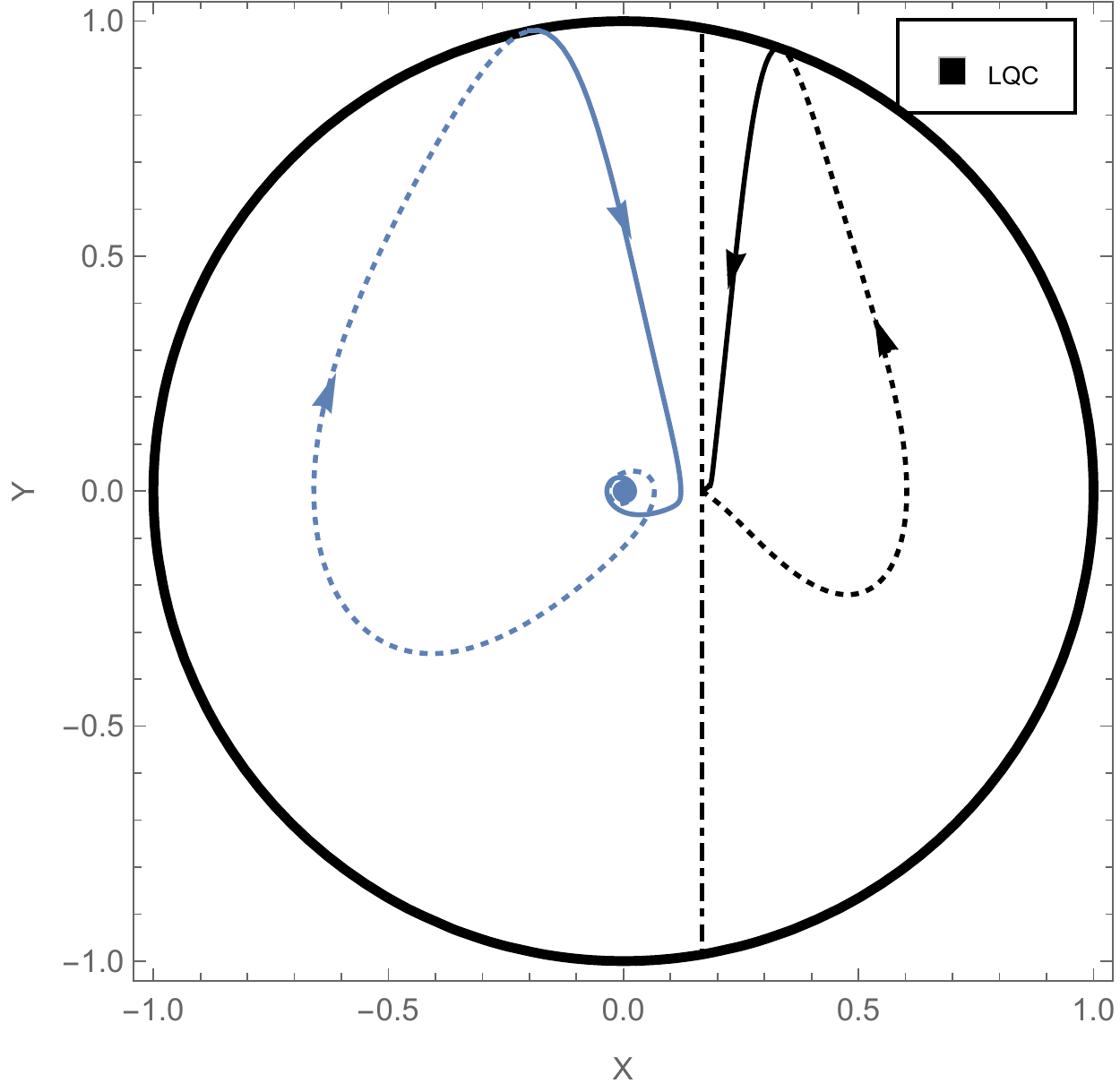}
\includegraphics[width=6cm]{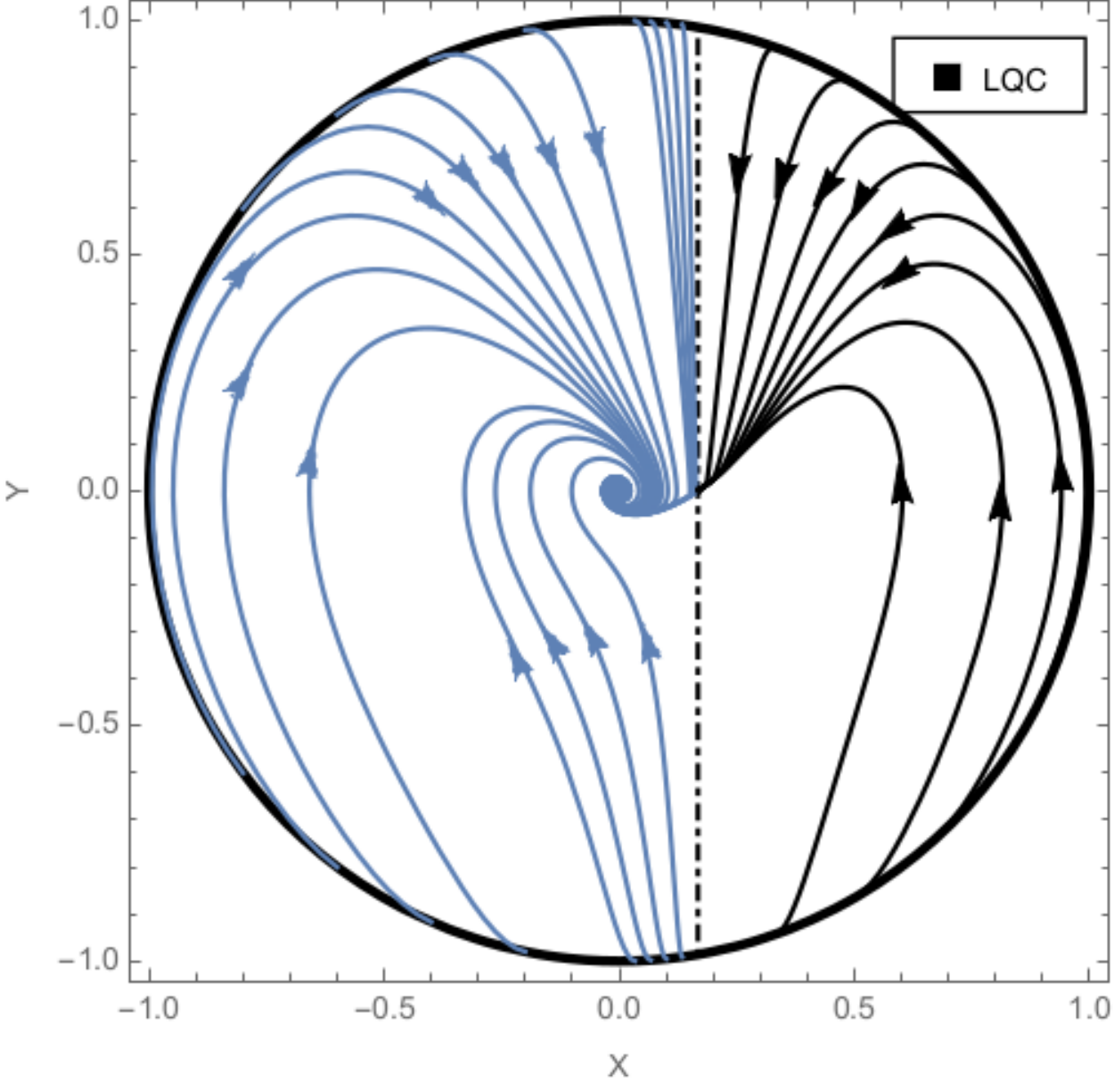}
\includegraphics[width=6cm]{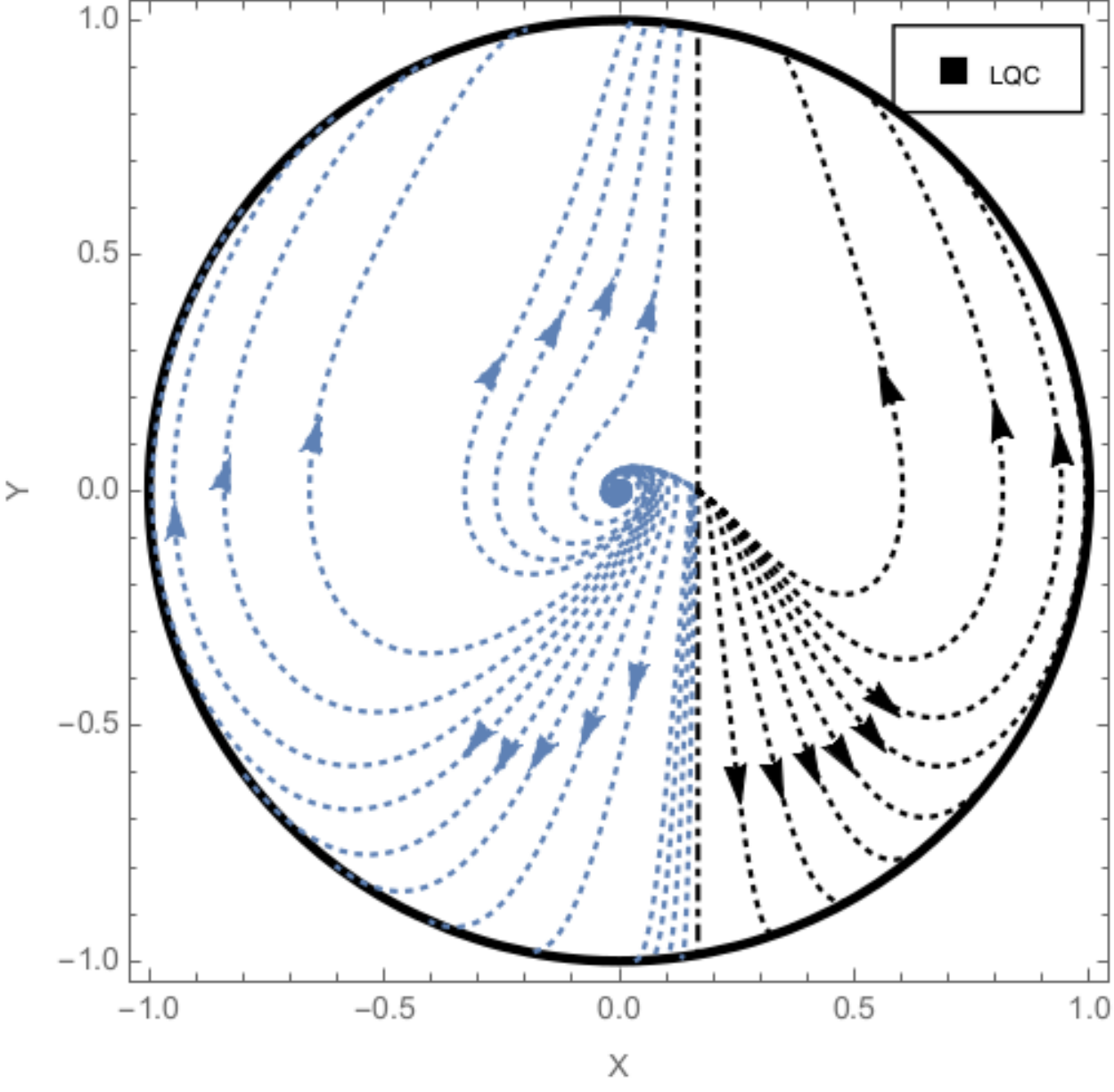}
}
\caption{ Phase portraits in LQC: (a) the left half circles for the Starobinsky potential (\ref{3.19}); and (b) the right half circles  for  the potential (\ref{SP2}).  The arrows indicate the direction of the forward-time evolution. The solid/dotted lines represent the evolution of phase space trajectories in the post/pre-bounce phase. Mass is set to $0.62$.  In the figures, the black dot-dashed vertical lines located at $X=\chi_0=0.167$ divide the whole phase portraits into two distinguishable regions.
}
\label{fig3.4b}
\end{figure}

\begin{figure}[h!] 
{
\includegraphics[width=6cm]{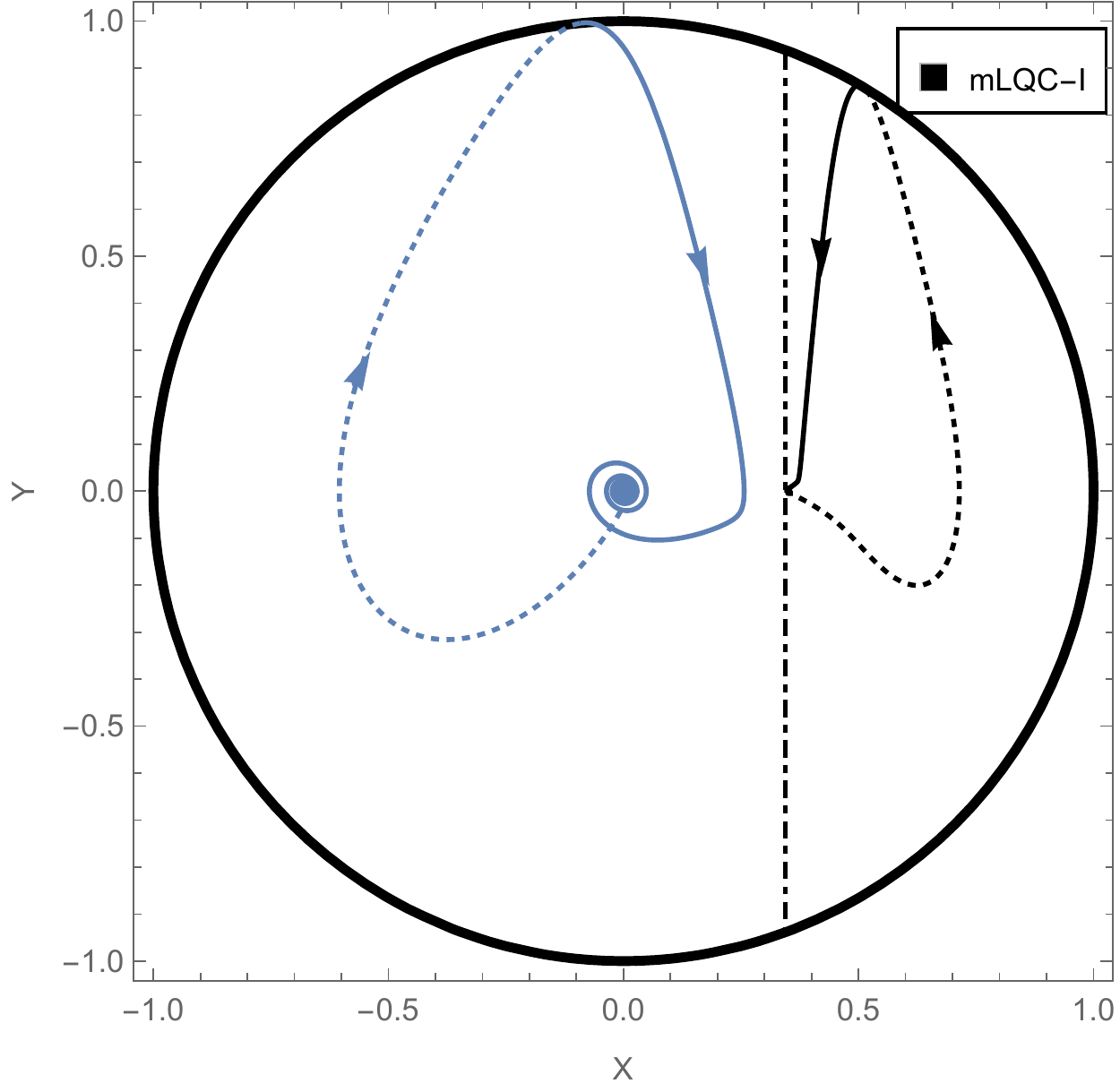}
\includegraphics[width=6cm]{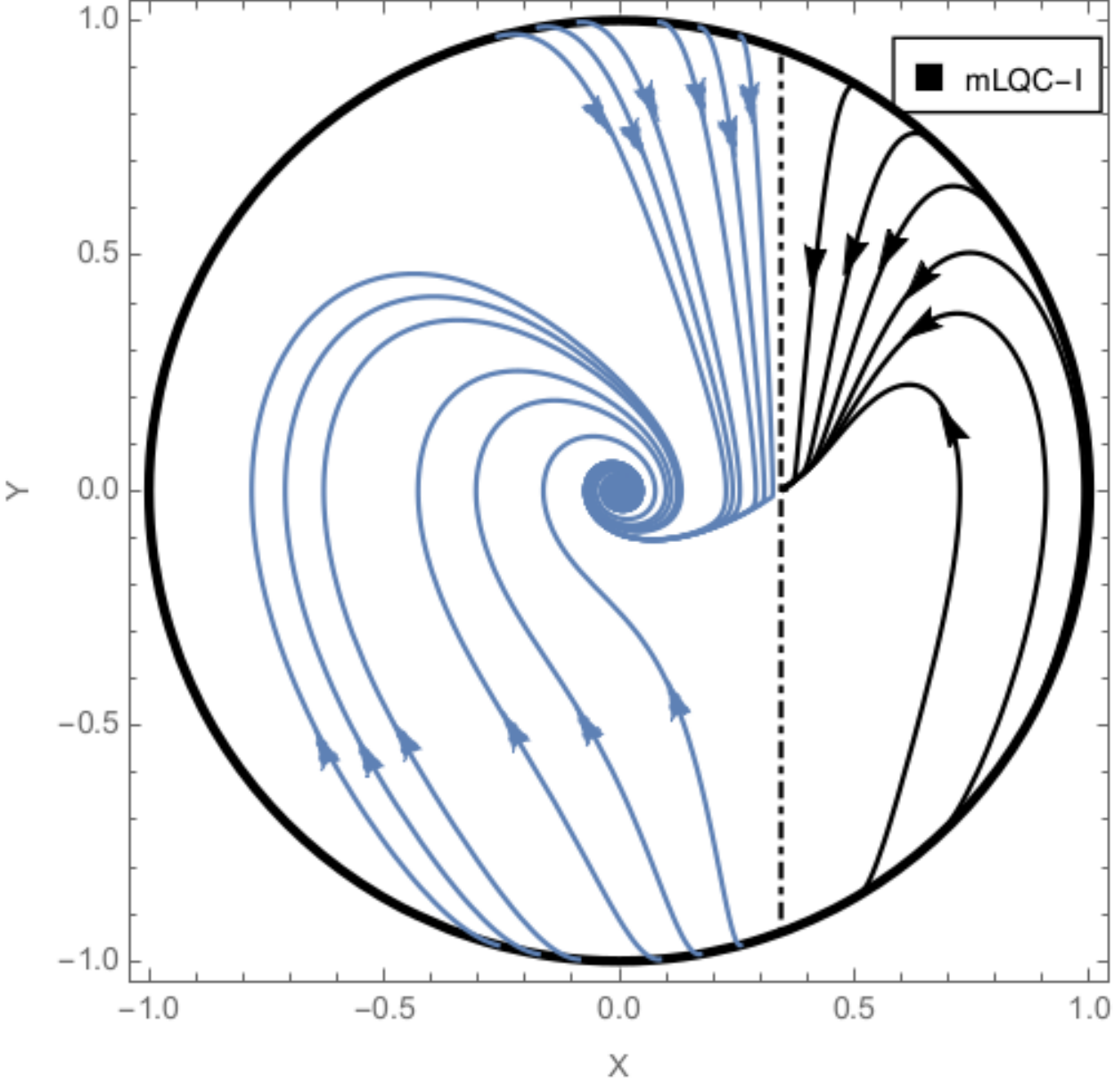}
\includegraphics[width=6cm]{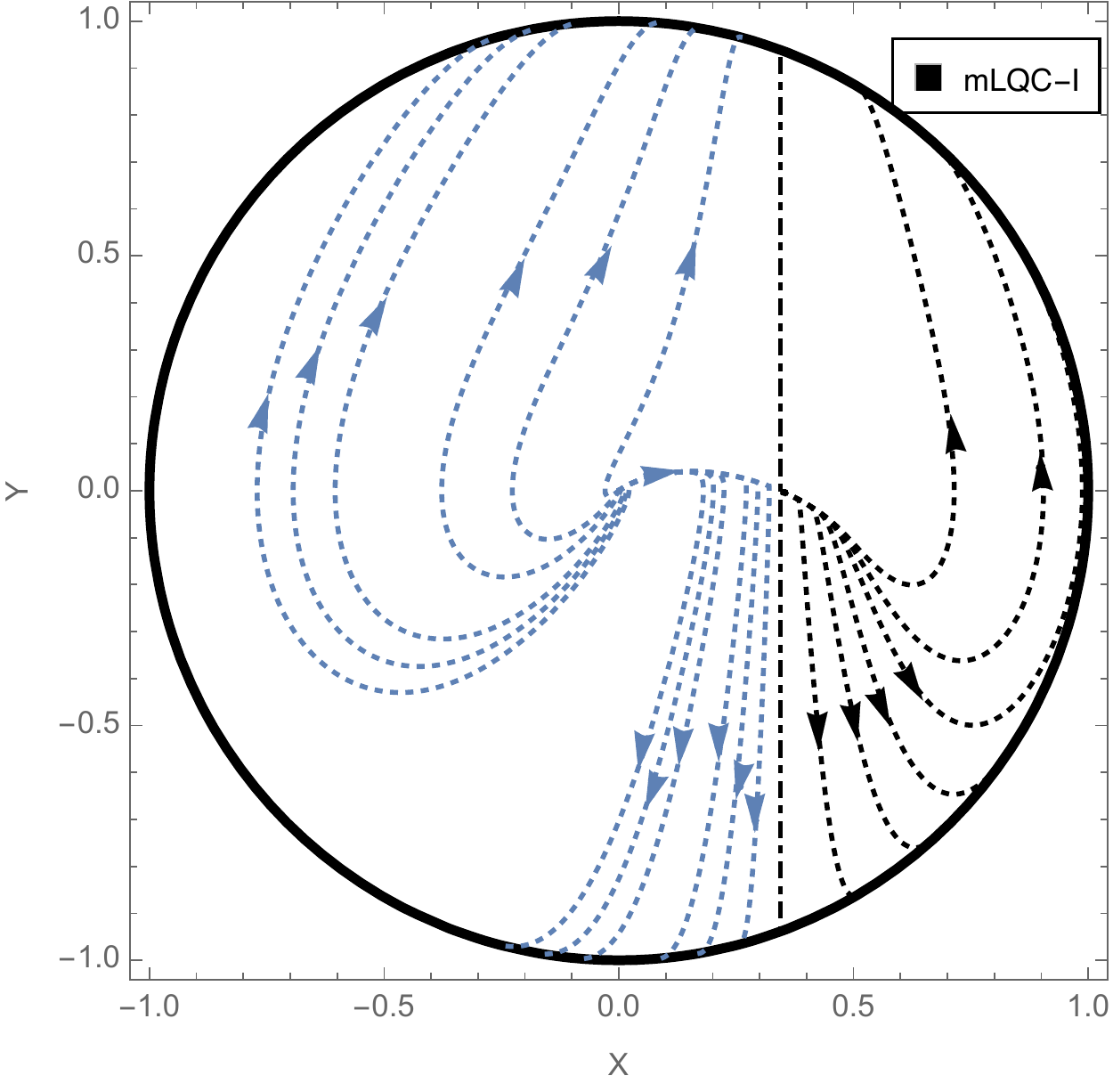}
}
\caption{Phase portraits in mLQC-I: (a) the left half circles for the Starobinsky potential (\ref{3.19}); and (b) the right half circles  for  the potential (\ref{SP2}).  
The black dot-dashed vertical lines now are located at $X=\chi_0=0.344$. The same notations, symbols and value of mass  as in Fig. \ref{fig3.4b} are used. 
}
\label{fig3.4a}
\end{figure}

\begin{figure}[h!]  
{
\includegraphics[width=6cm]{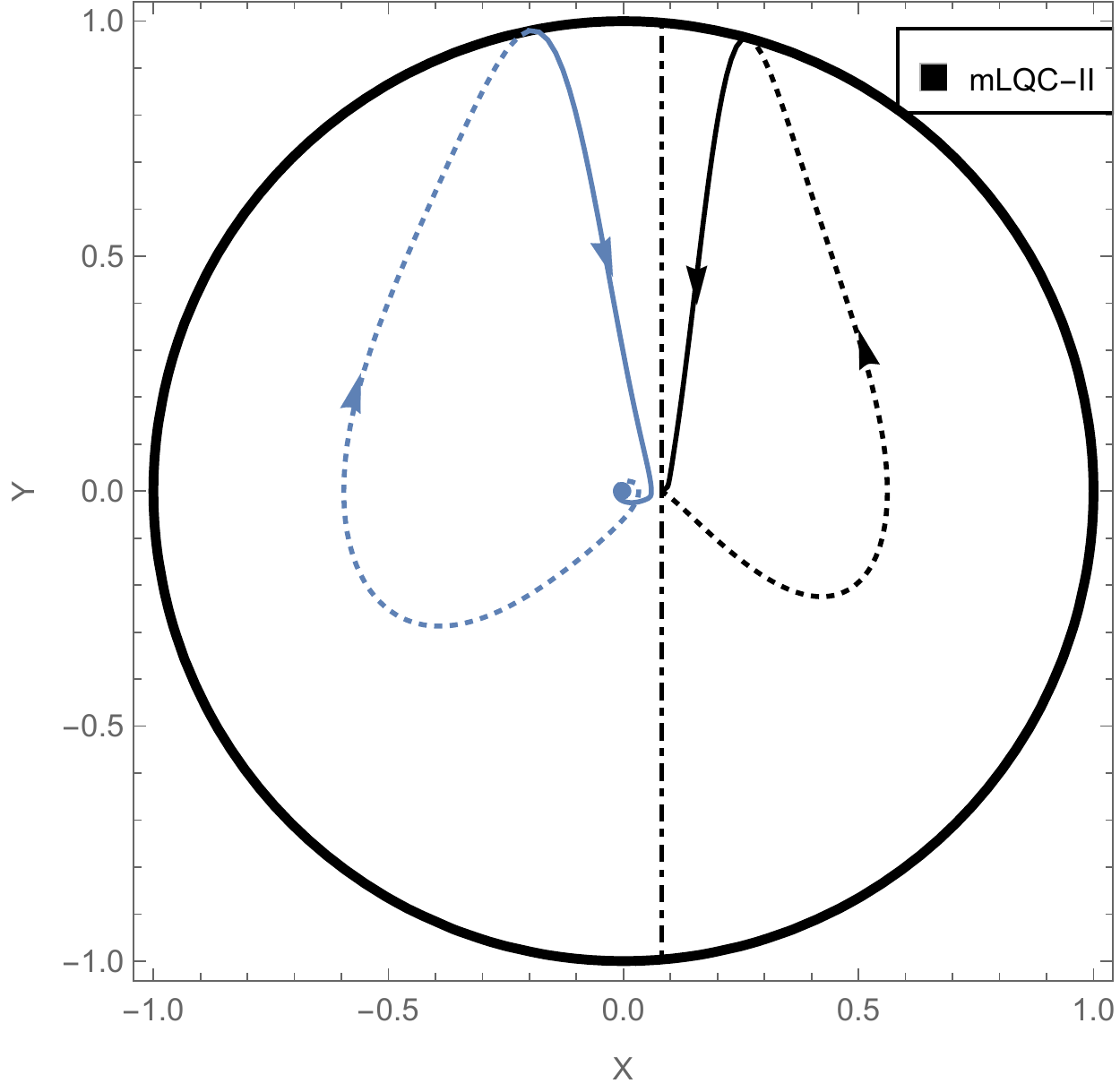}
\includegraphics[width=6cm]{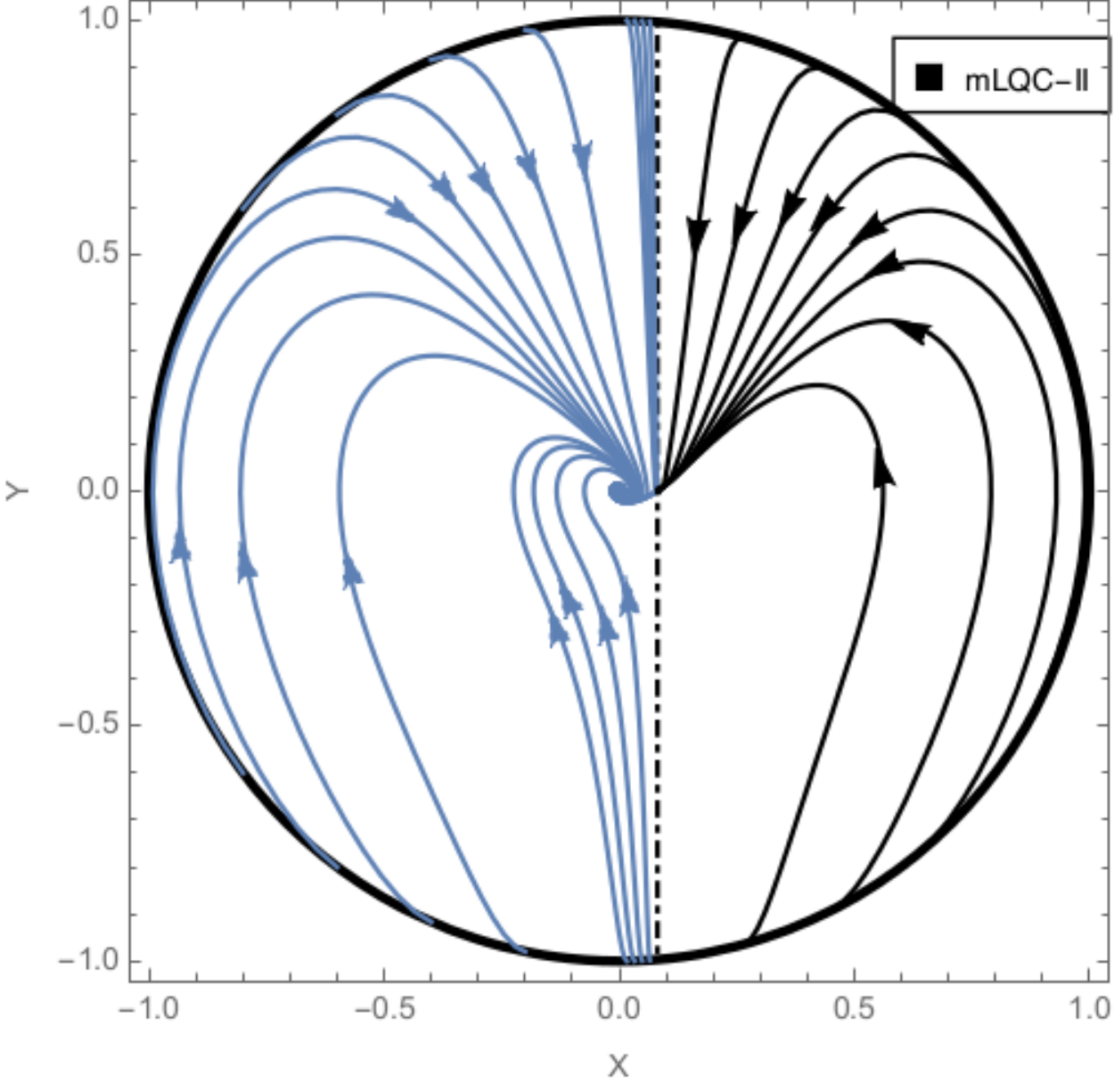}
\includegraphics[width=6cm]{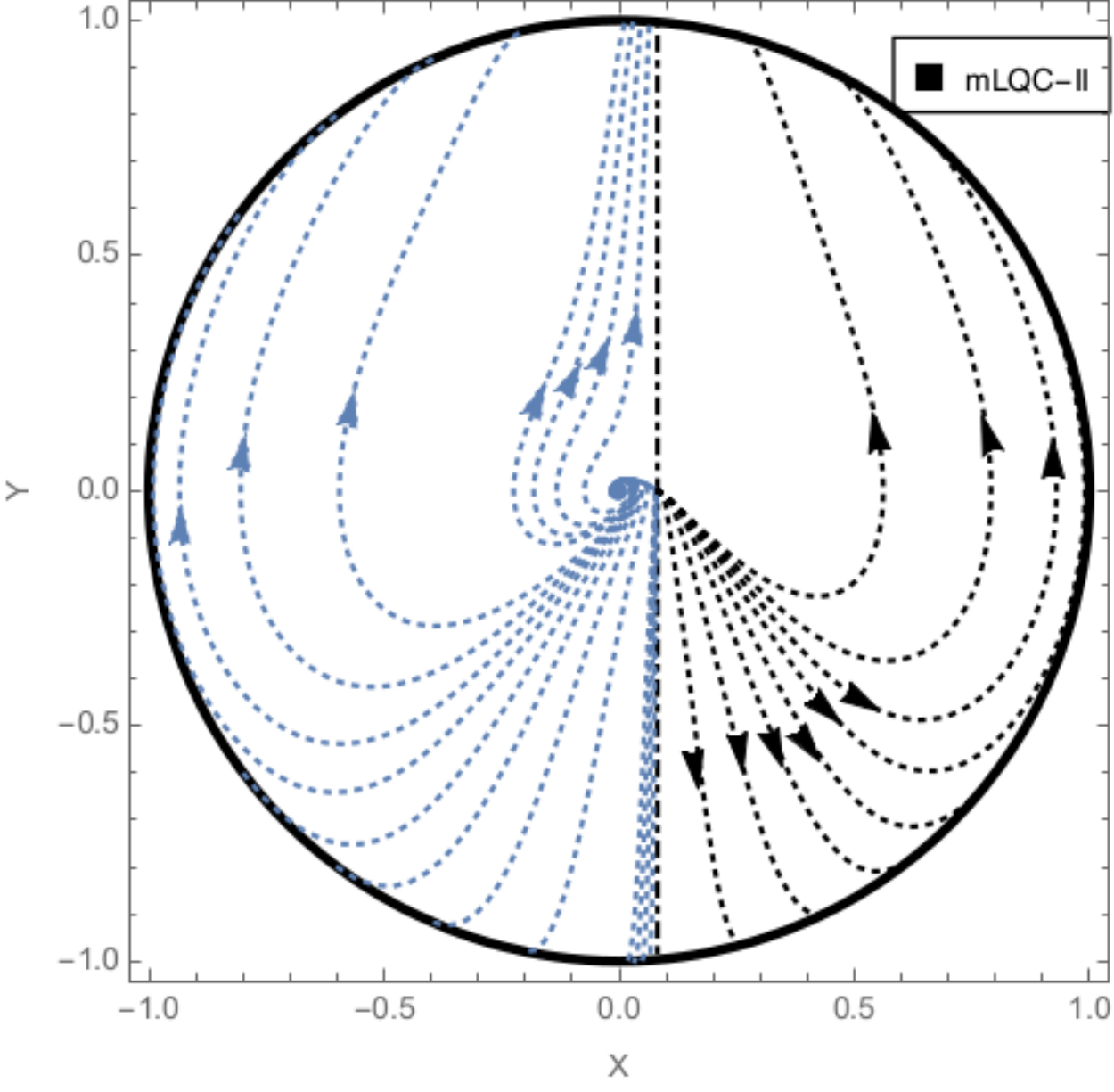}
}
\caption{Phase portraits in mLQC-II: (a) the left half circles for the Starobinsky potential (\ref{3.19}); and (b) the right half circles  for  the potential (\ref{SP2}).  
Mass is also set to $0.62$, and the black dot-dashed vertical lines now are located at $X=\chi_0=0.081$. The same notations and symbols as in Fig. \ref{fig3.4b} are used. 
}
\label{fig3.4c}
\end{figure}

\subsection{Starobinsky Potential}

In classical cosmology, Starobinsky inflation is based on adding an $R^2$ term in the gravity action, which is equivalent to adding the following potential in the Einstein frame \cite{s1980}:
\bq
\lb{3.19}
V=\frac{3m^2}{32\pi G}\left(1-e^{-\sqrt{16\pi G/3} \phi}\right)^2,
\eq
 where $m$ represents the mass of the scalar field.  The potential (\ref{3.19}) is asymmetric about $\phi=0$, unbounded for negative $\phi$ on one hand but no larger than $\frac{3m^2}{32\pi G}$ for positive $\phi$ on the other hand. An analysis of the slow-roll parameters implies that a slow-roll inflationary phase can only exist in the range of positive $\phi \ge 0$ \cite{mst2018}. It should be noted that in LQC (and in mLQC-I and mLQC-II), the starting point of the dynamics is an effective Hamiltonian and not an action. In LQC, an effective action does exist \cite{olmo}, but it contains an infinite number of higher curvature terms. If one implements $R^2$ inflation in LQC, mLQC-I and mLQC-II, the resulting potential will be different from the above expression (\ref{3.19}) in the quantum regime where there are non-trivial differences of these models from GR. However, one can phenomenologically investigate effects of Starobinsky potential in these models by assuming above form in the effective Hamiltonian of these models. In LQC\footnote{ A recent investigation on the pre-inflationary dynamics of Starobinsky potential in the context of loop quantum Brans-Dick cosmology can be found in \cite{JMZ18}.}, this phenomenological approach was first used in Ref. \cite{bg2016} where some aspects of pre-inflationary dynamics and resulting signatures in CMB were discussed. 
 
 Similar to the power law potential, together with the phase space variables defined by 
 \bqn
\lb{starx}
X&=&\chi_0 \left(1-e^{-\sqrt{16\pi G/3}\phi}\right),\\
Y&=&\frac{\dot \phi}{\sqrt{2\rho^i_c}},
\eqn
with $\chi_0 :=  \sqrt{\frac{3m^2}{32\pi G \rho^i_c}}$, the Klein-Gordon equation can be transformed into a set of the first-order differential equations:
\bqn
\lb{3.20a}
\dot X&=&m Y \left( 1-X/\chi_0\right),\\
\lb{3.20b}
\dot Y&=&-3 H^i Y-m X \left(1-X/\chi_0\right).
\eqn
As a result, all the trajectories in the phase space are confined within the unit circle: $X^2 + Y^2 \leq 1$.  Dynamical equations (\ref{3.20a})-(\ref{3.20b}) have two fixed points: the origin and ($\chi_0$, 0). In order to analyze the stability of point  ($\chi_0$, 0), one can plug perturbations around this point: 
\bq
\lb{3.21}
X=\chi_0+\mu, \quad \quad Y=\nu,
\eq
into the dynamical equations (\ref{3.20a})-(\ref{3.20b}).  In LQC, mLQC-I, and  mLQC-II, this leads to the following  equations of motion of the perturbations: 
\begin{equation}
\lb{starc1}
\dot \mu = 0, ~~~~
\dot \nu = m \mu \mp 3  \tilde H^i \nu, 
\end{equation}
where $\tilde H^i =H^i|_{\rho=\rho^i_c \chi_0^2}$ and $\mp$ stands for expanding/contracting phase, respectively. At the fixed point, the Hubble rate does not vanish as
\bqn
\rho&=&\rho^i_c (X^2+Y^2)=\rho^i_c \chi^2_0 + {\cal{O}}(\mu,\nu),
\eqn 
where  higher-order terms in $\mu$ and $\nu$ are suppressed. Correspondingly, the eigenvalues of Eq. (\ref{starc1}) become
\bqn
m_-=0 \quad \text{and} \quad m_+=\mp 3 \tilde H^i.
\eqn
Since one of the eigenvalues is zero, the fixed point is non-simple (degenerate). We note that 
  in the contracting phase, one of the eigenvalues $m_+$ is positive and the fixed point is unstable in the pre-bounce phase of LQC, mLQC-I and mLQC-II. On the other hand, in the expanding phase, $m_-=0$ and $m_+=-3  \tilde H^i<0$, one thus comes up with a critical problem of stability which cannot be solved by simply linearizing the system (\ref{3.20a})-(\ref{3.20b}). In this particular situation, it is the higher-order terms that determine the stability of the equilibrium. Detailed analysis using center manifold theory in the Appendix shows that the fixed point $(\chi_0, 0)$ is also unstable in the post-bounce phase which is consistent with the observations in the phase space  portraits. It is to be noted that the phase portraits are in $X$ and $Y$, not $\mu$ and $\nu$ which if plotted would yield a line of fixed points due to zero eigenvalue at the linearized order. Due to this zero eigenvalue a more careful analysis performed in the Appendix becomes necessary.
 
 Following Appendix, from Eq.  (\ref{app10}), one can find the approximate solution of $X$ near the above fixed point: 
 \bq
 \lb{dis}
 X=\chi_0+\left(c_0 t- c_1\right)^{-1},
 \eq
where $c_0=\frac{m^2}{3 \tilde H^i \chi_0}>0$ and $c_1$ is an integration constant. At the initial time $t=0$, the points on the left-hand side of the fixed point acquires a positive $c_1$ while the  points on the other side acquires a negative $c_1$.  
This difference results in the distinctive behavior of the trajectories on the two sides of  $X=\chi_0$, depicted by the vertical lines in Figs. \ref{fig3.4b}-\ref{fig3.4c}. As one can see in the second subfigures of Figs. \ref{fig3.4b}-\ref{fig3.4c}, all the trajectories on the right-hand sides of $X=\chi_0$ tend to converge to the fixed point since in these regions $c_1<0$, and as a result, the second term in the approximate solution (\ref{dis}) decreases with  time. On the other hand, all the trajectories on the left-hand sides of $X=\chi_0$ move away from the fixed point, since the positivity of $c_1$ makes the second term of Eq. (\ref{dis}) grow up quickly when time elapses. Moreover, as the second eigenvalue of the system in the expanding phase is negative, the $Y$ components of all the trajectories quickly converge to the $X$-axis. 

It is interesting to note that in each of the phase portraits of Figs. \ref{fig3.4b}-\ref{fig3.4c}, it is the left-hand side of the vertical line ($X < \chi_0$) that corresponds to  the Starobinsky potential (\ref{3.19}).  In the right-hand side of the vertical line ($X > \chi_0$), 
 the autonomous system of Eqs.(\ref{3.20a}) and (\ref{3.20b})   corresponds to a scalar field $\tilde\phi$ with a potential 
\bq
\lb{SP2}
{V}(\tilde\phi) \equiv \frac{3m^2}{32\pi G}\left(1+e^{-\sqrt{16\pi G/3} \tilde\phi}\right)^2,
\eq
where  $\tilde\phi$ in terms of $X$ is given by, 
\bq
\lb{SP2a}
\tilde\phi \equiv -\sqrt{\frac{3}{16\pi G}}\ln\left(\left|1-\frac{X}{\chi_0}\right|\right).
\eq
Therefore, the phase portraits of Figs. \ref{fig3.4b}-\ref{fig3.4c} actually correspond to two different scalar fields, one with the  Starobinsky potential (\ref{3.19}) and one with the potential of Eq.(\ref{SP2}), which can be obtained from  Eq.(\ref{3.19})
by setting $\phi  = \tilde\phi + \phi_0$ with $\phi_0 \equiv  -i \pi \sqrt{\frac{3}{16\pi G}}$. So, the two disconnected regions (divided by the vertical line $X = \chi_0$ or $\phi = + \infty = \tilde\phi$) actually represent  two different inflationary models with distinct potentials.

For the other fixed point, the origin, since the  dynamical equations (\ref{3.20a})-(\ref{3.20b}) reduce exactly to  (\ref{3.10a})-(\ref{3.10b}) of the power law potential, if the second-order terms in { $X$ and $Y$} are discarded, all the conclusions about the origin in the chaotic potential are still valid for the Starobinsky potential.    In particular,  the origin is still a late-time (early-time) attractor (repeller) in the post-bounce (pre-bounce) phase of LQC, mLQC-I and mLQC-II. In the pre-bounce phase of LQC and mLQC-II, there is an unstable spiral at the origin as evident from the last panels of Figs. \ref{fig3.4b} and \ref{fig3.4c}. While in the pre-bounce phase of mLQC-I,  depending on the value of the mass, the origin can be either an unstable spiral or an unstable node. The characteristic eigenvalues of the evolution equations for linear perturbations about the origin are still given by Eq. (\ref{node1}). In the chaotic potential, the mass is set to 0.2 which generates two real positive eigenvalues. In the current case, we can choose a small mass parameter to make both  of eigenvalues real and positive, this sets an upper limit on the mass which in Planck units is,
\bq
\lb{ST}
m\le \frac{3}{2\lambda (1+\gamma^2)}\approx 0.62.
\eq
Therefore,  in the phase space portraits Figs. \ref{fig3.4b}-\ref{fig3.4c},  where we have chosen $m=0.62$, the origin in the pre-bounce phase of mLQC-I becomes an unstable node, as can be seen from the last panel in Fig. \ref{fig3.4a}. All the trajectories in the phase space portraits Fig. \ref{fig3.4b}-\ref{fig3.4c} are confined within the region $-1\le X<\chi_0$ as expected from the foregoing analysis. Besides,  it is obvious that in contrast to the symmetric potentials  discussed earlier, there is only one slow-roll inflationary separatrix in the pre-bounce/post-bounce phase portraits of the Starobinsky potential for LQC and mLQC-II as well as in the post-bounce phase of mLQC-II,  since the inflation can only take place on the right branch of the potential with $\phi>0$.  Most trajectories starting from   $(X \textless0, Y \textless0)$, i.e. $(\phi_B\textless 0, \dot \phi_B\textless 0)$  on the unit circle do not converge to the inflationary separatrix   and thus do not experience the inflationary phase before entering into the reheating phase. As shown in the first panels of Figs. \ref{fig3.4b}-\ref{fig3.4c}, in all three models, generic solutions start from the unstable origin in the contracting phase, and approach the stable spiral at the origin after the quantum bounce.

\begin{figure}[h!] 
{
\includegraphics[width=6cm]{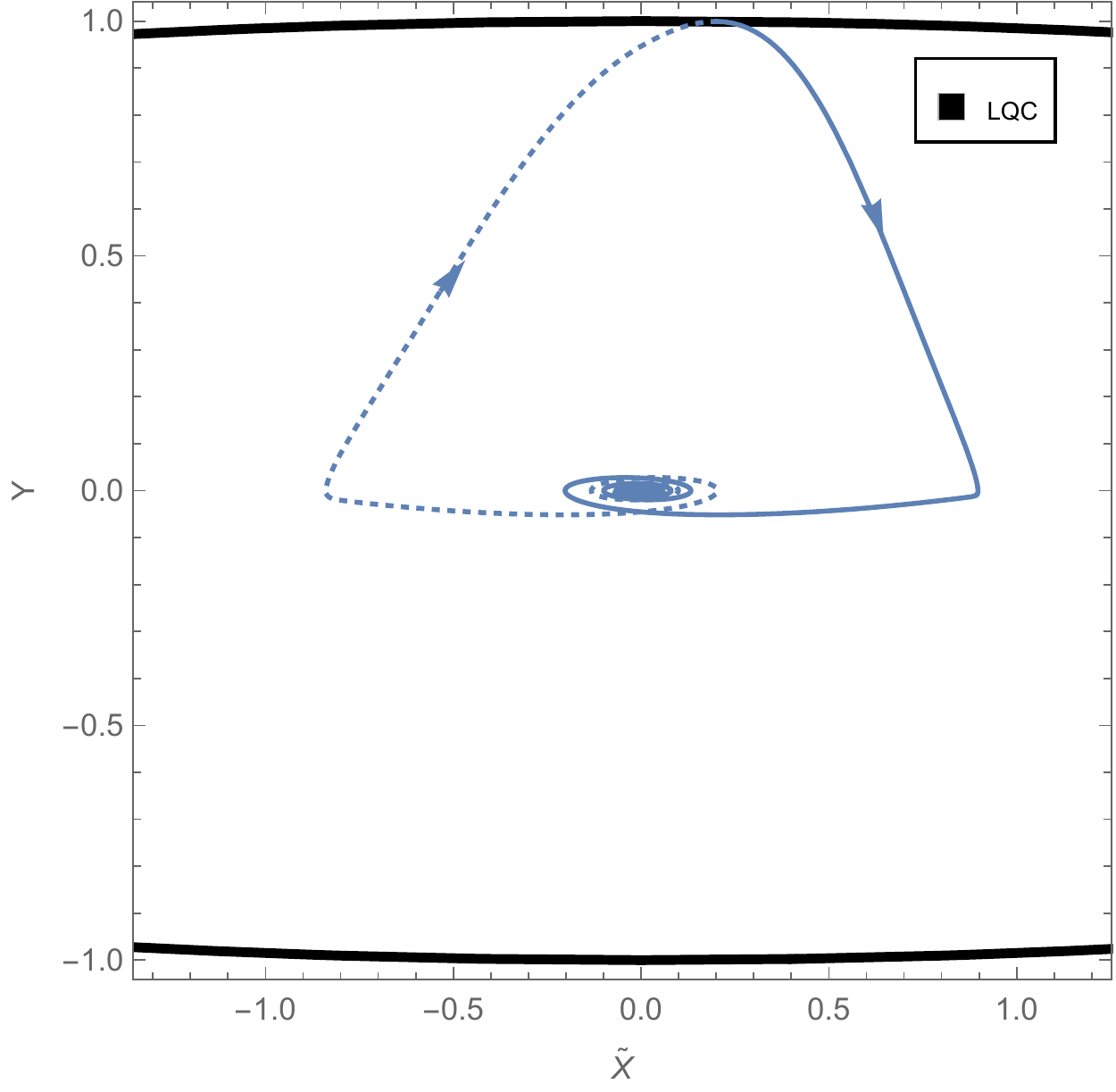}
\includegraphics[width=6cm]{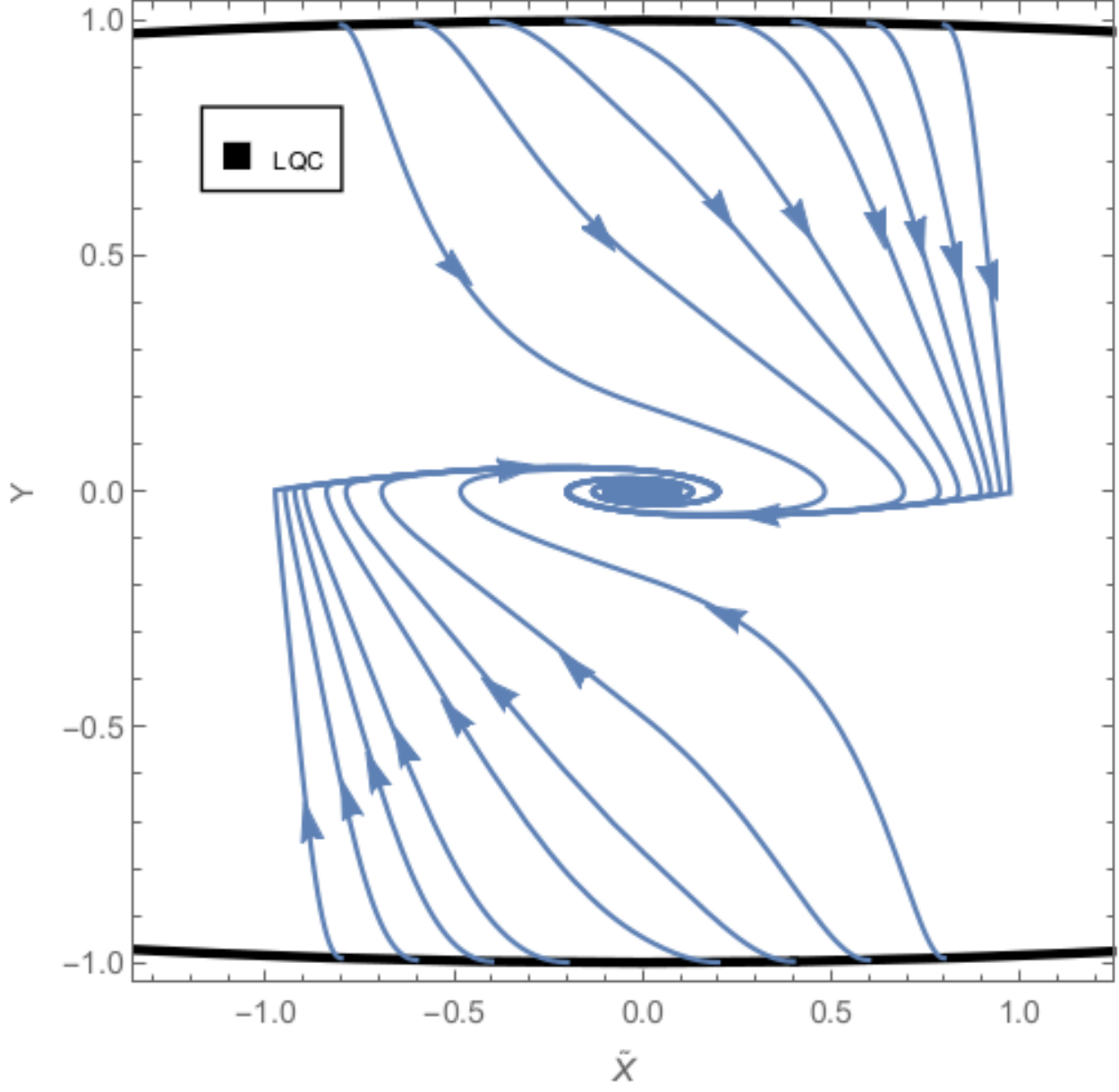}
\includegraphics[width=6cm]{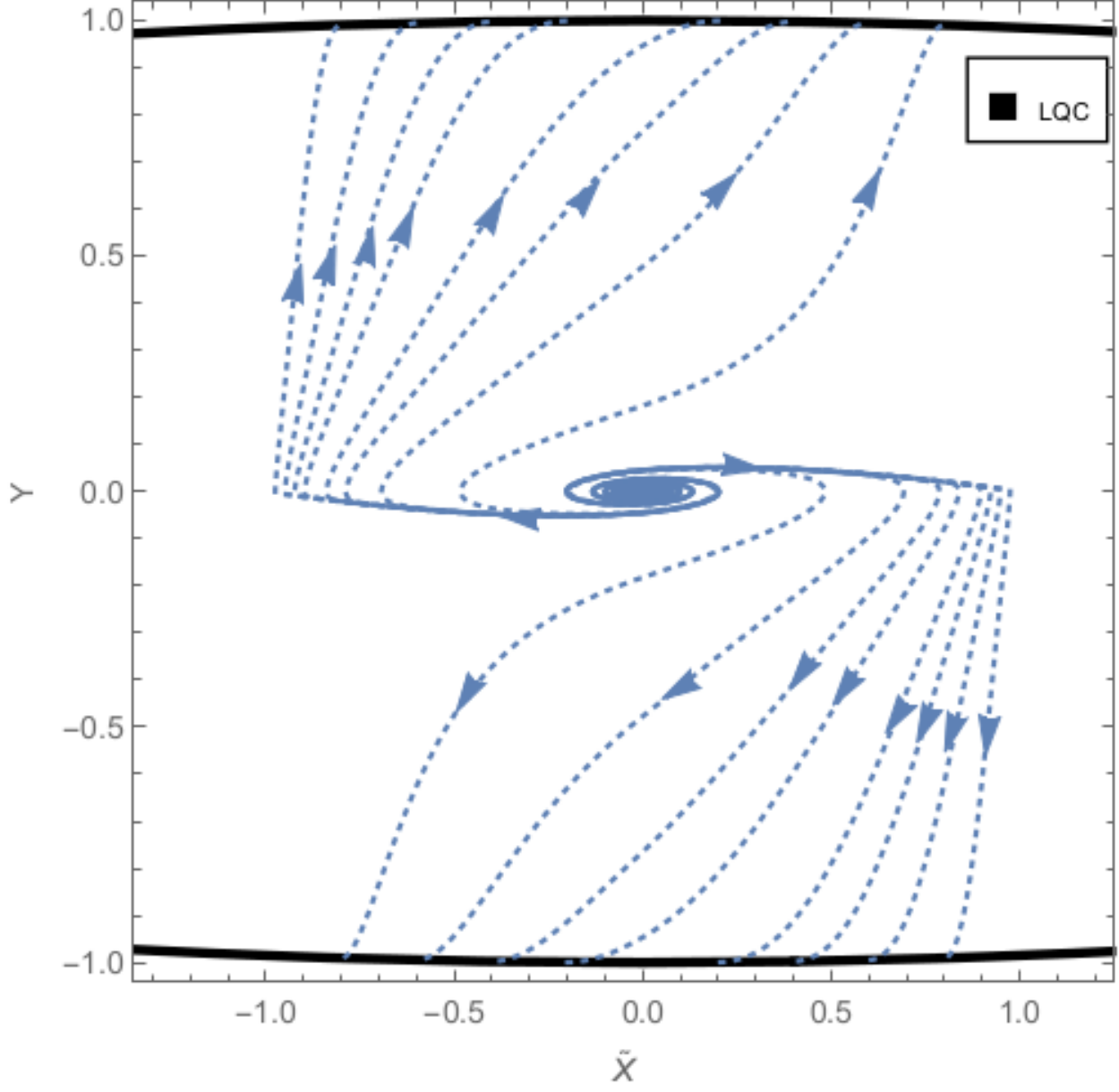}
}
\caption{Phase space portrait for non-minimal Higgs potential in LQC. $\hat V_0$ is set to $0.012$ in Eq.(\ref{3.37}). The arrows indicate the direction of forward-time evolution. The solid/dotted lines represent the evolution of phase space trajectories in the post/pre bounce phase. The quantum bounce takes place on the black solid line defined by the ellipse $\hat \chi^2_0 \tilde X^2+Y^2=1$. All the trajectories are confined within the region $|\tilde X|<1$.}
\label{fig37}
\end{figure}

\begin{figure}[h!]  
{
\includegraphics[width=6cm]{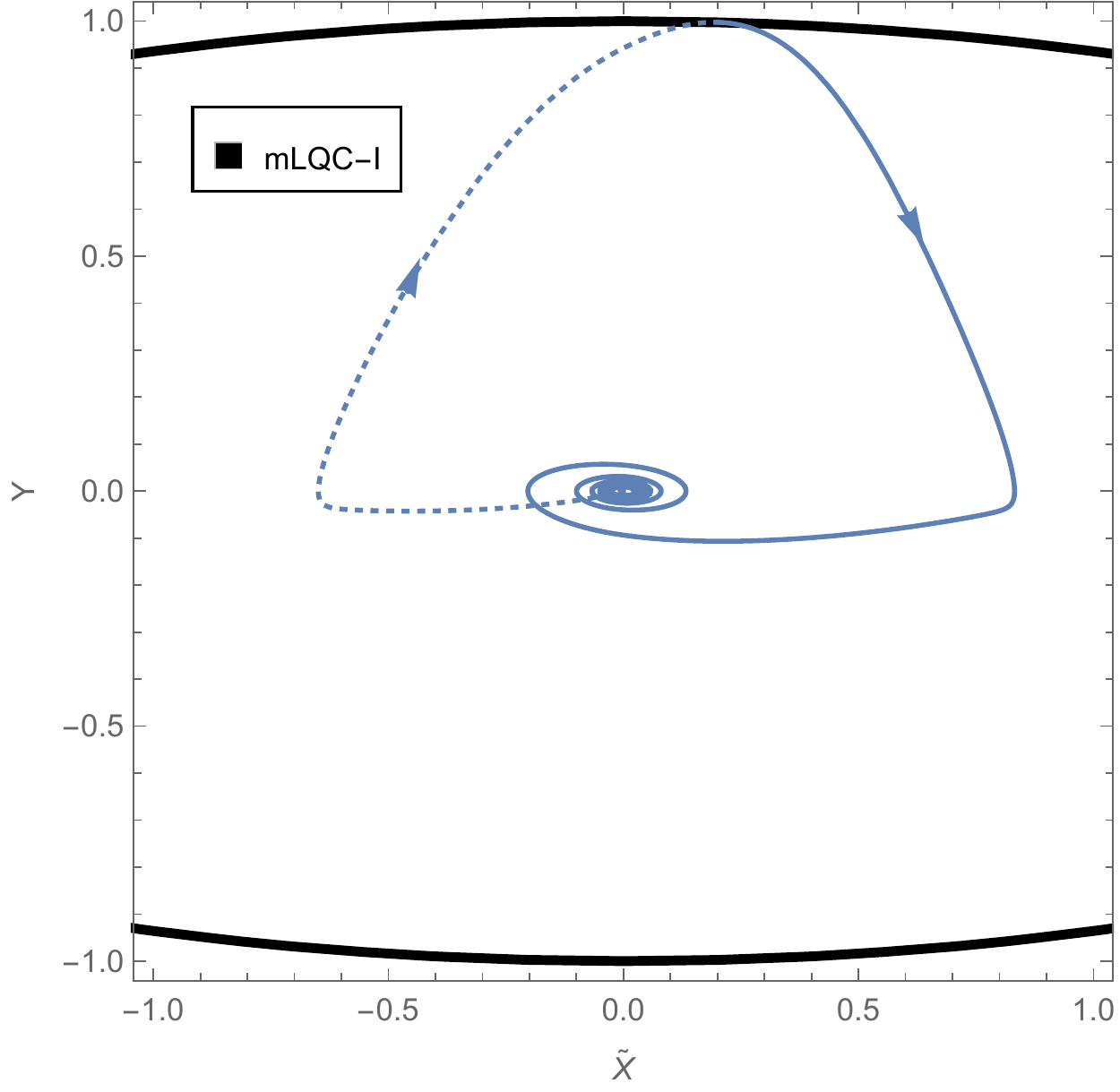}
\includegraphics[width=6cm]{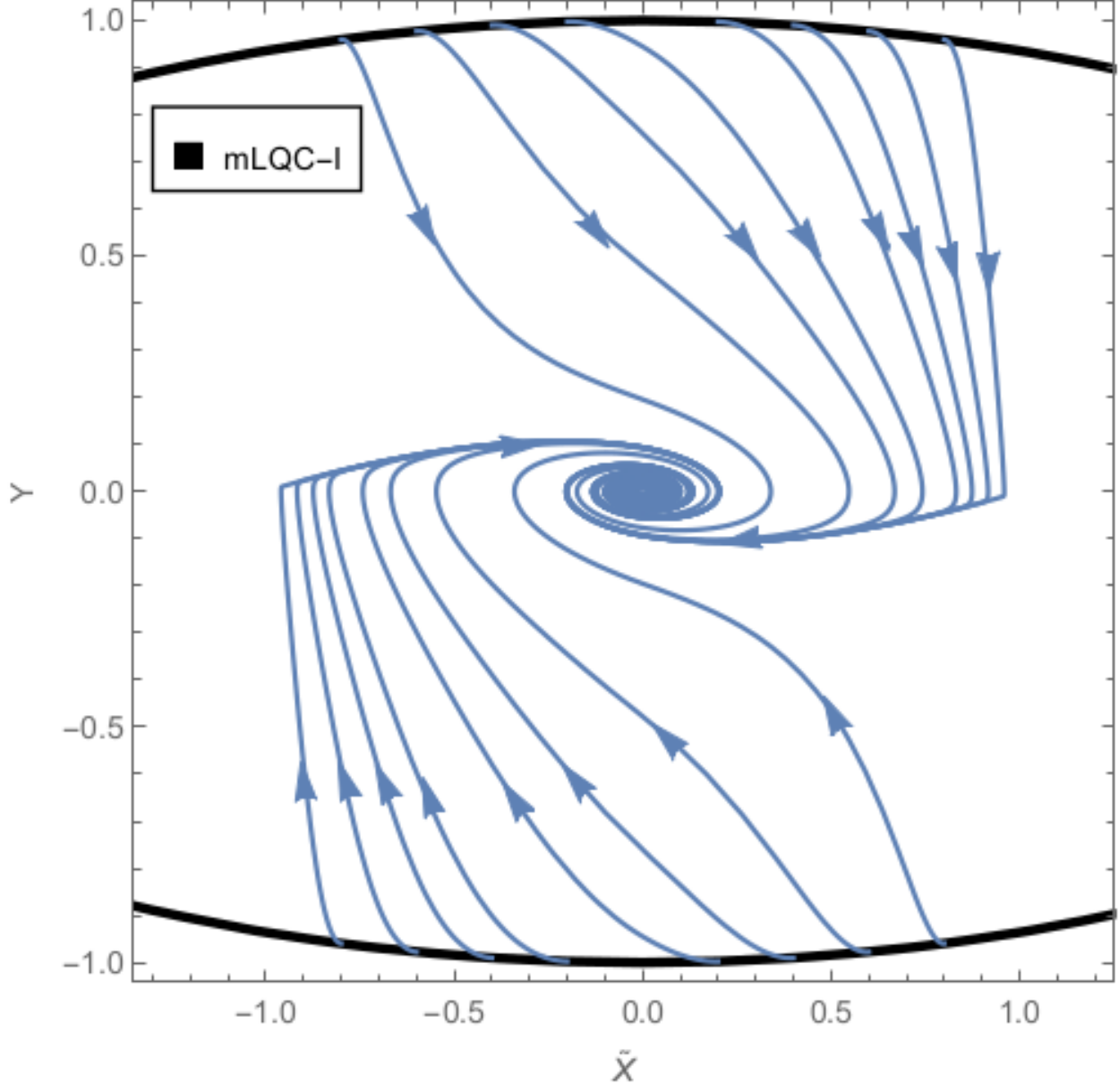}
\includegraphics[width=6cm]{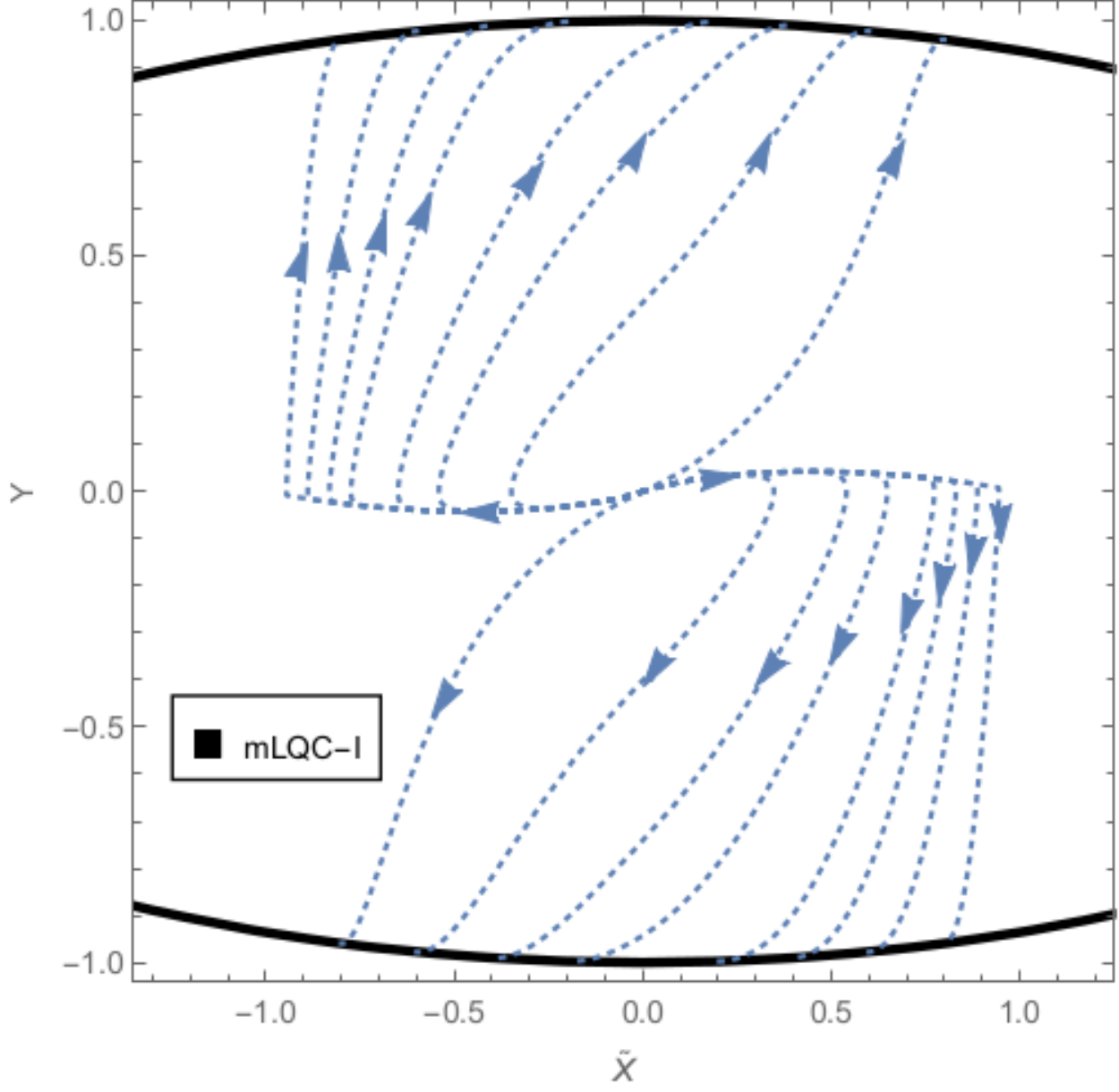}
}
\caption{Phase space portrait for non-minimal Higgs potential in mLQC-I.  All the parameters are chosen as the same as in Fig. \ref{fig37}. Qualitative evolution of generic solutions are depicted in the first subfigure. In the second subfigure, there are two inflationary separatrices and a stable spiral at the origin. The last subfigure exhibits an unstable node at the origin due to the non-vanishing Hubble rate at early times in the pre-bounce phase. }
\label{fig36}
\end{figure}

\begin{figure}[h!] 
{
\includegraphics[width=6cm]{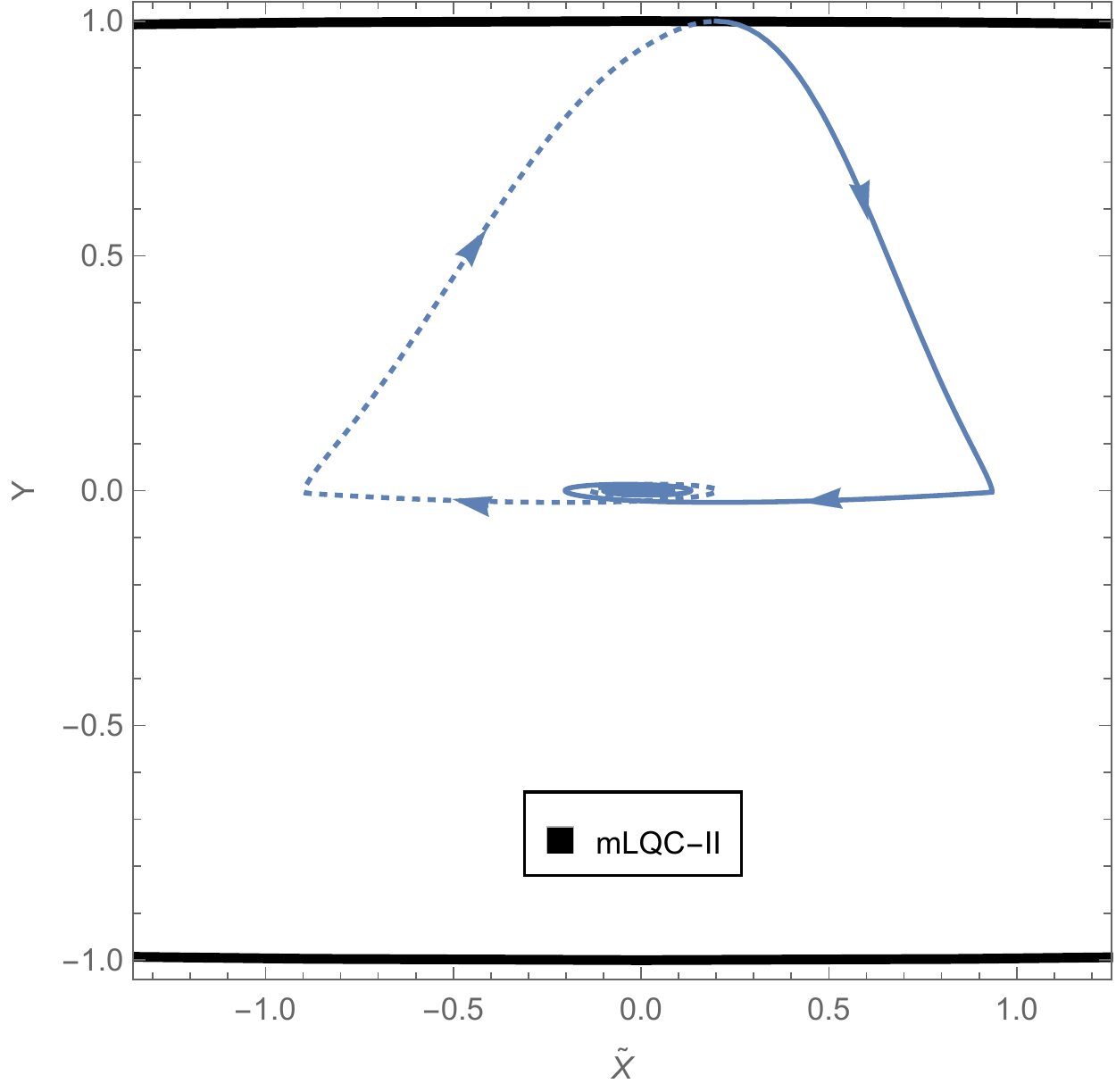}
\includegraphics[width=6cm]{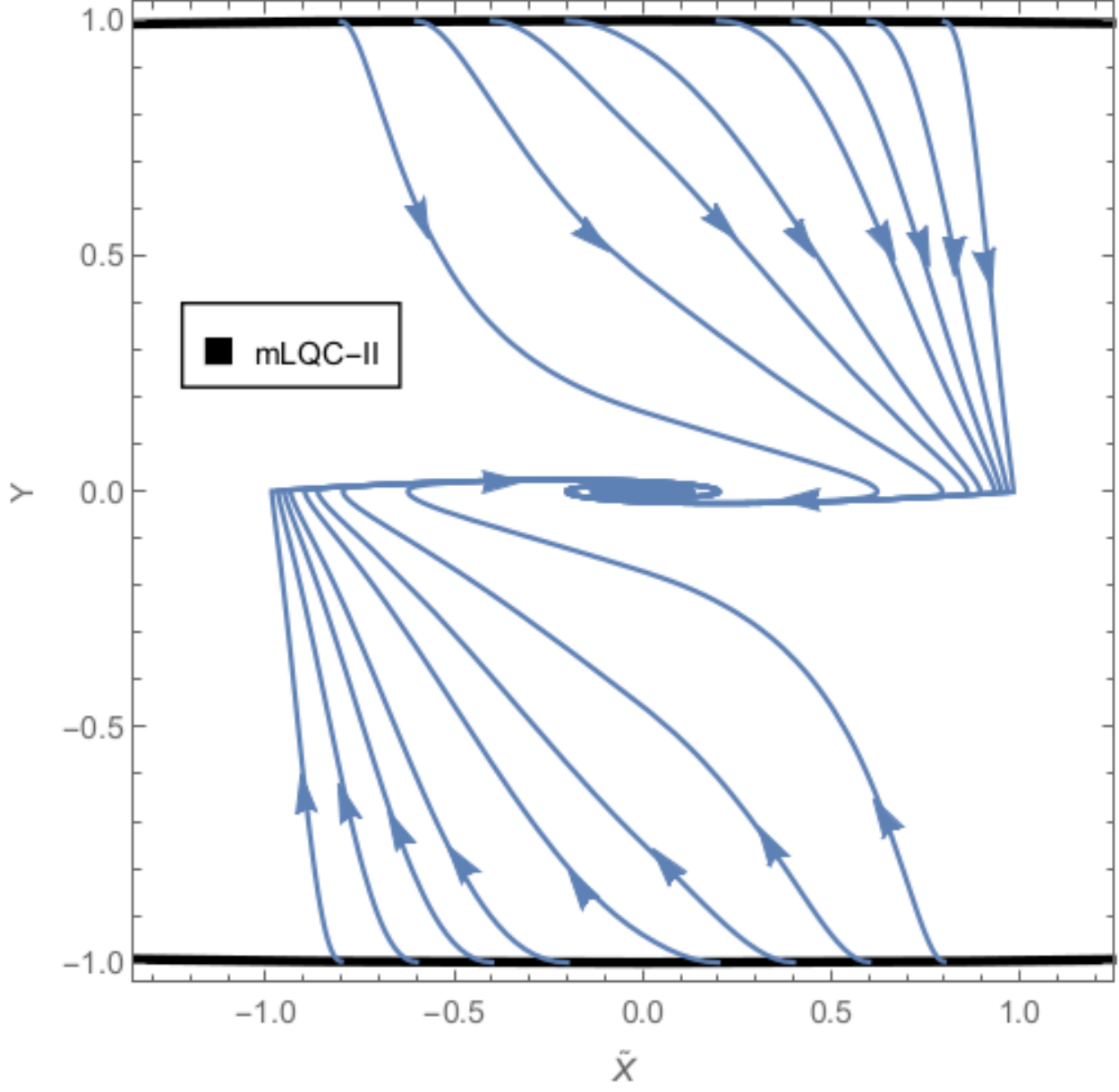}
\includegraphics[width=6cm]{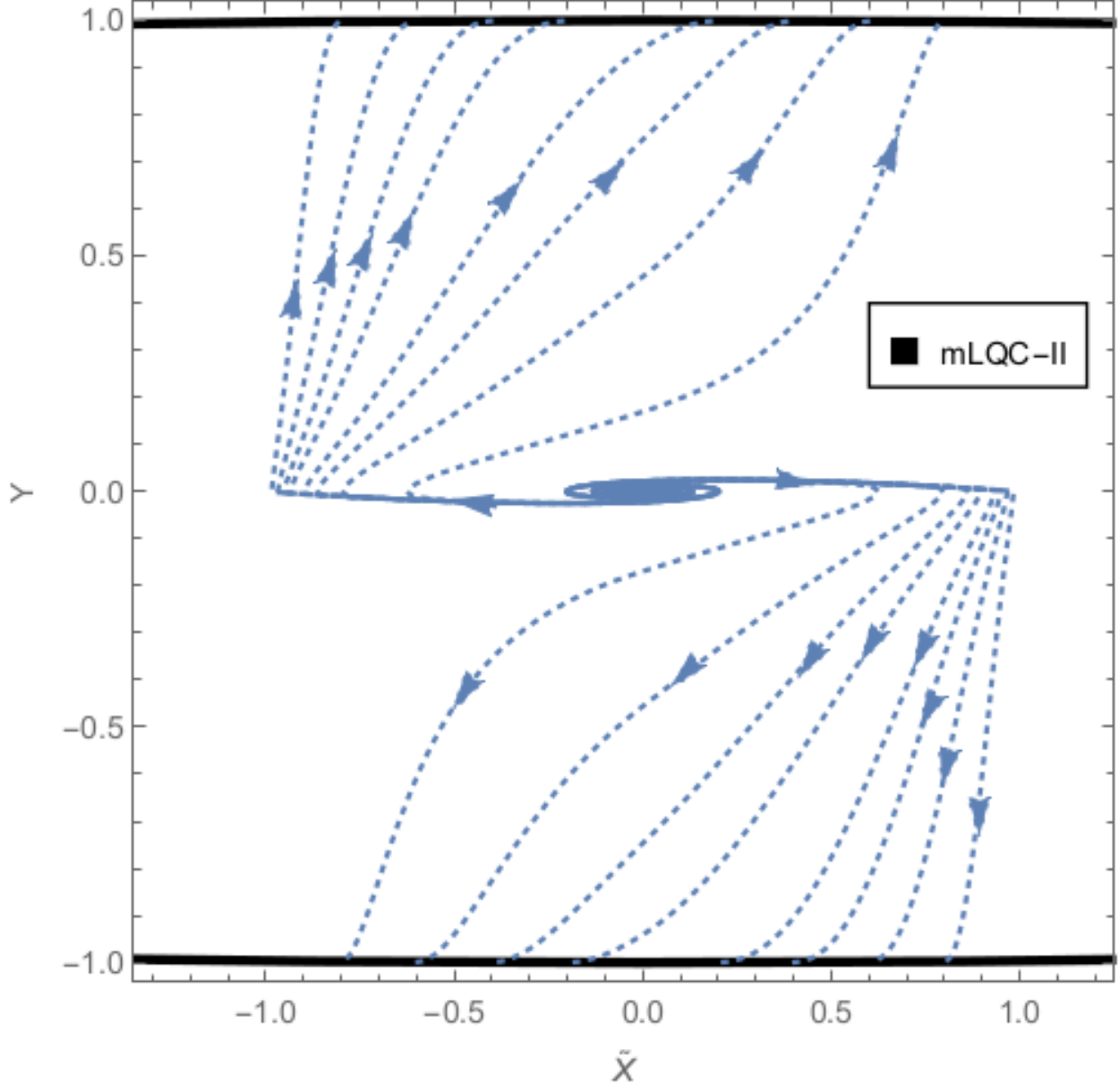}
}
\caption{Phase space portrait for non-minimal Higgs potential in mLQC-II.  Here all the parameters are chosen as the same as in Fig. \ref{fig37}. The evolution of the universe is symmetric  about the bounce just like in LQC. All the trajectories start from the unstable spiral in the contracting phase, across the quantum bounce at the boundary, and then converge to inflationary separatrices in the expanding phase before approaching the stable spiral at the origin.}
\label{fig38}
\end{figure}

\subsection{Non-Minimal Higgs Potential}

 For the minimal coupling between the gravitational and Higgs sectors,  the self-interaction potential is given by
\bq
\lb{3.22}
U=\frac{\alpha}{4}\left(h^2-\sigma^2\right)^2,
\eq
where $h$ represents the Higgs field,  $\alpha$  is the coupling constant\footnote{In particle physics, the self-interaction coupling constant is  generally  denoted by a $\lambda$. In this manuscript, following LQC conventions $\lambda$ is already designated as the square root of the smallest nonzero eigenvalue of the area operator in LQG. Therefore, we use $\alpha$ for the coupling constant to avoid any confusion.} and $\sigma$ is the vacuum expectation value of the Higgs field. { Since} 
restriction on $\alpha$ from particle physics cannot  be reconciled with { CMB observations}, hence one considers a non-minimal coupling and drop the naive ansatz of the minimal coupling. Insights on resolving this issue were first achieved in \cite{bs2008} where the non-minimal coupling was taken into consideration. Here we just summarize some of the main results. For the non-minimal coupling, in the Jordan frame the action for the Higgs boson coupled to gravity can be cast in the general form 
\bq
\lb{3.33}
S_J=\int d^4x \sqrt{-g} \Big(f(h)R-\frac{1}{2}g^{\mu\nu}\partial_\mu h \partial_\nu h -U \Big),
\eq
with
\bq
\lb{3.34}
f(h)=\frac{1}{2}\big(m^2+\xi h^2\big),
\eq
and 
\bq
\lb{3.35}
 m^2=M^2_{\text{Pl}}-\xi \sigma^2.
\eq
Here $M_{\text{Pl}}$($=1/\sqrt{8\pi G}$) stands for the reduced Planck mass, and $\xi$($=1.61\times 10^4$) { represents the non-minimal coupling strength.}  Considering the relatively small value of $\sigma$ ($=246\text{GeV})$,
 which is far below the energy scale of inflation, one can safely use the approximations $m^2\approx M^2_{\text{Pl}}$ in the following { discussion. Using the conformal transformation $
\hat g_{\mu\nu}=\Omega^2 g_{\mu\nu} $ 
with the conformal factor $\Omega^2=2 f(h)/M^2_{\text{Pl}}$, we get the following action in Einstein's frame:}
\bq
\lb{3.36}
S_E=\int d^4x \sqrt{-\hat g} \Big(\frac{M_{\text{Pl}}}{2}\hat R-\frac{1}{2}\hat g^{\mu\nu}\partial_\mu \phi \partial_\nu \phi -V \Big),
\eq
where 
\bq
V(\phi)=\frac{U}{\Omega^4},
\eq
and 
\bq
\frac{d\phi}{d h}=\sqrt{\frac{\Omega^2+6 \xi^2h^2/M^2_{\text{Pl}}}{\Omega^4}}.
\eq

It should be noted that in the small field limit, $\phi \approx h$,  $\omega_\phi  \approx 1$ and $U \approx V$, { thus} the quartic self-interaction is recovered. On the other hand, 
for  $h \gg \sqrt{\frac{2}{3}} \frac{M_{\text{Pl}}}{\xi}$ , the potential can be well approximated by \cite{mst2018}
\bq
\lb{3.37}
V \approx \hat V_0 \left(1-e^{- \sqrt{\frac{16 \pi G}{3}}|\phi|}\right)^2,
\eq
where the magnitude $\hat V_0$ is determined by the CMB observations \cite{Planck2015}
\bq
\lb{3.38}
\hat V_0=\frac{\alpha M^4_{\text{Pl}} }{4\xi^2}=9.6\times 10^{-11} M^4_{\text{Pl}}.
\eq
Now one can see the way the non-minimal  coupling can help { alleviate  the incompatibility problem of the minimal  coupling Higgs potential mentioned earlier.} In Eq.(\ref{3.37}), $\alpha$ is still fixed from the experimental data in particle physics with $\alpha=0.1$, while $\hat V_0=9.6\times 10^{-11} M^4_{\text{Pl}}$ ensures the predictions from the model matched with the  CMB data \cite{Planck2015}. { To fit both observations,} the value of $\xi$ can thus be fixed, which turns out to be $\xi =1.61\times 10^4$. 

At large value of $\phi$,  the effective potential 
Eq.(\ref{3.38})  comes into effect and sources the  slow-roll inflation. At the end of inflation, the scalar field $\phi$ quickly diminishes to a small value, which justifies the small field limit. One thus recovers quartic self-interaction of the Higgs boson  in the reheating phase. Since we are interested in the inflationary phase, we shall work in the Einstein frame with the effective potential (\ref{3.37}). This potential is symmetric about $\phi=0$, and at the positive side of the scalar field, it is similar to the Starobinsky potential. Therefore, we shall adopt  a similar parametrization of the phase space variables for the phase space portraits with  
\bqn
\lb{3.39}
X&=&\frac{\phi}{|\phi|}  \hat \chi_0 \left(1-e^{-\sqrt{16\pi G/3}|\phi|}\right),\\
Y&=&\frac{\dot \phi}{\sqrt{2\rho^i_c}},
\eqn
in which $\hat \chi_0=\sqrt{\hat V_0/\rho^i_c}$. The resultant equations of motion simply become 
\bqn
\lb{3.40a}
\dot X&=&\sqrt{\frac{32\pi G \hat V_0}{3}} Y \left( 1\mp  X/\hat \chi_0\right),\\
\lb{3.40b}
\dot Y&=&-3 H^i Y-\sqrt{\frac{32\pi G \hat V_0}{3}}X\left(1\mp X/\hat \chi_0\right),
\eqn
where $\mp$ correspond to the positive/negative $\phi$, respectively.
As a result,  the dynamical equations (\ref{3.40a})-(\ref{3.40b}) have three fixed points: the origin and the points  $(\pm \hat \chi_0, 0)$. 

The analysis of the properties of these fixed points can be carried out in the same way as in the Starobinsky potential,  since Eqs.(\ref{3.20a})-(\ref{3.20b}) reduce to Eqs.(\ref{3.40a})-(\ref{3.40b}) for positive $\phi$ if the mass  is set to {  $\sqrt{\frac{32\pi G \hat V_0}{3}}$ }. In Figs. \ref{fig37}-\ref{fig38}, we have chosen $\hat V = 0.012$. As it is a small number, if plotted in the variables $X$ and $Y$, the phase portraits will be cornered in a small region near the origin. To better demonstrate the patterns of the portraits, we use $\tilde X= X/ \hat\chi_0$.  Following the analysis for Starobinsky potential, several points in the following can be easily justified: (i) As can be seen from the phase space portraits given in Fig. \ref{fig37}-\ref{fig38}, all the physical trajectories are confined within the region $|X| < \hat \chi_0$. (ii) Points $(\pm \hat \chi_0,0)$ are fixed points with one eigenvalue vanishing in the linear stability analysis. As in Starobinsky potential, these turn out to be unstable. of the dynamical system which can only be reached in the limit $\phi \rightarrow \pm \infty$. Therefore, $|\bar X| = 1$ are boundaries of the system. (iii) Outside the region bounded by $|\tilde X|<1$, there exist solutions corresponding to an analogous scalar field $\tilde \phi$ for a different potential (as in Sec. IIIC).  The fixed points $(\pm \chi_0, 0)$ behave as an/a attractor/repeller of  the generic solutions in  $|\tilde X|>1$ in the post-/pre-bounce stage in all three models. However, all these solutions can not be mapped to the real $X-Y$ plane due to the term $\phi/|\phi|$ in the transformation equation (\ref{3.39}). Besides, as viewed in the whole phase space, similar to the point $(\chi_0,0)$ in Starobinsky potential, the fixed points $(\pm \chi_0, 0)$ are not stable either. (iv) And finally, the origin still plays the same role as in the Starobinsky potential or any other potentials studied in the above subsections. More specifically, it is  an attractor/repeller in the post/pre-bounce phase of LQC, mLQC-I and mLQC-II.  In the pre-bounce phase of mLQC-I, the origin can be either an unstable node or an unstable spiral depending on the value of the magnitude of the potential $\hat V_0$. It can be found directly from Eq. (\ref{ST}) that the condition for two real positive characteristic eigenvalues of the evolution equations for   the linear perturbations about the origin is 
\bq
\hat V_0 = \frac{3m^2}{32 \pi G} \le 0.012.
\eq
Therefore,   if $\hat V_0$ is less than or equal to $0.012$, the origin in the pre-bounce phase of mLQC-I is an unstable node, while if $\hat V_0$ is larger than $0.012$, there is an imaginary part in the eigenvalues of the evolution equations which makes the origin an unstable spiral. Since in our simulations we choose $\hat V_0=0.012$, thus, there is no spiral structure showing up in the third subfigure of Fig. \ref{fig36}. 
 
   In Figs. \ref{fig37}-\ref{fig38}, the boundaries outlined by the black solid curves are the ellipses  $\hat \chi^2_0 \tilde X^2+Y^2=1$. In the plots, only the central parts of the ellipses with $|\tilde X|<1$ are depicted. The first panels of Figs. \ref{fig37}-\ref{fig38} show us the qualitative evolution of generic solutions which start from the unstable origin in the pre-bounce phase, across the quantum bounce at the boundary, and then finally approach the stable origin in the post-bounce phase. In the second panels,  two inflationary separatrices appear in all three models in contrast to the Starobinsky potential, since the non-minimal Higgs potential is symmetric about $\phi=0$. Therefore, the inflationary phase can be sourced by two sides of the $\phi$ axis. There is also a stable spiral at the origin in each of the portraits for all three models. The last subfigures in Figs. \ref{fig37} and \ref{fig38} contain an unstable spiral at the origin, while in Fig. \ref{fig36} the origin is an unstable node. 
 
\begin{widetext}

\begin{figure} 
{
\includegraphics[width=5.5cm]{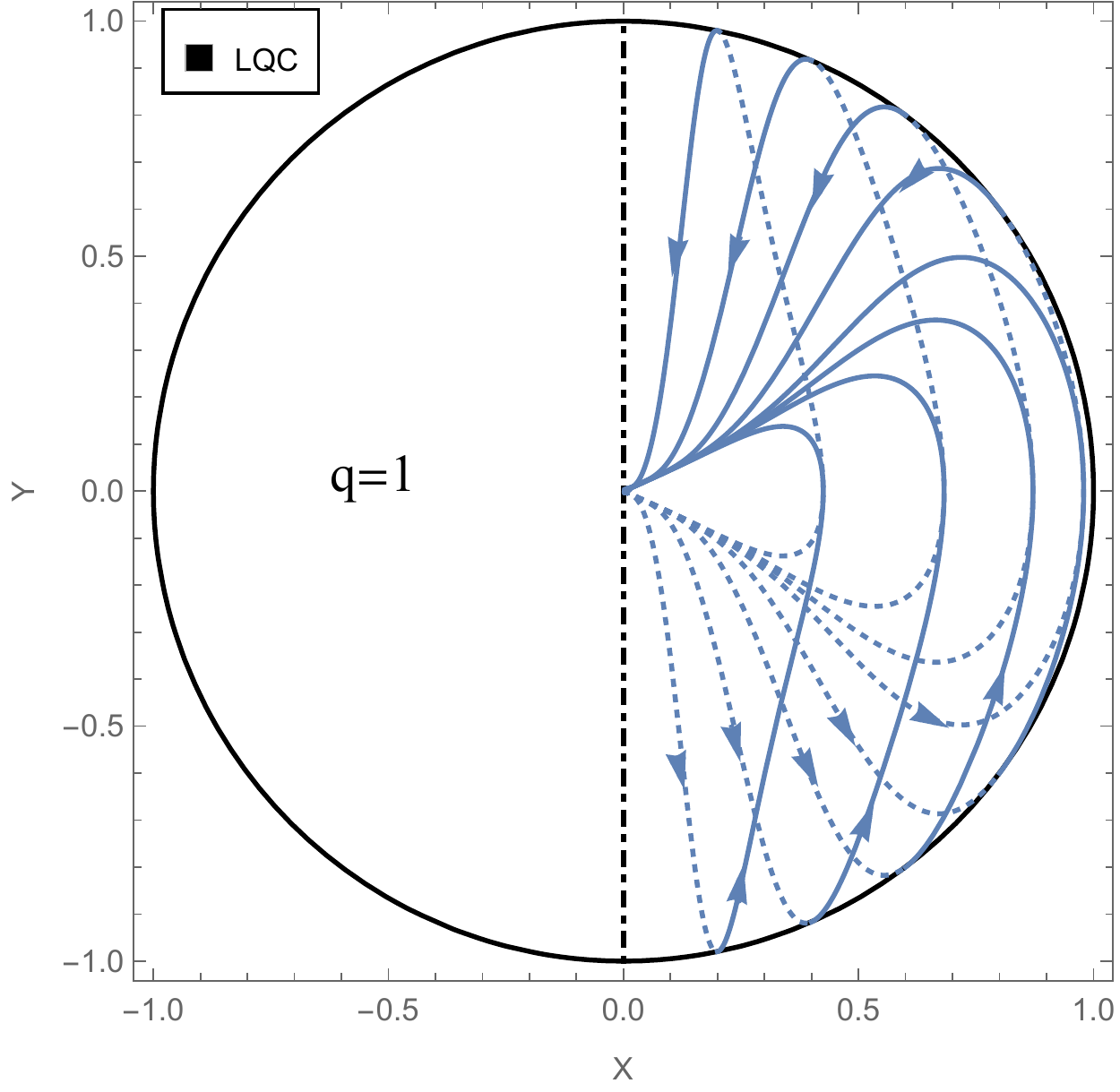}
\includegraphics[width=5.5cm]{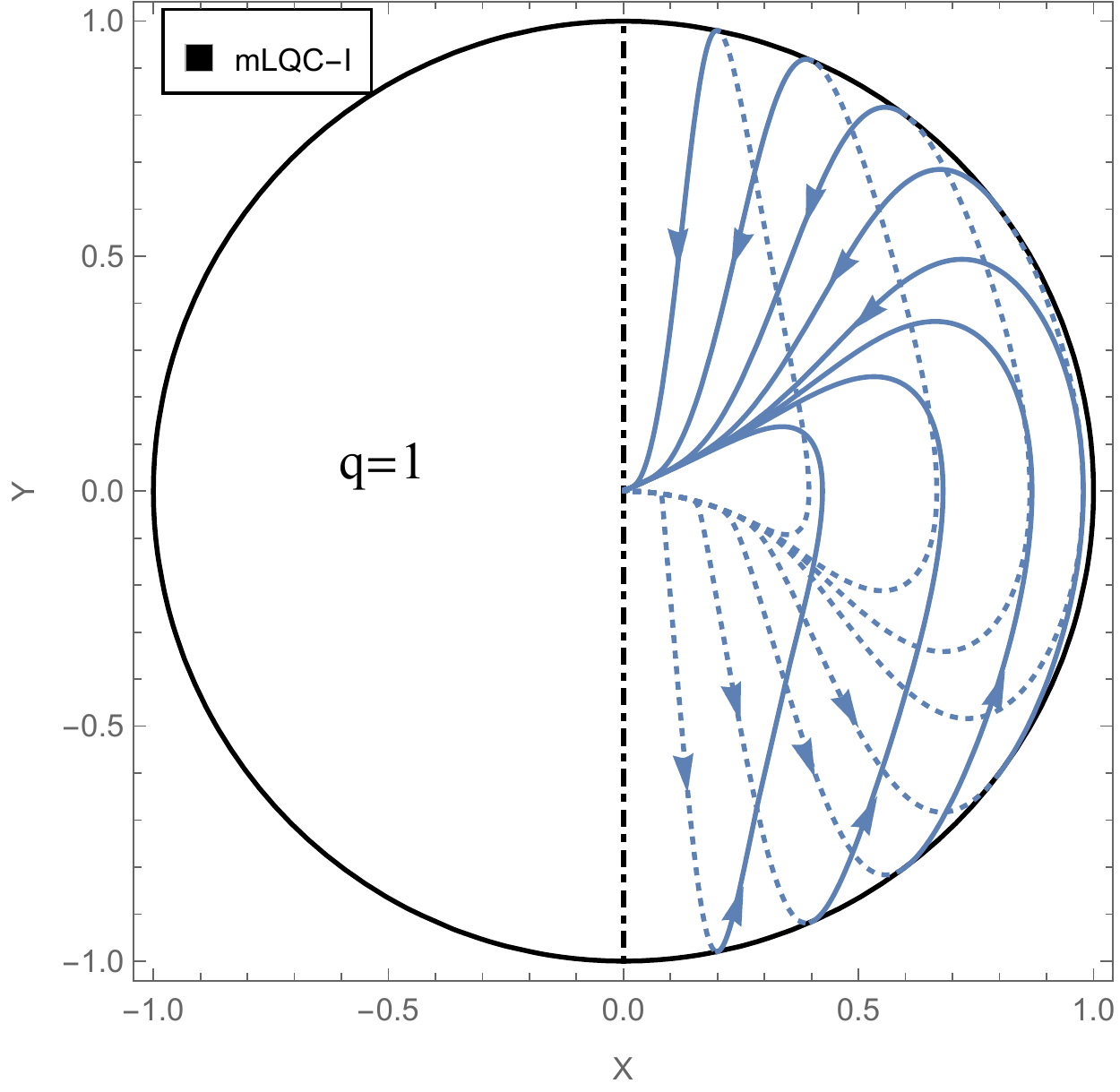}
\includegraphics[width=5.5cm]{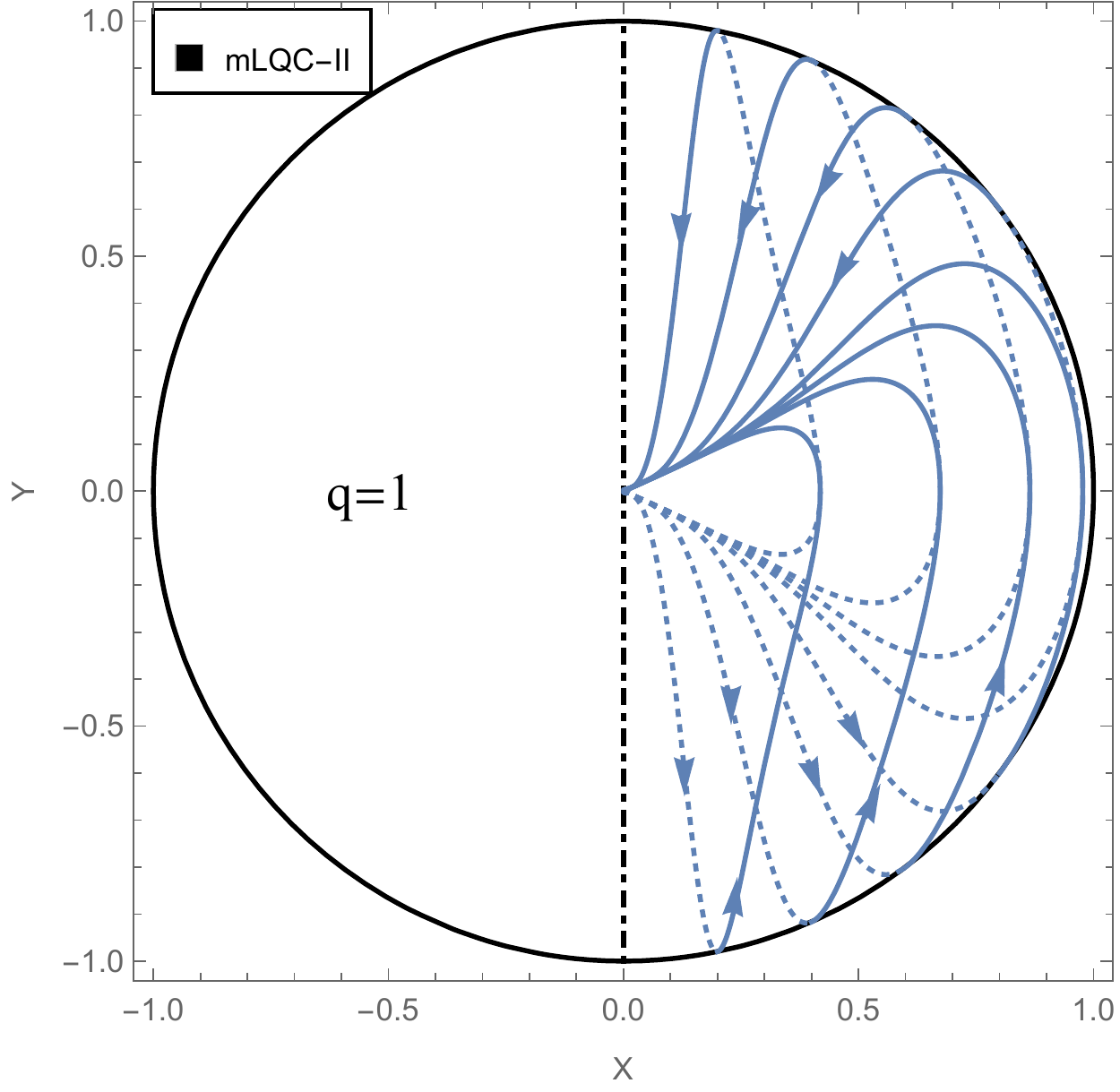}\\
\includegraphics[width=5.5cm]{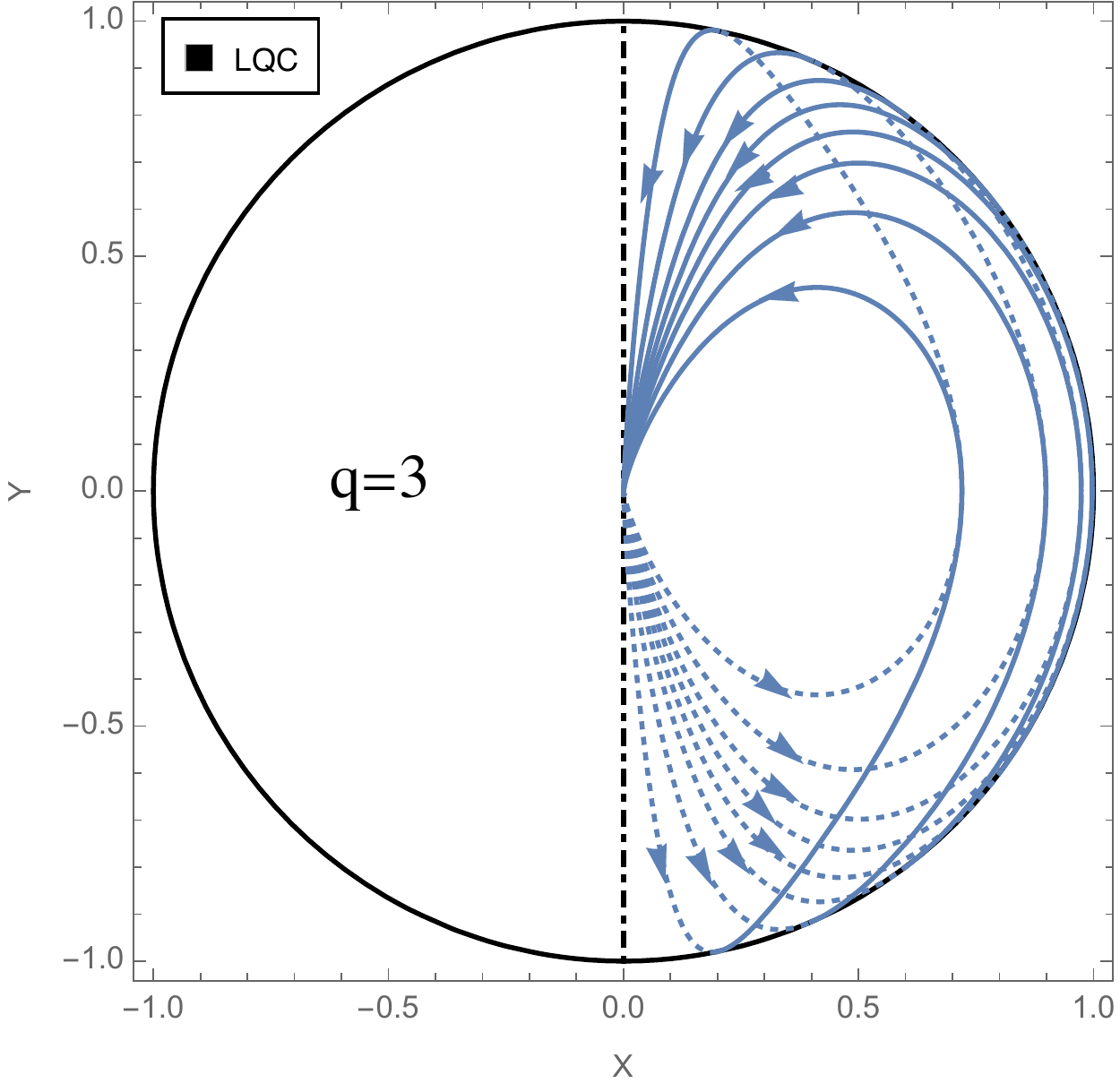}
\includegraphics[width=5.5cm]{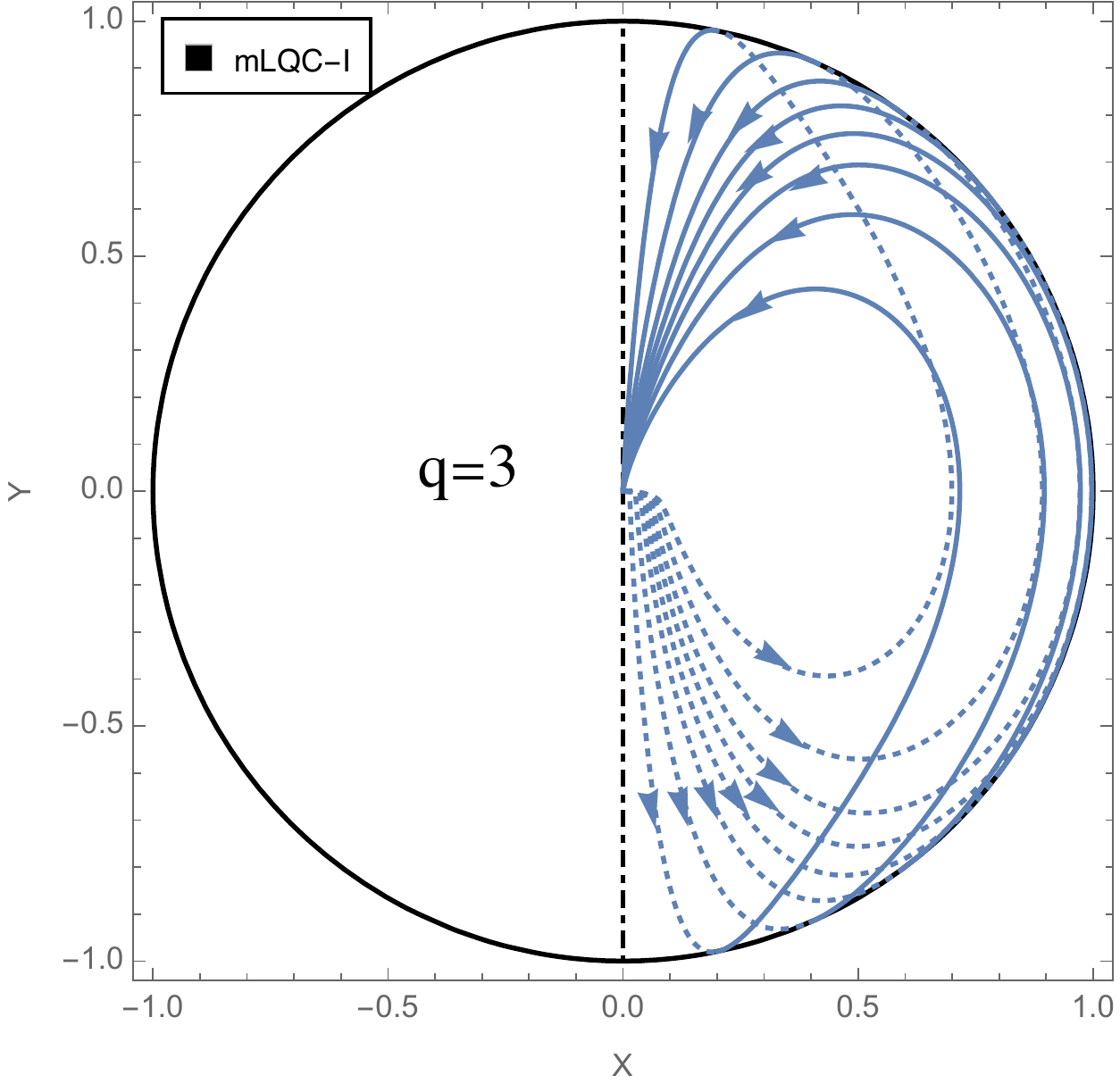}
\includegraphics[width=5.5cm]{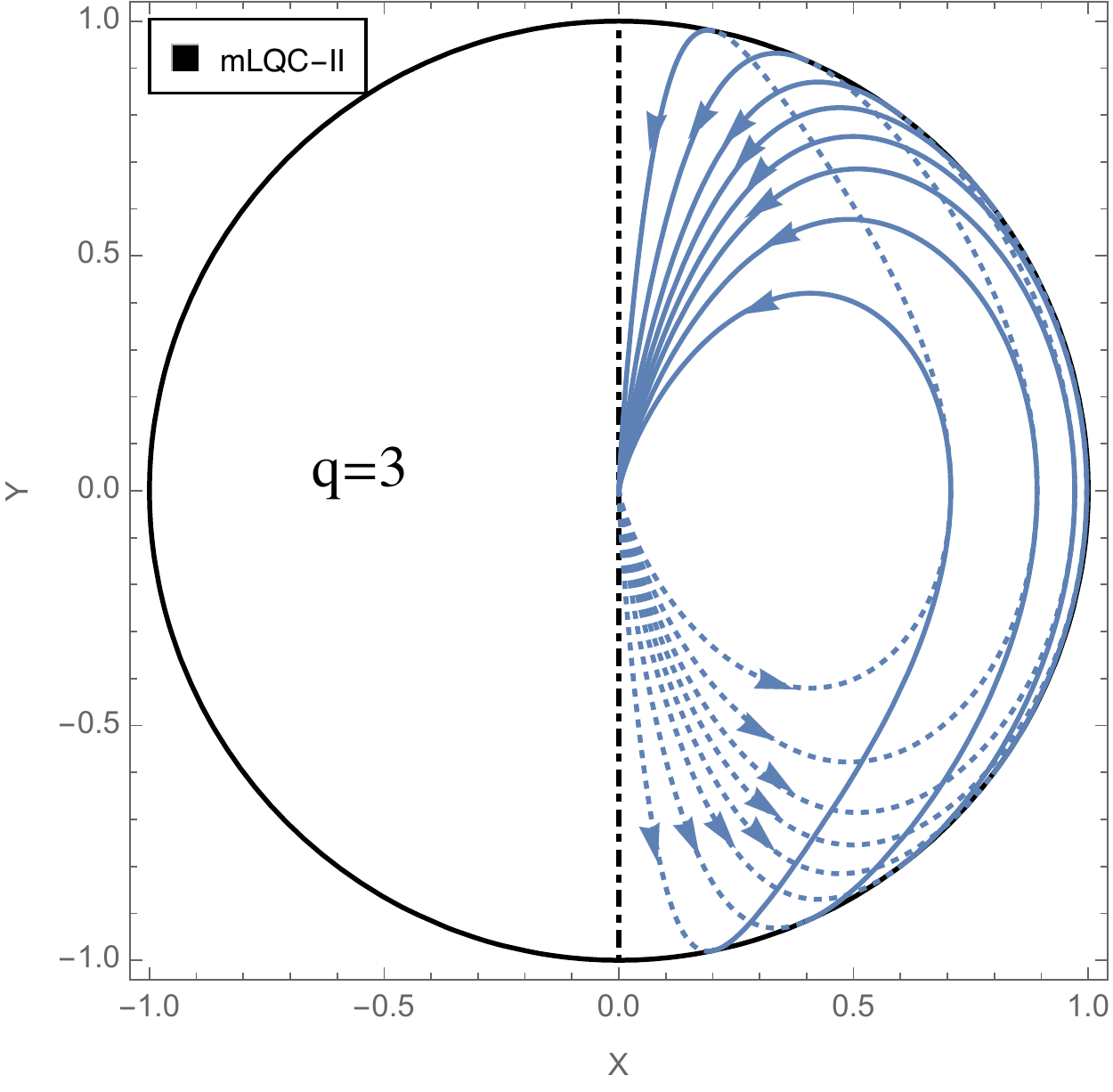}
}
\caption{Phase space portraits of the scalar field in the exponential potential Eq.(\ref{3.11}) using phase space variables (\ref{3.18}).  The initial data is picked on the boundary with the arrow indicating the forward evolution of the system. The solid/dotted lines describe the trajectories in the post/pre- bounce stage. In bottom panels,  $V_0=0.01$ and $q=3$. In top panels, $V_0=0.01$ and $q=1$. The black dot-dashed line at $X=0$ is the boundary of all the trajectories.}
\label{fig22}
\end{figure}

\begin{figure} 
{
\includegraphics[width=8cm]{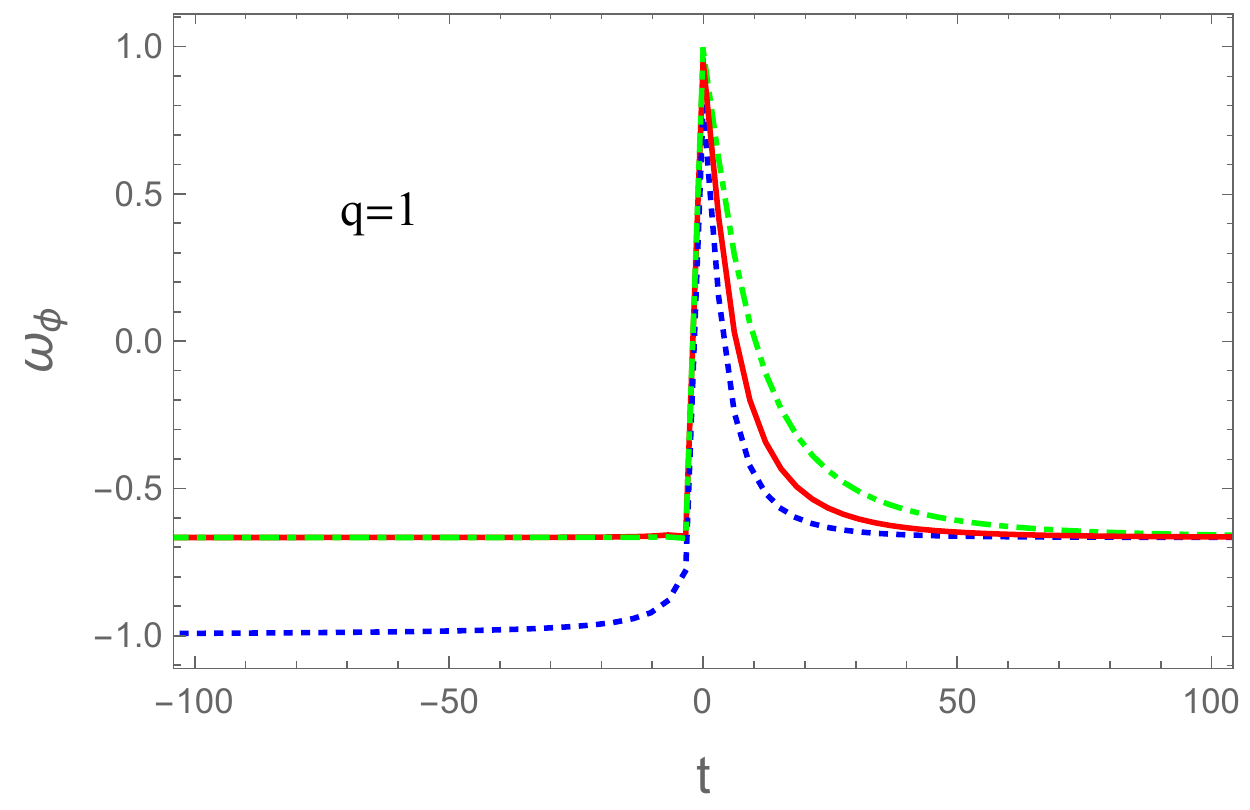}
\includegraphics[width=8cm]{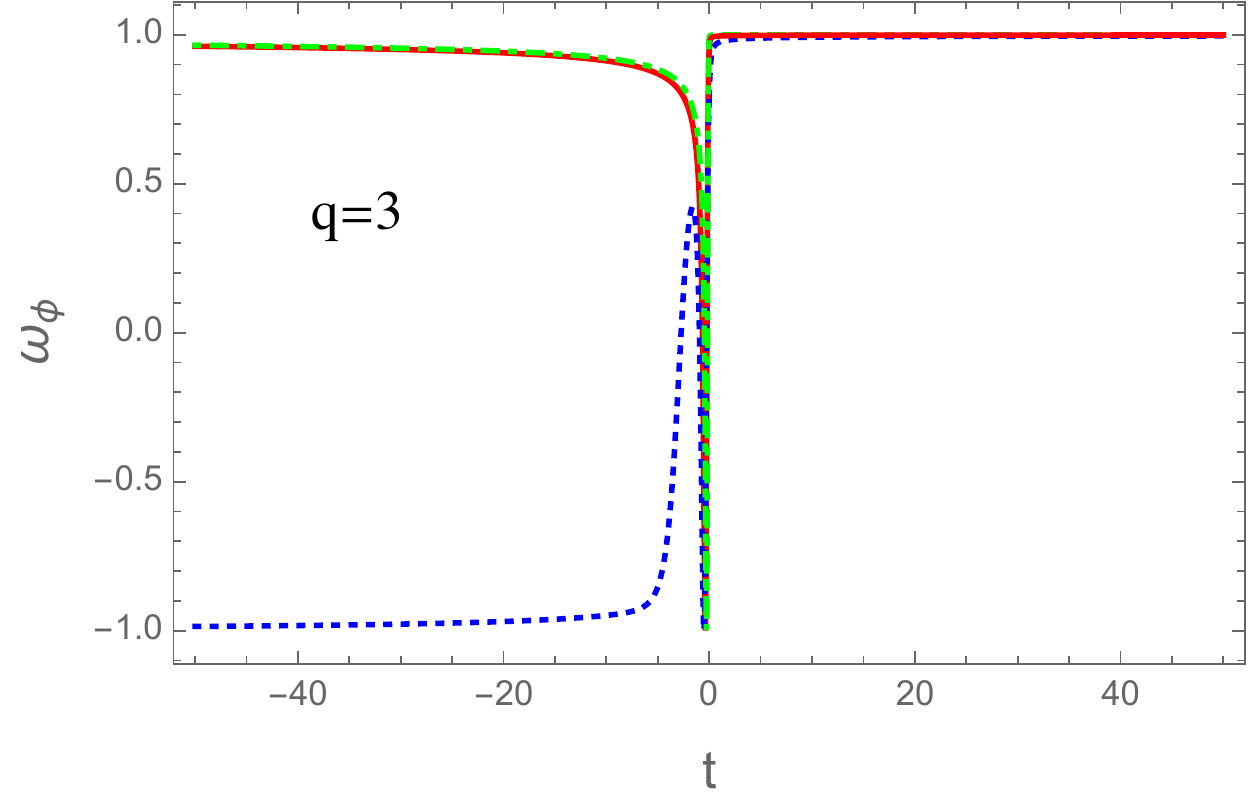}
}
\caption{Evolution of the equation of state for the exponential potential in LQC (red), mLQC-I (blue) and mLQC-II (green) is depicted with the parameters $V_0=0.01$, $q=1$ and $q=3$. Initial conditions   $\phi_B=0$ are chosen at the bounce which occurs at $t=0$. In the figures, we show the scaling solutions for $q=1$ in LQC, mLQC-II and the post-bounce phase of mLQC-I. No scaling solutions exist in the pre-bounce phase of mLQC-II. For $q=3$, which is greater than the required value ($q \leq \sqrt{6}$) for existence scaling solutions, no scaling solutions are present in these models. Instead, the kinetic-dominated solutions show up in LQC, mLQC-II and the post-bounce phase of mLQC-I.   }
\label{fig23}
\end{figure}

\begin{figure} 
{
\includegraphics[width=5.5cm]{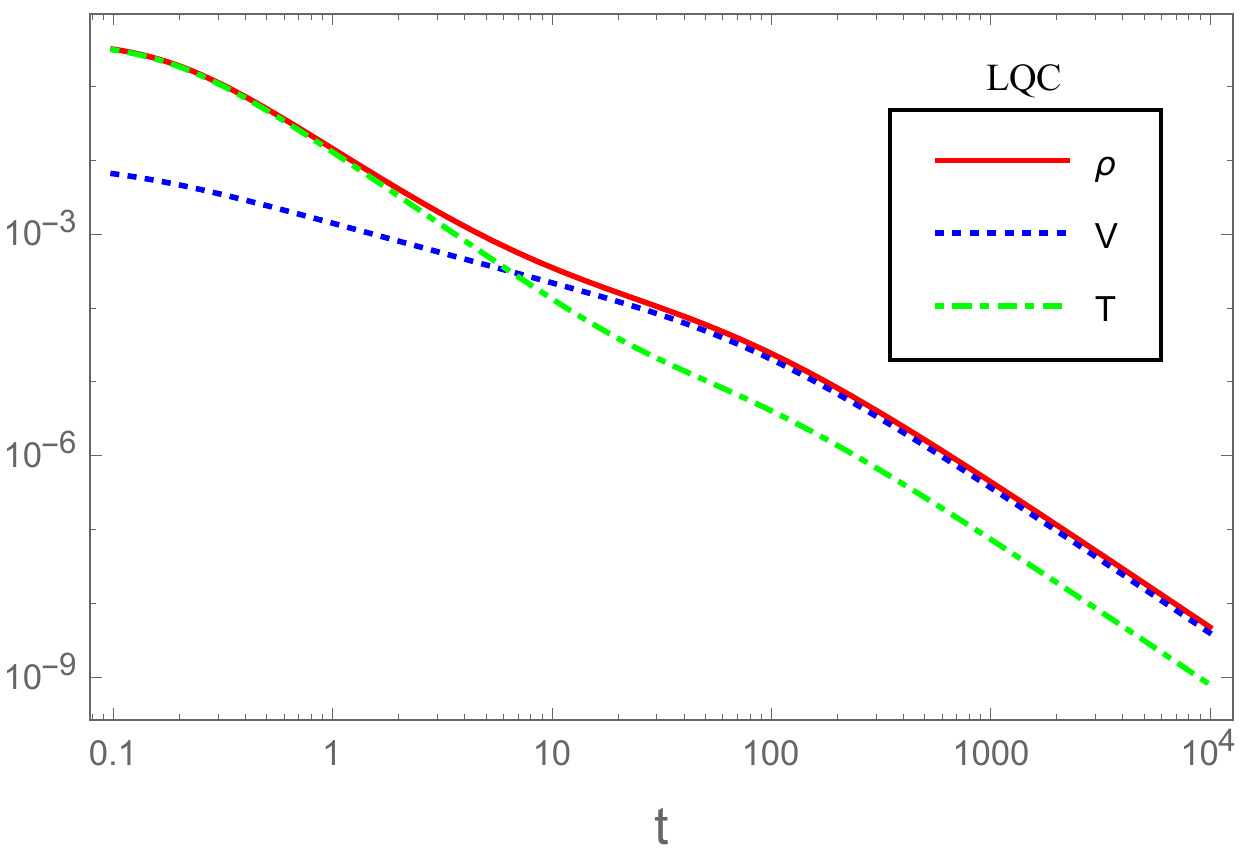}
\includegraphics[width=5.5cm]{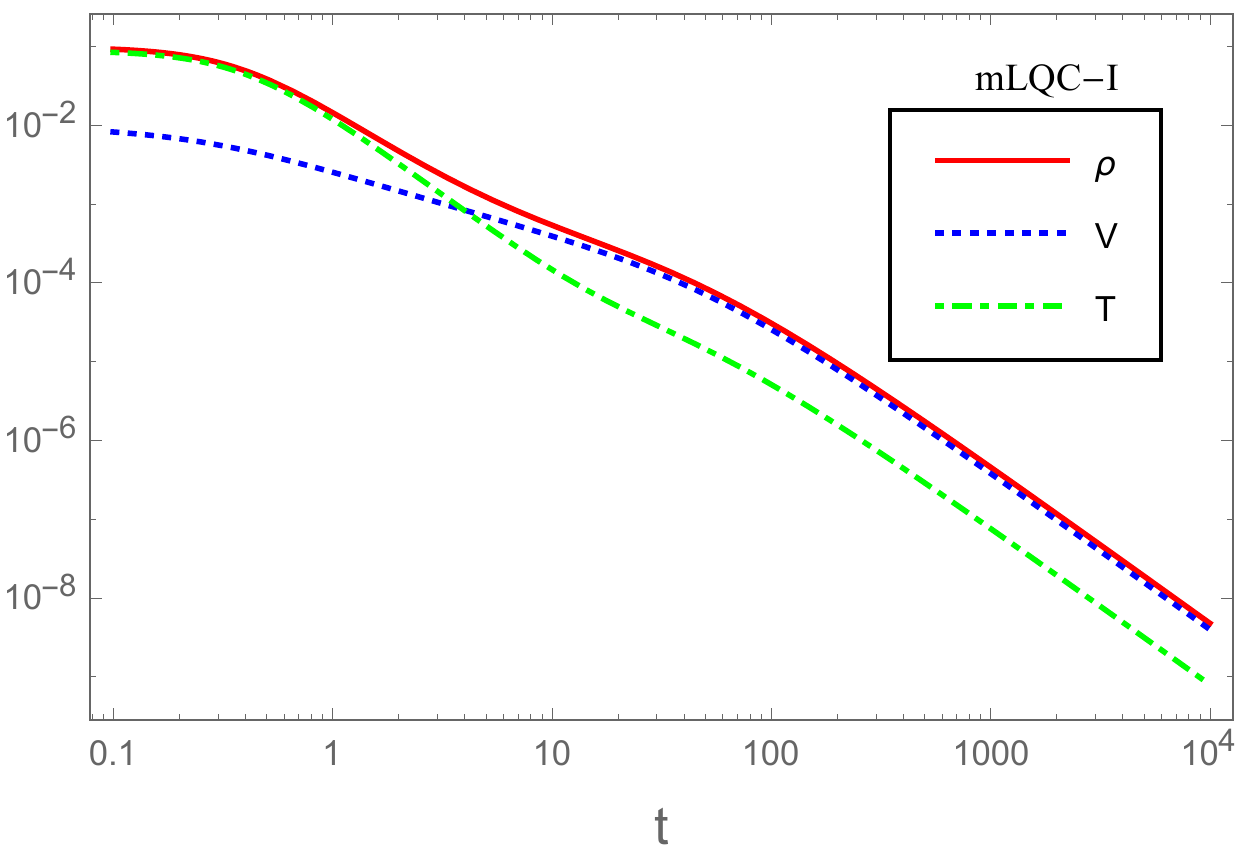}
\includegraphics[width=5.5cm]{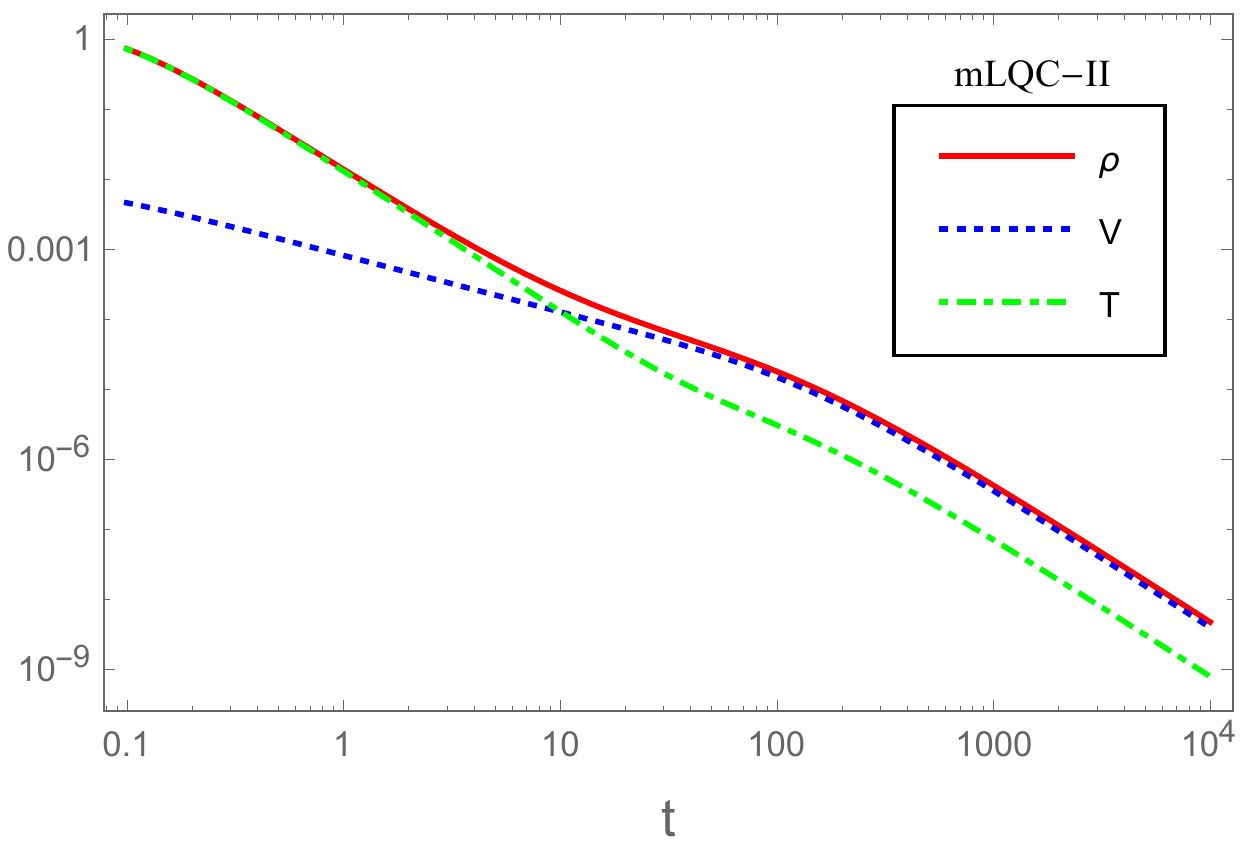}\\
\includegraphics[width=5.5cm]{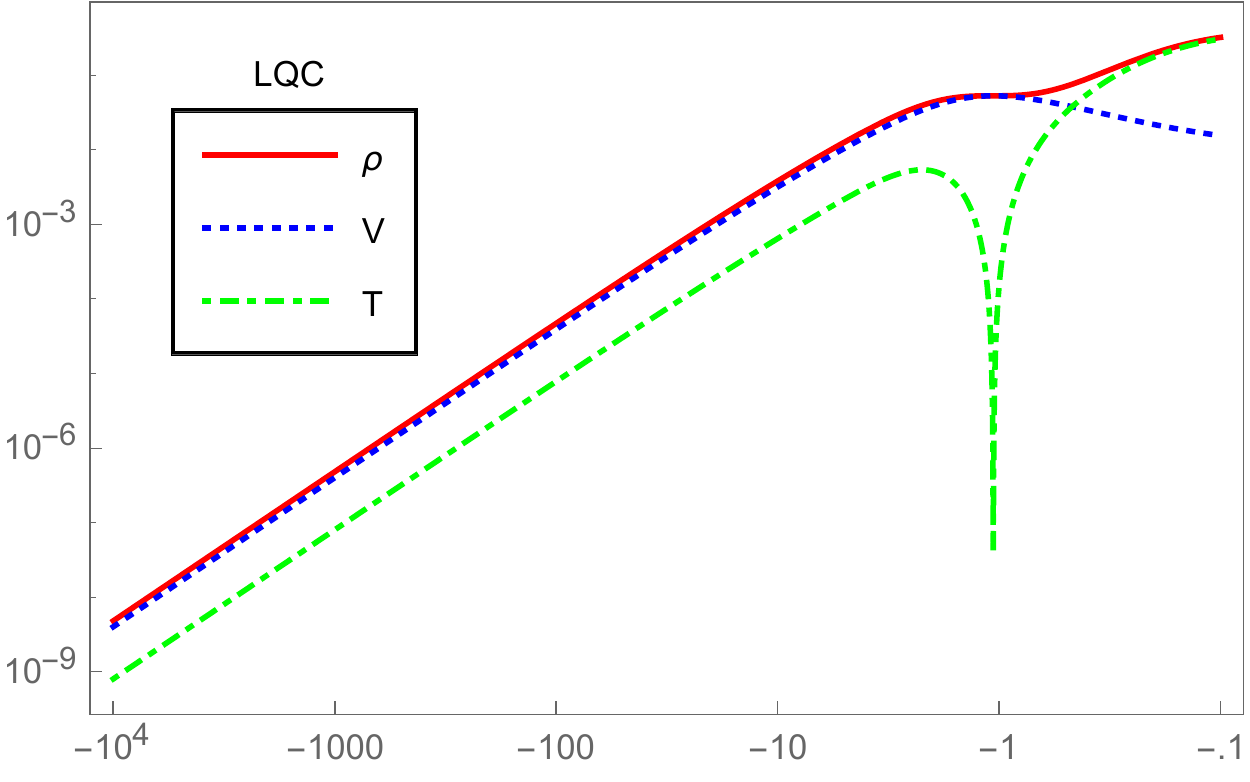}
\includegraphics[width=5.5cm]{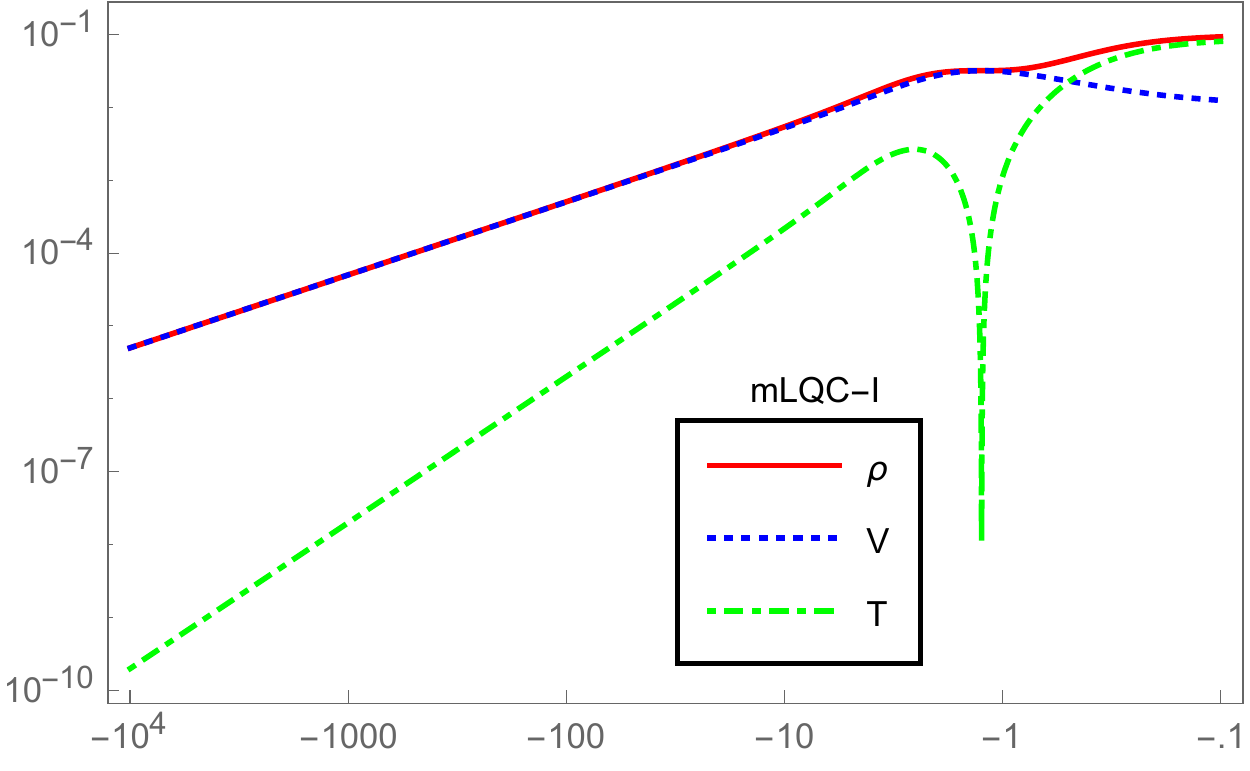}
\includegraphics[width=5.5cm]{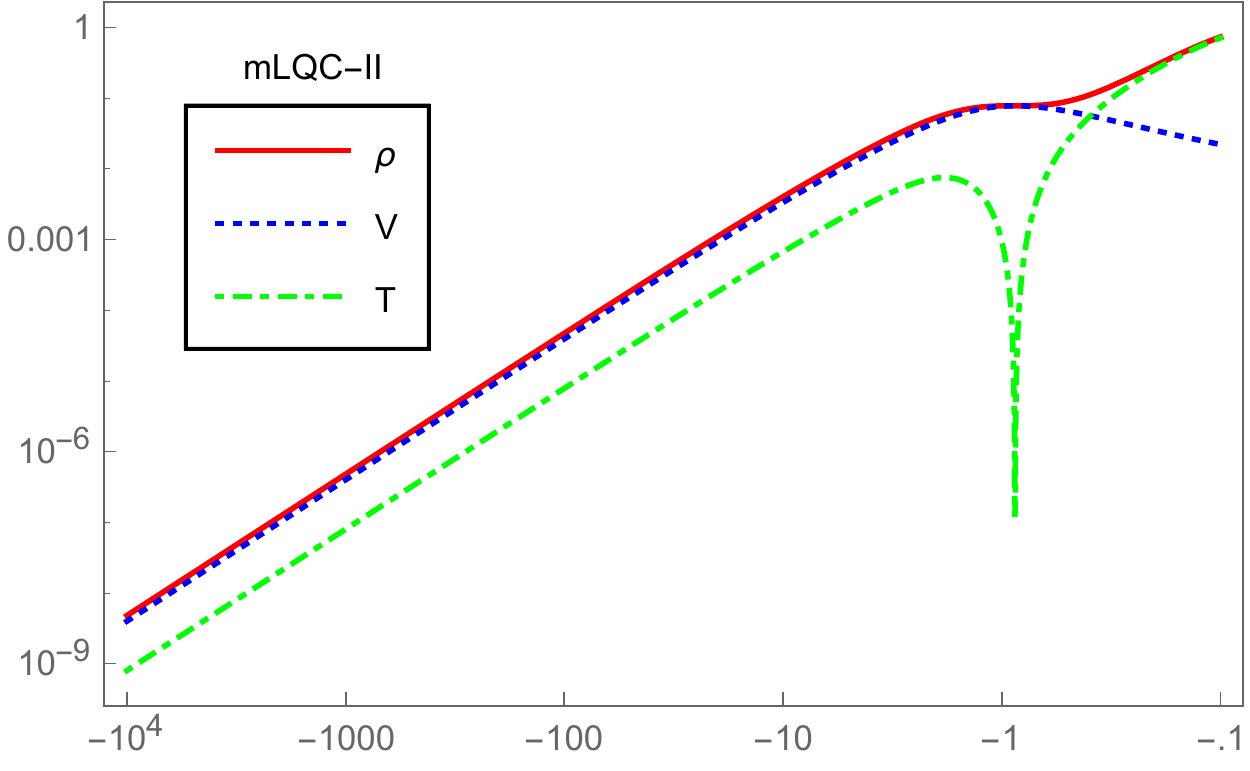}
}
\caption{Evolution of the energy density, kinetic energy $(T)$ and potential energy $(V)$ of the scalar field in LQC, mLQC-I and mLQC-II.  
 Initial condition is chosen at bounce with $\phi_B=0$ which occurs at $t=0$. The steepness of the potential $q$ is set to $1$. Scaling 
solutions can be easily seen from the post-bounce phase of the three models where the potential energy comes into dominance.}
\label{fig24}
\end{figure}

\begin{figure} 
{
\includegraphics[width=5.5cm]{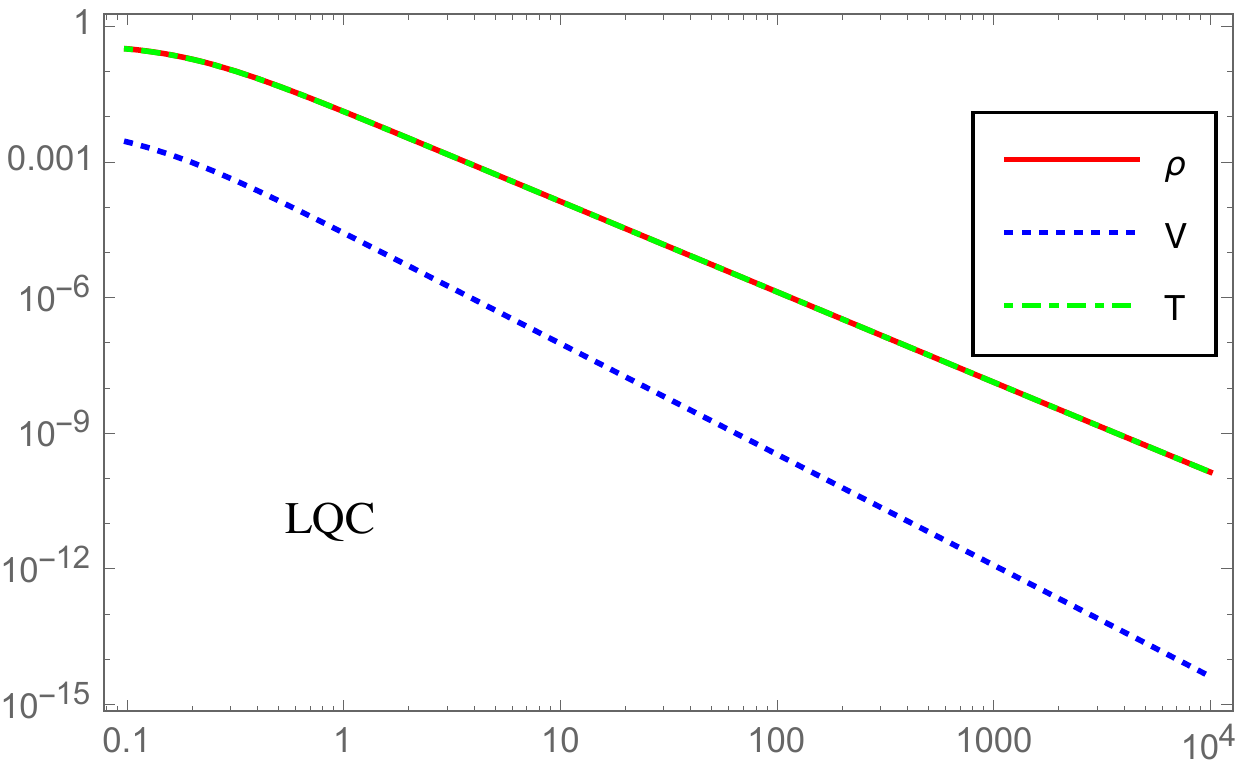}
\includegraphics[width=5.5cm]{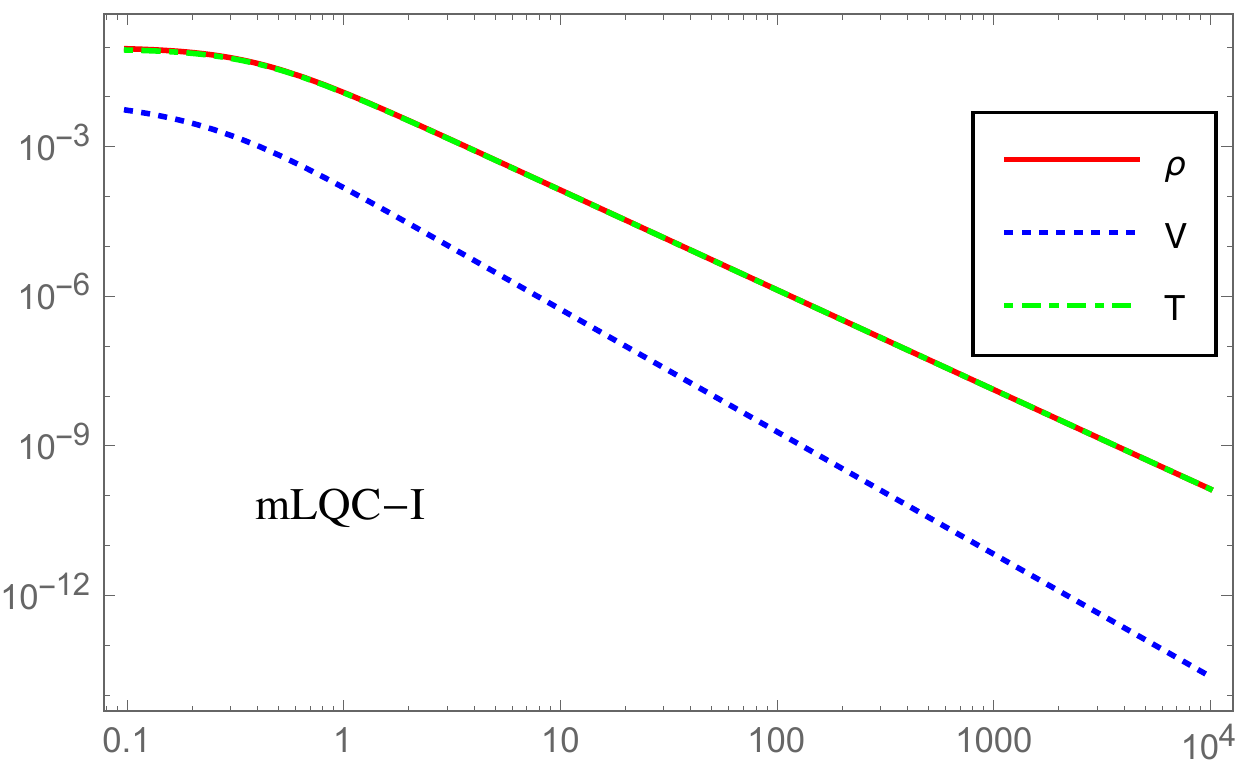}
\includegraphics[width=5.5cm]{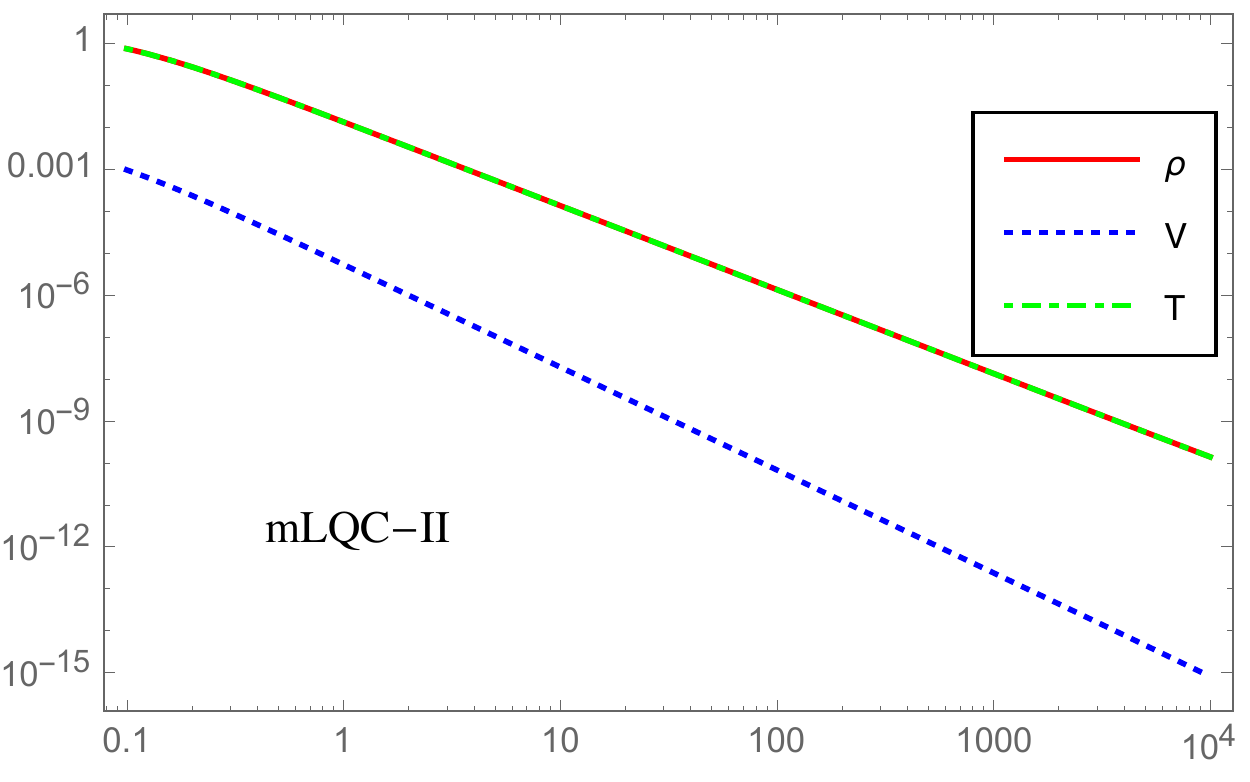}\\
\includegraphics[width=5.5cm]{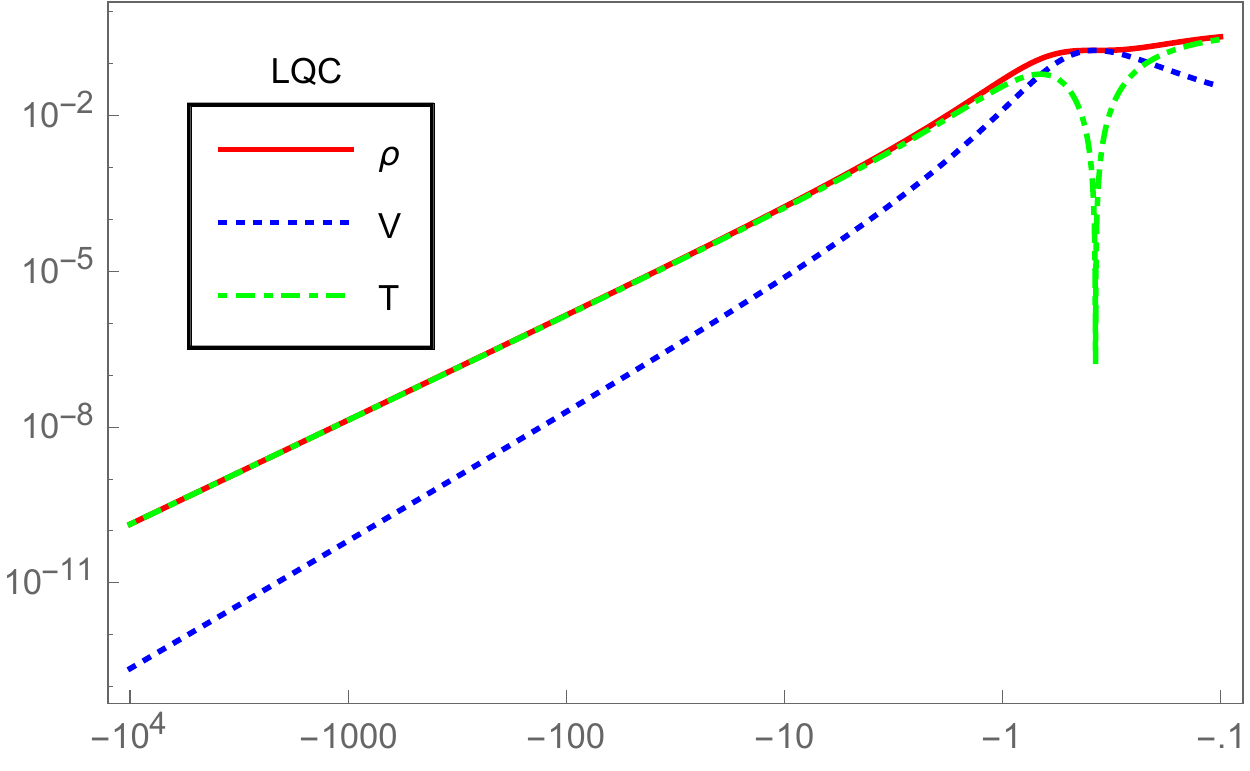}
\includegraphics[width=5.5cm]{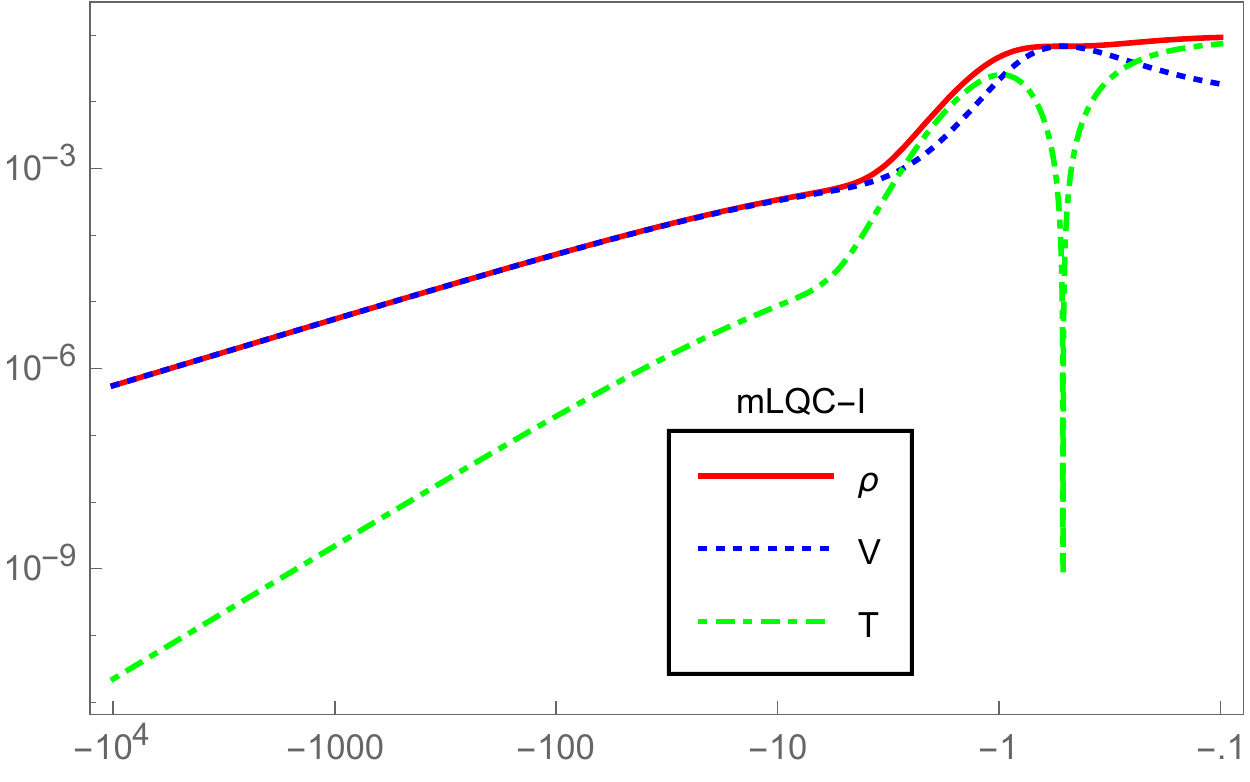}
\includegraphics[width=5.5cm]{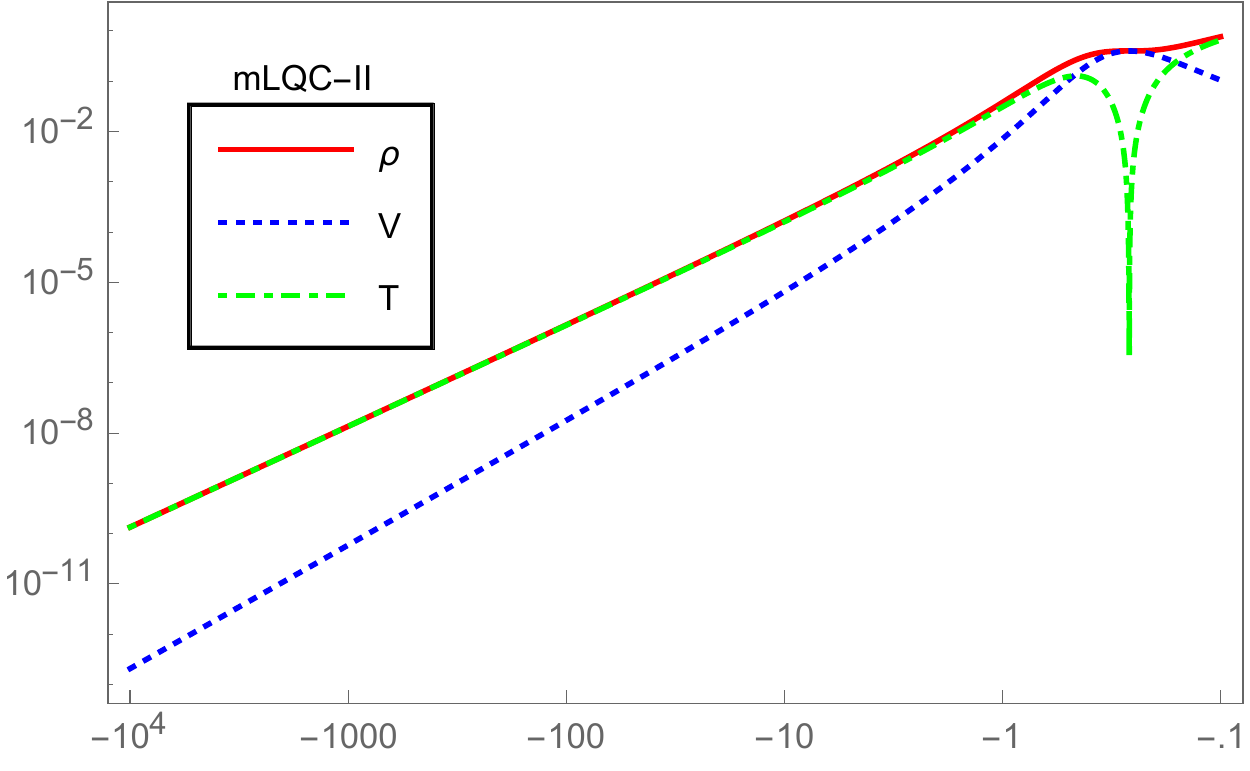}
}
\caption{With the same initial condition as in Fig. \ref{fig24}, $q$ is set to $3$ in the figures. Only the pre-bounce phase of mLQC-I is dominated by the potential energy, all of the other regimes are dominated by the kinetic energy.}
\label{fig25}
\end{figure}

\end {widetext}

\subsection{Exponential Potential}

In classical theory,  power law inflation  is realized by the exponential potential  given by  \cite{lm1985}, 
\bq
\lb{3.11}
V=V_0 e^{-\sqrt{8\pi G} q \phi},
\eq
where $V_0$ and $q$ are two free parameters characterizing the magnitude and steepness of the potential. The slow-roll parameters can be shown as  $ \epsilon_V=\epsilon_H=\frac{q^2}{2}$. Hence, the slow-roll inflation is possible when $q<\sqrt{2}$. Since the slow-roll condition is determined by a constant $q$, there is no graceful exit unless there is a complementary mechanism.
 Moreover,  the predicted tensor to scalar ratio by a canonical scalar field  with the exponential potential  is larger than constraints from CMB observations \cite{Planck2015}. Although the model is observationally disfavored \footnote{Note that with a non-canonical scalar field, power law inflation can be made observationally viable but in that case the potential ceases to be exponential.}, it is still interesting because of the presence of late time scaling solutions, and capturing the universal behavior of the scale factor at the kinetic dominated bounce \cite{lsw2018c}.

In order to show the existence of the scaling solutions in the exponential potential, two new variables are introduced as follows \cite{clw1998}: 
\bq
\lb{3.12}
X=\frac{\dot \phi}{H}\sqrt{\frac{4\pi G}{3}}\quad \quad \quad Y=\frac{1}{H}\sqrt{\frac{8 \pi G V}{3}}. 
\eq
Here we only consider the positive potential with $V_0>0$.  As a result, the energy density and the pressure of the scalar field can be expressed in terms of the new variables as
\bqn
\lb{3.13}
\rho&=&\frac{3H^2}{8\pi G}\left(X^2+Y^2\right), \\
P&=&\frac{3H^2}{8\pi G}\left(X^2-Y^2\right) ~.
\eqn
 Therefore, the effective equation of state of the scalar field is 
\bq
\omega_\phi=\frac{X^2-Y^2}{X^2+Y^2}.
\eq
On the other hand, if we assume a general form of the Friedmann equation for different models:
\bq
\lb{3.14}
H^2=\frac{8 \pi G}{3}L^i(\rho), 
\eq
where $L^i(\rho)$  is a positively well-defined function of the energy density and $i$ stands for three different models, LQC, mLQC-I and mLQC-II. Using \eqref{3.13} along with the above equation we find that the trajectories of the solutions should satisfy the condition  
\bq
\lb{3.15}
\rho= L^i(\rho) \, \left(X^2+Y^2 \right).
\eq
The Klein-Gordon equation (\ref{3.1}), along with the definitions (\ref{3.2}) can be written into an autonomous system governed by a set of coupled first-order differential equations, given by:\footnote{In contrast to potentials considered so far, we perform stability analysis using in terms of number of e-foldings $N$. This is to assist a comparison with earlier works on exponential potential \cite{hw2002,clw1998}.} 
\bqn
\lb{3.16a}
\frac{d X}{d N}&=&-3 X\left(1-X^2 \partial_\rho L^i\right)+\sqrt{\frac{3}{2}}q Y^2,\\
\lb{3.16b}
\frac{d Y}{d N}&=&XY\left(3X\partial_\rho L^i-\sqrt{\frac{3}{2}}q \right),
\eqn
where  $\partial_\rho $ represents  the differentiation with respect to $\rho$ and $N$ denotes the e-folds defined by $N=\ln(a)$. The system is closed since the energy density is an  implicit function of $X$ and $Y$ due to Eq. (\ref{3.15}). Thus, the fixed points can be deduced by $ X'= Y'=0 $, from which one can find that, based on the values of the equation of state $\omega_\phi$, there are three types of solutions shown as below: 

1.  The origin  $X=Y=0$ with $\omega_\phi$ undefined. This fixed point can only exist in the pre-bounce phase of mLQC-I  where $L^{{\scriptscriptstyle{\mathrm{I}}}}(\rho=0) \neq 0$. Note that due to Eq. (\ref{3.15}), $\rho=0$ as $X=Y=0$ but due to the effective cosmological constant in the pre-bounce branch of mLQC-I, $L^{{\scriptscriptstyle{\mathrm{I}}}}(\rho=0) \neq 0 $. As a result, the solution $X=Y=0$ corresponds to the asymptotic de-Sitter spacetime in the contracting phase of mLQC-I. The eigenvalues at the origin can be found from the linearization of Eqs. (\ref{3.16a})-(\ref{3.16b}) in which the perturbations $X\rightarrow X+\mu$ and $Y\rightarrow Y+\nu$ follow the evolution equations
\bqn
\mu'&=&-3 \mu,\\
\nu'&=&0.
\eqn
Therefore, the eigenvalues are $m_-=0$ and $m_+=-3$. In contracting phase, since $N$ is a decreasing function in the forward evolution,  the negative eigenvalue indicates the point $X=Y=0$ is unstable.

2.   The kinetic-dominated solutions  $X=\pm \left(\partial_\rho L^i\right)^{-1/2}$  and $Y=0$ which results in $\omega_\phi=1$.  
In the classical theory, $L^i=\rho$, and the solutions are simply $X=\pm 1$, $Y=0$. At large volumes, in LQC, mLQC-II and the post-bounce phase of mLQC-I, 
\bq
\lb{3.16d}
L^i=\rho +\mathcal O\left(\rho^2\right),
\eq
which, once plugged into Eq. (\ref{3.15}),  gives 
\bq
1=1+\mathcal O(\rho).
\eq
As a result, at the fixed point, the energy density has to vanish and the kinetic-dominated fixed points are $X=\pm 1, Y=0$ in LQC, mLQC-II and post-bounce phase of mLQC-I (at large volumes). In the pre-bounce phase of mLQC-I, one has to solve Eq. (\ref{3.15}) with $L^i=L^{{\scriptscriptstyle{\mathrm{I}}}}$. But there is no kinetic-dominated solution in this branch, since the dynamics is dictated by an effective cosmological constant.  To analyze the stability of kinetic-dominated solutions in LQC, mLQC-II and post-bounce phase of mLQC-I, one can consider the linear perturbations $(X \rightarrow \pm1+\mu, Y\rightarrow \nu)$ and their evolution equations
\bqn
\mu'&=& 6\mu, \nb\\
\nu'&=&\left(3\mp \sqrt{\frac{3}{2}}q\right)\nu,
\eqn
where a prime denotes a derivative with respect to $N$. As a result, in the expanding phase, the kinetic-dominated solutions are unstable (since $\mu \propto e^{6 N}$) while in the contracting phase, the solution $(1,0)$ is stable when $q<\sqrt{6}$, and the solution $(-1,0)$ is stable, when $q>-\sqrt{6}$.

3. The last but perhaps the most interesting types of fixed points correspond to the kinetic-potential  scaling solutions given by 
\bqn
\lb{scaling}
X&=&\frac{q}{\sqrt{6}\partial_\rho L^i},\nb\\
Y&=&\pm \Big\{\frac{1}{\partial_\rho L^i}\left(1-\frac{q^2}{6\partial_\rho L^i}\right)\Big\}^{1/2},
\eqn
where the plus (minus) sign in $Y$ applies to the expanding (contracting) phase.  Plugging the scaling solutions into Eq. (\ref{3.15}), one can solve for the energy density from 
\bq
\lb{rho}
\rho  \partial_\rho L^i=L^i .
\eq
At large volumes, in LQC, mLQC-II and the post-bounce phase of mLQC-I, $L^i$ assumes the form $L^i= \rho +\mathcal O(\rho^2)$, the only solution for the energy density from Eq. (\ref{rho}) is $\rho=0$. Therefore, for  $0<q<\sqrt{6}$,  the scaling solutions in these cases are 
\bq
\lb{scaling1}
 X=\frac{q}{\sqrt{6}}, \quad Y=\pm \left( 1-\frac{q^2}{6}\right)^{1/2} .
\eq
 These are the same scaling solutions which  exist in the classical theory. In particular, they are termed as scalar field dominated solution in \cite{clw1998} and potential-kinetic-scaling solution in \cite{hw2002}. As in the case of kinetic-dominated solutions, kinetic-potential scaling solutions do not exist in the pre-bounce phase of mLQC-I.

Considering linear perturbations  $X\rightarrow X+\mu$ and $Y\rightarrow Y+\nu$  about the scaling solutions (\ref{scaling1}), we get
\bqn
\mu'&=&\left(\frac{3q^2}{2}-3\right)\mu \pm \sqrt{6}q\sqrt{1-\frac{q^2}{6}}\nu,\nb\\ 
\nu'&=&\pm \frac{3q}{\sqrt{6}}\sqrt{1-\frac{q^2}{6}}\mu,
\eqn
where $\pm$ sign denotes the expanding/contracting phase.  As a result, the eigenvalues of the Jacobian matrix for the scaling solutions are:
\bq
m_+=q^2 \quad  \text{and} \quad m_-=\left(\frac{q^2}{2}-3\right).
\eq
Therefore, $m_-$ is always negative as long as the scaling solution exists.  This identifies the scaling solutions as the saddle points of the dynamical system. 

We now discuss the phase portraits and scaling solutions for this potential. Noting that at the bounce the definition in Eq. (\ref{3.12})  breaks down, in the phase portraits 
we employ another set of variables 
\bq
\lb{3.18}
X= \sqrt{\frac{V_0}{\rho^i_c}}e^{-\sqrt{2\pi G}q \phi},\quad \quad \quad Y=\frac{\dot \phi}{\sqrt{2\rho^i_c}},
\eq
so that all the trajectories are confined within the half circle as evident in Fig. \ref{fig22}.  We have chosen two sets of parameters to show the way steepness of the potential affects phase space portraits. We first discuss the case of $q=1$. Since $q < \sqrt{6}$, kinetic-potential scaling solutions exist in this case. From Fig. \ref{fig22} we find that in all three models, two separatrices are present. Using Eq.(\ref{scaling1}), it is straightforward to see that the equation of state for scaling solutions in this case corresponds to $\omega_\phi= \left(\frac{q^2}{3}-1\right) \approx -2/3$. This is confirmed by Fig. \ref{fig23}, where the equation of state approximates this value in the post-bounce regime for all the models, and in the pre-bounce regime for LQC and  mLQC-II. For the pre-bounce stage of mLQC-I, the equation of state approaches $-1$ and this indicates the dotted separatrix in the top middle panel of Fig. \ref{fig22} is a de-Sitter spacetime.  Earlier we showed that this corresponds to  the fixed point $(0,0)$ which is unstable. Kinetic-potential scaling solutions in the post-bounce phase are clearly visible for three models as shown in Fig. \ref{fig24}. It can be clearly seen that in the post-bounce phase of all three models, scaling solutions are dominated by the potential energy. Further, scaling solutions are also present in the contracting phase of LQC and mLQC-II as evident in the bottom panels of Fig.\ref{fig24}, where the kinetic energy  is about ten times larger than the potential energy. The equation of state again approximately equals $-2/3$ which confirms the scaling solutions. However, in mLQC-I, it is the de-Sitter phase that occupies the pre-bounce regime as the potential energy is about $10^5$ times larger than the kinetic energy in most of the evolution which leads to effective equation of state as  $\omega \approx -1$.

For the case of $q=3$, kinetic-potential scaling solutions are absent in all the models. The contrast of the phase space portraits can be seen from Fig. \ref{fig22}. We see that in LQC and mLQC-II, all the trajectories start and end  at the origin after the bounce. In the case of mLQC-I, a short separatrix in dotted line can still be observed near the origin. This corresponds to the de Sitter spacetime in the pre-bounce epoch of mLQC-I, which is confirmed by the equation of state plot (Fig. \ref{fig23}). Moreover, in the same figure, at very early (pre-bounce) and late (post-bounce) times, solutions in LQC and mLQC-II tend to be the kinetic-dominated solutions. Such solutions can be seen in Fig. \ref{fig25}, in which except the pre-bounce phase of mLQC-I, all the other regimes in the three models  are dominated by the kinetic energy. This observation is  consistent with Fig. \ref{fig23}, in which the blue dotted line tends to $\omega_\phi=-1$ at early times when $q=3$, while both green and red lines are approaching $\omega_\phi=1$.

To conclude our discussions on qualitative dynamics, we present Table. \ref{table1}-\ref{table2} in which all the fixed points and their properties discussed above in LQC, mLQC-I and mLQC-II are summarized. 

\begin{widetext}

\begin{table}[h!]
\renewcommand{\tabularxcolumn}[1]{m{#1}}
\centering
\caption{We summarize all the fixed points in the chaotic, fractional monodromy, Starobinsky  and non-minimal Higgs potentials in LQC, mLQC-I and mLQC-II. }
\label{table1}
\begin{tabularx}{\textwidth}{|c|*{6}{Y|}}
\toprule
\hline
\multirow{2}{*}{\text{Potentials}}
 & \multicolumn{3}{c|}{\text{LQC/mLQC-II} }  
 & \multicolumn{3}{c|}{\text{mLQC-I}} \\ \cline{2-7} 
\cmidrule(rl){2-4} \cmidrule(l){5-7}
  \midrule   
  & \text{Fixed Points} & \text{Region}  & \text{Stability}  & \text{Fixed Points}  &  \text{Region} &  \text{Stability}  \\ \hline
\multirow{3}{*}{Chaotic} & origin & pre-bounce & unstable spiral & origin & pre-bounce & \small{unstable node for $m \leq 0.62 m_{\rm{Pl}}$, unstable spiral for $m > 0.62  m_{\rm{Pl}}$}   \\ \cline{2-7} 
   &origin & post-bounce & stable spiral & origin & post-bounce & stable spiral        \\ \hline     
\multirow{3}{*}{Fractional Monodromy} & origin & pre-bounce & unstable spiral & origin & pre-bounce & \small{unstable node for $\phi_0 \geq 0.59 m_{\rm{Pl}}$, unstable spiral for $\phi_0 < 0.59  m_{\rm{Pl}}$}   \\ \cline{2-7} 
   &origin & post-bounce & stable spiral & origin & post-bounce & stable spiral        \\ \hline     
\multirow{5}{*}{Starobinsky}      & origin & pre-bounce & unstable spiral & origin & pre-bounce &  \small{unstable node for $m \leq 0.62 m_{\rm{Pl}}$, unstable spiral for $m > 0.62  m_{\rm{Pl}}$}    \\ \cline{2-7} 
                   &origin & post-bounce & stable spiral &origin & post-bounce & stable spiral      \\  \cline{2-7}          
                              & $(\chi_0,0)$ & pre/post-bounce & unstable non-simple fixed point & $(\chi_0,0)$ & pre/post-bounce & unstable non-simple fixed point \\ \hline                                 
\multirow{5}{*}{Non-minimal Higgs}& origin & pre-bounce & unstable spiral & origin & pre-bounce &  unstable node for $\hat V \leq 0.012$, unstable spiral for  $\hat V > 0.012$   \\ \cline{2-7} 
                      &origin & post-bounce & stable spiral &origin & post-bounce & stable spiral       \\  \cline{2-7}    
                                     & $(\pm\hat\chi_0,0)$ & pre/post-bounce &unstable  non-simple fixed point& $(\pm\hat\chi_0,0)$ & pre/post-bounce & unstable non-simple fixed point \\ \hline      
                          
\bottomrule
\end{tabularx}
\end{table}

\begin{table}[h!]
\centering
\caption{All the fixed points are listed in the table for exponential potential in LQC, mLQC-I and mLQC-II.  The fixed point is labeled in its coordinates (X, Y). }
\label{table2}
\begin{tabularx}{\textwidth}{|c|*{6}{Y|}}
\toprule
\hline
 \multicolumn{3}{|c|}{\text{LQC/mLQC-II} }  
 & \multicolumn{3}{c|}{\text{mLQC-I}}
\\ \cline{1-6} 
\cmidrule(l){1-3} \cmidrule(l){4-6}
  \midrule   
   \text{Fixed Points} & \text{Region}  & \text{Stability}  & \text{Fixed Points}  &  \text{Region} &  \text{Stability}  \\ \hline
\multirow{2}{*}{ $(1, 0)$} & \multirow{2}{*}{pre-bounce} & stable node for $q<\sqrt{6}$ and saddle point for $q>\sqrt{6}$ & \multirow{2}{*}{$(0, 0)$} & \multirow{2}{*}{pre-bounce} & \multirow{2}{*}{unstable node} \\  \hline 
 \multirow{2}{*}{$  (1, 0) $} & \multirow{2}{*}{post-bounce} &  unstable node for $q<\sqrt{6}$ and saddle point for $q>\sqrt{6}$ & \multirow{2}{*}{$(1, 0)$} & \multirow{2}{*}{post-bounce} & unstable node for $q<\sqrt{6}$ and saddle point for $q>\sqrt{6}$      \\ \hline                   
\multirow{2}{*}{ $ (-1, 0) $} & \multirow{2}{*}{pre-bounce} &  saddle point for $q<-\sqrt{6}$ and stable node for  $q>-\sqrt{6}$  & \multirow{2}{*}{N/A} & \multirow{2}{*}{N/A} &\multirow{2}{*}{N/A}  \\  \hline
 \multirow{2}{*}{ $(-1, 0) $} &\multirow{2}{*}{ post-bounce} & saddle point for $q<-\sqrt{6}$ and unstable node for  $q>-\sqrt{6}$  & \multirow{2}{*}{$(-1, 0)$} & \multirow{2}{*}{post-bounce} & saddle point for $q<-\sqrt{6}$ and unstable node for  $q>-\sqrt{6}$    \\   \hline         
                             $\Big(\frac{q}{\sqrt{6}},  \pm \sqrt{1-\frac{q^2}{6}}\Big)$ & post/pre-bounce & saddle point &   $\Big(\frac{q}{\sqrt{6}}, \sqrt{1-\frac{q^2}{6}}\Big)$ & post-bounce & saddle point \\  \hline

\bottomrule
\end{tabularx}
\end{table}

\end{widetext}

\section{Summary}
\renewcommand{\theequation}{6.\arabic{equation}}\setcounter{equation}{0}

In the quantization procedure, several ambiguities can modify the form of effective Hamiltonian in loop cosmologies. The main goal of this paper was to understand the physical implications for treating Lorentzian term in isotropic spacetimes in loop cosmology in detail using dynamical system analysis. We used three models, LQC \cite{asrev}, mLQC-I \cite{YDM09,DL17} and mLQC-II \cite{YDM09}. The modified dynamics of LQC has been extensively studied, and recently modified FR equations in mLQC-I were derived \cite{lsw2018}. Similar equations for mLQC-II were not known so far.  In this paper, we have first derived the FR equations in mLQC-II and then found that the evolution of the universe is symmetric with respect to the quantum bounce, although the critical density is  larger by a factor of $4\gamma^2+4$ than that of LQC. Though some of the properties of effective dynamics for this model were first studied in Ref. \cite{YDM09}, we have derived for the first time modified FR equations and discussed in detail comparison with LQC and mLQC-I.

With details of effective dynamics of three models available,  we have studied  general properties of  the  evolution of the universe 
driven by a single scalar field  for several inflationary potentials. They include the chaotic potential, the fractional monodromy 
potential, the non-minimal  Higgs potential, the Starobinsky potential, and the exponential potential.  Among them, the first three 
potentials are symmetric about  $\phi=0$, while the last two are not. As a result, inflation can occur for positive as well as negative 
values of $\phi$  for chaotic, fractional monodromy and non-minimal Higgs potentials. While in the 
Starobinsky potential, only positive part of the $\phi$ axis is able to drive inflation. Correspondingly, in the phase portraits of chaotic, 
fractional monodromy and non-minimal Higgs potentials, there are two slow-roll inflationary separatrices in the pre- and post-bounce stages 
for LQC and mLQC-II, while there only exists one inflationary separatrix in the phase portrait of the Starobinsky potential. In the 
exponential potential, power law inflation is just a special case of the scaling solutions when the steepness parameter $q < \sqrt{2}$.  In 
the post-bounce phase, all the three models discussed here result in an inflationary attractor starting from the bounce. One thus achieves 
a non-singular dynamical evolution of inflationary spacetime thanks to quantum gravitational effects of LQG. Starting from the initial data 
set at the bounce, the scalar field continues to grow to some values until the slow-roll inflationary parameter drops below unity. After a 
short super-inflationary phase, slow-roll inflation is triggered and driven by the nearly constant potential. Near the end of inflation, the 
potential drops quickly towards its minimal value and keeps oscillating around it in the reheating phase, except for the exponential 
potential which requires a separate mechanism for graceful exit. In the phase portraits, for all potentials except the exponential one, this 
process is reflected by the fact that regardless of the nature of inflationary potentials, the origin is always a stable spiral (a late-time 
attractor) in the post-bounce phase in all the three models. The spiral structure of the separatrices, when they are moving toward the 
origin, reflects the oscillations of the scalar field during reheating. 

In LQC and mLQC-II, the evolution of the universe is symmetric about the quantum bounce. The notable difference lies in mLQC-I which due to asymmetric evolution and a quantum geometric cosmological constant in the pre-bounce stage does not allow inflation sourced by the inflationary potential in the pre-bounce phase, but dynamics approximates that of a de Sitter spacetime.   
Besides, regardless of the nature of inflationary potentials, in the contracting phase, there always exists an unstable spiral at the center of the phase portraits for LQC and mLQC-II. In particular, the spiral structure in LQC and mLQC-II can be accounted for by the imaginary eigenvalues of the linear perturbations. On the contrary, in the contracting phase of mLQC-I, depending on the magnitude of inflationary potential, there can be either an unstable spiral or an unstable node at the origin, as the characteristic eigenvalues of the evolution equations for linear perturbations about the origin always have a real positive part which makes the perturbations grow at an exponential rate. In the phase portraits, we have carefully chosen the mass and the amplitude of the potential, so that the unique feature of the pre-bounce of mLQC-I is aways highlighted by the unstable node at the center of the figures. In the Starobinsky and non-minimal Higgs potentials,  unstable fixed points show up as the boundaries for real solutions, these points  appear in the extreme situations when the real scalar field blows up.  For these potentials, there exist fixed points which are non-simple, for which we use center manifold theory to understand stability properties. 

Furthermore, analogous to the classical theory, scaling solutions also appear for the exponential potential. Despite the holonomy corrections at the quantum bounce, these scaling solutions in LQC, mLQC-I and mLQC-II have the same properties as in the classical theory: they are the saddle points which only exist for $q^2<6$. Besides the scaling solutions, we have also found one unstable fixed point that only shows up in the contracting phase of mLQC-I and is later identified with the unique asymptotic de-Sitter spacetime. It turns out that no other fixed points can ever exist in the pre-bounce phase of mLQC-I. In the phase space portraits for the exponential potential, the scaling solutions as well as the asymptotic de-Sitter spacetime show themselves in the form of the separatrices toward/from which all the trajectories converge/diverge. The qualitative evolution of generic solutions in the exponential potential depends on the steepness of the potential. In LQC and mLQC-II, for $q^2<6$, generic solutions start from the scaling solutions at early times, as the Universe is in a state of deflation, the energy density becomes larger and larger until the critical energy density is reached, then the bounce occurs which brings the Universe into a state of expansion. Finally, the Universe reaches the same scaling solution as it starts from at late times. When $q^2>6$, the Universe starts from the kinetic-dominated solution in the pre-bounce stage and end up with the same kinetic-dominated solution in the post-bounce stage. In mLQC-I, for any $q$, the Universe starts from the unique de-Sitter spacetime in the contracting phase and arrives at either scaling solutions ($q^2<6$) or kinetic-dominated solutions ($q^2>6$).  

Our results establish various features of qualitative dynamics for LQC, mLQC-I and mLQC-II. On one hand they bring out various similarities between these models in the post-bounce epoch, but on the other hand contrasting differences in the nature if fixed points is revealed between LQC and mLQC-II on one hand, and mLQC-I on the other hand. Occurrence of attractors demonstrated for various different potentials shows that despite quantization ambiguities, inflation is natural in loop cosmology. It will be interesting to understand various details of the pre-inflationary dynamics in these models, which will be addressed in a future work \cite{lsw2018c}.

\section*{Acknowledgements}

A.W. and B.F.L. are supported in part by the National Natural Science Foundation of China (NNSFC) 
with the Grants Nos. 11375153 and 11675145. P.S. is supported by NSF grant PHY-1454832.

\appendix
\section{Stability Analysis of the Equilibrium Points for Planar Nonlinear Systems}
\renewcommand{\theequation}{A.\arabic{equation}} \setcounter{equation}{0}

In this appendix, we outline a specific procedure to determine the stability of the fixed points for a planar system  defined by\footnote{The conclusions here are not restricted to a 2D system but can be generalized to any N-dimensional system.}
\bqn
\lb{app1}
\dot X&=& f(X, Y), \\
\dot Y&= &g(X, Y).
\eqn
Here `dot' denotes derivative with respect to time $t$. 
For simplicity, we can always assume  that there is a fixed point at the origin, that is, $f(0,0)=g(0,0)=0$. The first important object of the system is its Jacobian matrix defined by
\bq
 J=
\left|\begin{array}{cc}
     \partial_x f     &  \partial_y f \\ 
    \partial_x g   & \partial_y g
\end{array}\right|,
\eq
from  which the eigenvalues  at the origin can be computed. Once  eigenvalues are known,  the stability of the origin can be determined according to the following criteria:

a). If the real parts of all the eigenvalues are negative, then the origin is locally stable.

b). If the real part of at least one of the eigenvalues is positive, then the origin is locally unstable. 

One should notice that a particular situation which is called the critical problem of stability remains unaddressed in the above two statements. This problem arises when at least one of the eigenvalues has a zero real part. Then,  the linearization  of nonlinear system is not enough to determine the stability of the fixed points as it is the higher-order terms that ultimately decide the behavior of the solutions around the equilibrium. Generally speaking, there are two candidates at hand to deal with the critical problem, the center manifold theory and the Lyapunov theorem.   The center manifold theory is tailored to the situation when some of the eigenvalues have zero real parts and the rest have negative real parts. 
 In our first example, this theory is applied to the fixed point $(\chi_0, 0)$ in Starobinsky potential. On the other hand, the Lyapunov theorem is more general, it is suited for both critical and non-critical problems. In our second example, we apply this theorem to the origin in the chaotic potential as both of the eigenvalues at the origin are purely imaginary.

When determining the stability of the fixed points by using center manifold theory,  one needs to find out the local center manifold at the fixed point and then reduce the critical problem  to a low-dimensional system. In a 2D planar system,  reduction to a one-dimensional system is usually straightforward as it only concerns simple algebra. However, in higher dimensional nonlinear systems, things can become complicated and subtle \cite{ar2001}. To illustrate the basic steps, we would like to take for an example the system governed by Eqs. (\ref{3.20a})-(\ref{3.20b}). First shifting the fixed point $(\chi_0, 0)$ to the origin by redefining  $ X\rightarrow X-\chi_0$, thus the system is turned into
\bqn
\lb{app3}
\dot { X}&=&-\frac{mX Y}{\chi_0}, \\
\lb{app4}
\dot Y&=& m  X-3 H^i Y+\frac{mX^2}{\chi_0}.
\eqn
Correspondingly, the Jacobian matrix at the origin is  given by 
\bq
\lb{app2}
J=
\left[\begin{array}{cc}
    0  & 0\\ 
    m   & -3 \tilde H^i 
\end{array}\right],
\eq
where $\tilde H^i$ is given in Eq. (\ref{starc1}). Now one is required to work in the special coordinates in which the Jacobian matrix (\ref{app2})  is diagonalized. With the help of  transformation matrix 
\bq
Q=
\left[\begin{array}{cc}
    1  & 0\\ 
    -m/(3 \tilde H^i)   & 1
\end{array}\right],
\eq
 the special coordinates $(\mu, \nu)$  can be shown as 
 \bqn
 \mu&=&X, \\
 \nu&=&Y-m \frac{X}{3 \tilde H^i}.
 \eqn
Then in terms of $\mu$ and $\nu$, the autonomous system (\ref{app3})-(\ref{app4}) are turned into its standard form
\bqn
\lb{app5}
\dot {\mu}&=&-\frac{m^2\mu^2}{3\tilde H^i \chi_0}-\frac{m\mu\nu}{\chi_0}, \\
\lb{app6}
\dot \nu&=&-3 H^i \nu+\left(1-\frac{H^i}{\tilde H^i}\right)m \nu+\frac{m\mu^2}{\chi_0}\nb\\
  &&+\frac{m^3 \mu^2}{9 \tilde {H^i}^2 \chi_0}+\frac{m^2 \mu \nu}{3 \tilde H^i\chi_0}.
\eqn
At this point, one is able to define local center manifold at the origin via 
\bq
\lb{app7}
\nu=\phi(\mu),
\eq
with the necessary  conditions $\phi(0)=\phi'(0)=0$. Near the origin, all the solutions with initial conditions on the manifold (\ref{app7}) would stay on the manifold for at least a finite period of time. As a result, once we plug Eqs. (\ref{app7}) and (\ref{app5}) into Eq. (\ref{app6}),  Eq. (\ref{app6}) can be transformed into a nonlinear equation for $\phi(\mu)$.  Although, on most occasions, the exact form of $\phi(\mu)$ is far from available since Eq. (\ref{app6}) is highly nonlinear, we still can find an approximate solution to $\phi$ by Taylor expansion 
\bq
\lb{app9}
\phi= a \mu^2+b \mu^3+c \mu^4+\mathcal O (\mu^5). 
\eq
Once coefficients  $a$, $b$ and $c$ are  determined order by order from Eq. (\ref{app6}), we can plug Eq. (\ref{app7}) into (\ref{app5}) and obtain a reduced system in one dimension, which is 
\bq
\lb{app8}
\dot {\mu}=-\frac{m^2\mu^2}{3\tilde H^i \chi_0}-\frac{m\mu\phi(\mu)}{\chi_0}.
\eq
The stability of the origin in this reduced system is identical to the stability of the origin in the original system. Thus, in the case of the Starobinsky potential, we do not  even need to know the coefficients in the expansion (\ref{app9}) as the lowest order term in Eq. (\ref{app8}) is $\mu$ squared term. Therefore, near the origin,  the essential part to determine the stability is 
\bq
\lb{app10}
\dot {\mu}=-\frac{m^2\mu^2}{3\tilde H^i \chi_0},
\eq
from which we immediately know $\mu=0, \nu=0$ is not stable and consequently $(\chi_0,0)$ is not stable in the expanding phase of LQC, mLQC-I and mLQC-II. \\

The second example we would like to deal with is the origin of the system defined in Eqs. (\ref{3.8})-(\ref{3.9}). To show that the origin is indeed stable in the expanding phase,  one need appeal to  the Lyapunov theorem \cite{william} which states:   

\text{\bf Theorem}: If there is a differentiable function $\mathcal V$ in the neighborhood of $z=0$ (the equilibrium point) denoted by $\mathcal D$ such that $\mathcal V(0)=0$, then the origin is asymptotically stable if $ \mathcal V(z)>0$ and  $\dot{\mathcal V}<0$ for all nonzero $z$ in $\mathcal D$. 

In the theorem, $z$ denotes collectively all the independent variables of the dynamical system and $\dot{\mathcal V}$ is defined by $\dot{\mathcal V}=   \dot z \partial_z \mathcal V$. In the case of  Eqs. (\ref{3.8})-(\ref{3.9}), one can simply choose $\mathcal V(z)$ as proportional to the energy density of the scalar field, that is 
\bq
\mathcal V(z)=X^2+Y^2,
\eq
then one can immediately  find 
\bqn
\dot{\mathcal V}=\dot X \partial_X \mathcal V+\dot Y \partial_Y \mathcal V=- 6 H^i Y^2.
\eqn
As a result, the origin is asymptotically stable in the expanding phase where $H^i>0$. Since $\mathcal V$ is the energy function of the system, $\dot{\mathcal V}<0$ along all the trajectories near the origin indicates the system is dissipative.

\end{document}